\documentclass[fleqn,usenatbib]{mnras}

\usepackage{newtxtext,newtxmath}

\usepackage[T1]{fontenc}

\DeclareRobustCommand{\VAN}[3]{#2}
\let\VANthebibliography\thebibliography
\def\thebibliography{\DeclareRobustCommand{\VAN}[3]{##3}\VANthebibliography}

\usepackage[tight,footnotesize]{subfigure}
\usepackage{color}
\usepackage{graphicx}	
\usepackage{amsmath}	

\usepackage{bm}
\usepackage{bbm}
\usepackage{latexsym}
\usepackage{gensymb} 

\usepackage{xfrac}
\usepackage{calc}
\usepackage{longtable}
\usepackage{dashrule}
\usepackage{dcolumn}
\usepackage{multirow}%
\usepackage{rotating}
\usepackage{adjustbox}
\usepackage[graphicx]{realboxes}
\usepackage{pdflscape}

\usepackage{float}
\usepackage{perpage}
\usepackage{morefloats}

\usepackage[dvipsnames]{xcolor}
\usepackage{colortbl}
\definecolor{Gray}{gray}{0.85}
\definecolor{LightRed}{rgb}{1,0.80,0.79}
\definecolor{LightCyan}{rgb}{0.88,1,1}
\definecolor{LightBlue}{rgb}{0.667, 0.847, 1}
\definecolor{LightGreen}{rgb}{0.564705882, 0.933333333, 0.564705882}

\makeatletter
\DeclareRobustCommand{\textsupsub}[2]{{%
		\m@th\ensuremath{%
			^{\mbox{\fontsize\sf@size\z@#1}}%
			_{\mbox{\fontsize\sf@size\z@#2}}%
		}%
}}
\makeatother

\title[NSs in the DM clumps]{Neutron star mass in dark matter clumps}

\author[M. Deliyergiyev et al.]{
	Maksym Deliyergiyev,$^{1,2}$\thanks{E-mail: maksym.deliyergiyev@hlrs.de}
	Antonino Del Popolo,$^{3,4,5}$\thanks{E-mail: adelpopolo@oact.inaf.it}
	Morgan Le Delliou,$^{6,7,8,9}$\thanks{E-mail: delliou@lzu.edu.cn}
	\\
	$^{1}$Department of Nuclear Particle Physics, University of Geneva, CH-1211, Switzerland\\
	$^{2}$High Performance Computing Center Stuttgart (HLRS), Universit\"{a}t Stuttgart, 70550 Stuttgart, Germany\\
	$^{3}$Dipartimento di Fisica e Astronomia, University Of Catania,
	Viale Andrea Doria 6, 95125, Catania, Italy\\
	$^{4}$ Institute of Astronomy and National Astronomical Observatory, 
	Bulgarian Academy of Sciences, 72, Tsarigradsko Shosse Blvd., 1784 Sofia, Bulgaria\\
	$^{5}$ Institute of Astronomy, Russian Academy of Sciences, Pyatnitskaya street, 48, 119017 Moscow, Russia\\	
	$^{6}$Institute of Theoretical Physics \& Research Center of Gravitation, Lanzhou University, Lanzhou 730000, China\\
	$^{7}$Key Laboratory of Quantum Theory and Applications of MoE, Lanzhou University, Lanzhou 730000, China\\
	$^{8}$Lanzhou Center for Theoretical Physics \& Key Laboratory of Theoretical Physics of Gansu Province, Lanzhou University, Lanzhou 730000, China\\ 	
	$^{9}$Instituto de Astrof\'isica e Ci\^encias do Espa\c co, Universidade de Lisboa,
	Faculdade de Ci\^encias, Ed.~C8, Campo Grande, 1769-016 Lisboa, Portugal\\
}

\date{Accepted 18 June 2024. Correction to 10.1093/mnras/stad3311 (09 November 2023)}

\pubyear{2024}

\begin{document}
	\label{firstpage}
	\pagerange{\pageref{firstpage}--\pageref{lastpage}}
	\maketitle
	
\begin{abstract}
	This paper investigates a hypothesis proposed in previous research relating neutron star (NS) mass and its dark matter (DM) accumulation. As DM accumulates, NS mass decreases, predicting lower NS masses toward the Galactic center. Due to limited NSs data near the galactic center, we examine NSs located within DM clumps. Using the \texttt{CLUMPY} code simulations, we determine the DM clumps distribution, with masses from 10 to $10^{8}$ $M_{\odot}$ and scales from $10^{-3}$ to 10 kpc. These clumps' DM exhibit a peak at the center, tapering toward the outskirts, resembling our Galaxy's DM distribution. 			
	We analyse these DM clumps' NS mass variations, considering diverse DM particle masses and galaxy types. We find relatively stable NS mass within 0.01 to 5 kpc from the clump center. This stability supports the initial hypothesis, particularly for NSs located beyond 0.01 kpc from the clump center, where NS mass reaches a plateau around 0.1 kpc. Nevertheless, NS mass near the clump's periphery reveals spatial dependence: NS position within DM clumps influences its mass in Milky Way-type galaxies. Moreover, this dependence varies with the DM model considered. 			
	In summary, our study investigates the proposed link between NS mass and DM accumulation by examining NSs within DM clumps. While NS mass remains stable at certain distances from the clump center, spatial dependencies arise near the clump's outer regions, contingent on the specific DM model.		
\end{abstract}
	
\begin{keywords}
neutron stars -- dark matter -- sky map -- dark matter halo -- clumps -- microhaloes -- minihaloes
\end{keywords}
	
\section{Introduction}
	
Dark matter (DM) is one of the main pillars of the current cosmological models to explain structure formation without modifying gravity \footnote{
\textcolor{blue}{In our original paper, which examined the relationship between neutron star (NS) mass and dark matter (DM) accumulation within DM clumps \citep{Deliyergiyev:2023uer}, we have identified an error in our code computing 3d-separation distances between NSs and DM clumps: while we had assumed that the distances for all sets of NSs were measured with respect to the Galactic center (GC), we realized they were given with respect to the Sun, namely using the stellar parallax method.}
	
\textcolor{blue}{In this erratum, we convert the reported distance values from the Sun to NSs to values from the GC to NSs, assuming that $Sgr A^{\star}$ is located in a nearly stationary position at the dynamical center of the Galaxy with the Galactic longitude $l = 359.944\degree$ and latitude $b = -0.046\degree$ \citep{Reid:2004rd}. Furthermore, we adopt the distance from the Sun to the Galactic center as $8.122 \pm 0.031$ kpc \citep{GRAVITY:2018ofz}.}
	
\textcolor{blue}{The resulting plots have changed, although the results and conclusions are qualitatively similar, as we discuss in the following. In Fig.~\ref{fig:GalaxyMap_NS} (top) we show a sky map of the examined NS with the color coded distances, while the size of the circle markers represents the NSs masses  \cite[see Fig.~2 of our original paper][]{Deliyergiyev:2023uer}. The bottom panel describes the average mass distribution of the NSs as a function of the distance from the Galactic center.}
}. 
Such an introduction induces well documented gravitational effects on structures \citep{SDSS:2014iwm,Planck:2013pxb}, however this dominant part of matter continues to elude detection of its constituting particles, whether by direct detection, through accelerators or nuclear recoil experiments \citep{DAMA:2010gpn, CoGeNT:2010ols,CMS:2012ucb,SuperCDMS:2013eoh,XENON100:2012itz,XENONCollaboration:2022kmb,CDEX:2022kcd,CMS:2020krr, CMS:2021dzg, ATLAS:2022bzt}, or by indirect searches, through scrutinizing WIMP annihilation detection \citep{Conrad:2014tla,Armand:2022sjf,PerezdelosHeros:2020qyt}, effects on DM stars \citep{Dai:2009ik,Kouvaris:2015rea} or through other indirect effects such as proposed in \citep{Bertolami:2007zm,
		LeDelliou:2007am,
		LeDelliou:2014fto,
		LeDelliou:2018vua,
		Bertolami:2008rz,
		Bertolami:2007tq,
		Bertolami:2012yp,
		Abdalla:2007rd,
		Abdalla:2009mt}. 
Given the uncertainty about the fundamental properties of DM, experimental and observational efforts must be multifaceted, and many are already probing mass and cross-section regimes relevant for theoretical models \citep[see, e.g., the recent reports]{LUX:2016ggv,PandaX-II:2016vec,XENON100:2016sjq,Fermi-LAT:2015att}.

~\\
Despite all efforts, to date, every attempt to directly detect DM through laboratory experiments on the Earth have come up empty. 
Direct detection experiments attempting to observe DM assume it is evenly distributed within each galaxy, however, many theoretical models propose that it may be lumped together in clusters the size of a regular solar system \citep{10.1142:97898127018480012,Alonso-Alvarez:2018tus}. 
At galactic levels, this would not be an issue (for instance, the galaxy rotation curves will remain unchanged), but on smaller scales, it could make a significant difference. 
	
One possibility for this discrepancy is alterations to the basic DM halo model -- the potential clumping of DM, which can drastically affect the expected signal. 
Although astrophysical uncertainties usually have a small influence on the typical direct detection experiments, they are more substantial for annual modulation searches \citep{Green:2017odb}. Clumping of DM would lead to a smaller number of expected interactions, thus reducing the amount of energy released by DM particles \citep{Berezinsky:2014wya}. The recent experimental estimation of the clumpiness in the current universe assuming the standard cosmological model anchored to the early universe was made in \citep{Hildebrandt:2016iqg,DES:2021wwk, PhysRevD.107.023531}.

This issue welcomes proposals for alternate testing strategies of DM effects. Constraints on DM can be obtained from stars that accrete DM during their lifetime and then collapse into a compact star, or into a black hole \citep{Kouvaris:2010jy,Bertone:2007ae,Kouvaris2008,McDermott:2011jp,Kouvaris:2011fi,Kouvaris:2015rea,Kouvaris:2016ltf}, inheriting the accumulated DM \citep{Kouvaris:2010jy}. Moreover, the cooling process of compact objects can be affected by the capture of DM, which subsequently annihilates \citep{Kouvaris2008,Bertone:2007ae,Kouvaris:2010vv,McCullough:2010ai,deLavallaz:2010wp,Sedrakian:2018kdm,Bhat:2019tnz}, releasing energy \citep{Pieri:2007ir}. 
At the same time, self-annihilating DM accreted onto neutron stars (NSs) may change significantly their kinematical properties \citep{PerezGarcia:2011hh} or provide a mechanism to seed compact objects with long-lived lumps of strangelets \citep{PerezGarcia:2010ap}, in increasingly DM rich environments, should accrete more DM and thus display the characteristic mass decrease, the closer to the galactic centre \citep{DelPopolo:2019nng,DelPopolo:2019rox,DelPopolo:2020hel}.

In last years several authors realized that admixing mirror matter with NS matter \citep{Sandin:2008db}, degenerate DM \citep{Leung:2011zz}, Asymmetric DM \citep[ADM][]{Li:2012ii, Karkevandi:2021ygv}, 
increasing the ratio of DM to normal matter, the stars had smaller radii and masses. 
As seen in Ref. \citep{Ciarcelluti:2010ji}'s Fig. 1 for the $M_{DM}-R$ relation, a DM ratio (DM over the total star mass) of the order of 20\%, or as shown in Ref. \citep{Goldman:2013qla} a DM ratio of 50\% 
can give rise to a $2M_{\odot}$ NS. 
More refinedly, the precise NS mass depends also on the DM particle mass \citep{Li:2012ii,Mukhopadhyay:2015xhs}, so the final NS mass results from a combination of relative acquired DM mass and DM particle mass \citep{Leung:2011zz}. 
The impact of various DM distribution regimes on observable quantities e.g, the maximum total gravitational mass and the tidal deformability has been considered in \citep{Karkevandi:2021ygv}.

In this context, extreme density environments, such as neutron stars (NSs), offer laboratories that can accrete DM. The presence of DM thus strains the NSs saturated neutron Fermi gas. In general, the amount of DM acquired by a given NS follows the determination of the Tolman-Oppenheimer-Volkoff (TOV) equation  \citep[Tolman-Oppenheimer-Volkoff]{Tolman:1939jz,Oppenheimer:1939ne}, which governs the amount of DM that can be acquired by a NS, as in e.g., \citep{Tolos:2015qra,Deliyergiyev:2019vti}. 
This can produce bounds on candidate DM masses, either directly
\citep{Goldman:1989nd,Kouvaris:2011fi}, or after a star that accreted asymmetric DM ends up as NS 
\citep{Kouvaris:2010jy}.
Self-annihilating DM produces some characteristic effects on NS \citep{Kouvaris2008,Bertone:2007ae,Kouvaris:2010vv,McCullough:2010ai,deLavallaz:2010wp,PerezGarcia:2011hh,PerezGarcia:2010ap}, while non-annihilating DM, among other results \citep{Li:2012ii,Sandin:2008db,Leung:2011zz,Xiang:2013xwa,Goldman:2013qla,Tolos:2015qra}, yields the counter-intuitive property of getting smaller \citep{Ciarcelluti:2010ji} and less massive NSs, the more DM they accrete \citep{Sandin:2008db,Tolos:2015qra}.
Whereas a typical NS has a mass of $\simeq 1.4 M_{\odot}$\citep{deLavallaz:2010wp}\footnote{We recall that NSs should have an upper limit $2.01^{+0.04}_{-0.04}-2.16^{+0.17}_{-0.15}$ \cite{Rezzolla2017}, coming from gravitational waves (GW) observations, and a theoretical minimum mass of 0.1 $M_{\odot}$. However, lepton-rich proto neutron stars are unbound below about 1 $M_{\odot}$ \cite{Lattimer2004}.} in recent years, some pulsars were measured at $2M_{\odot}$ -- 
PSR J1614-2230 with $M = 1.97 \pm 0.04 M_{\odot}$ \citep{Demorest:2010bx}, PSR J0348+0432 of $M = 2.01 \pm 0.04 M_{\odot}$ \citep{Antoniadis:2013pzd}, 4U 1700-37$^{\ast}$ of $M = 2.44 \pm 0.27 M_{\odot}$ \citep{Zhang:2010qr}, PSR J0952-0607 with $M = 2.35 \pm 0.17 M_{\odot}$ \citep{Romani:2022jhd}, and PSR J1748-2021B/NGC 6440B with $M = 2.74 \pm 0.21 M_{\odot}$ \citep{Freire:2007jd,Zhang:2010qr} 
-- larger than the typical observed pulsars. 
Changing the NS's EoS or adding DM to it can accommodate such large masses. For a stiff EoS, observations vs theory comparison can constrain the EoS and rule out NSs' exotic matter states \citep[e.g., quarks, mesons, hyperons,][]{Schaffner-Bielich2005}. 
In one of such studies \citep{Ellis:2018bkr} authors compared the result of LIGO/Virgo upper limit to the tidal deformability parameter, $\Lambda$, changes for a DM admixed NS, to constrain the EoS. 
However, the effect of the density dependent DM on the NS properties, such as a tidal deformability and the love number, in a two fluid approach \citep{Das:2020ecp} can be significantly different with respect to the results obtained in a single fluid picture in \citep{Das:2018frc}.

The significant list of quantities responsible for the behaviour of nuclear EoS in the presence of DM was examined in \citep{Panotopoulos:2017idn, Das:2020vng}.

The primary goal of the present work is to stress the idea that NS mass depends on the DM environment.
The basic idea dates back to the work of \citep{DelPopolo:2019nng,DelPopolo:2019rox}, which proposed an approximate evolution scheme for the old NS masses with galactocentric distance. 

To avoid experimental limitations, such as the limited number of the observed NSs close to the Galactic center, 
we are trying to check that model \citep{DelPopolo:2019nng,DelPopolo:2019rox}, by assuming that the same dependencies (e.g., the change of density) should be observed inside DM clumps, 
gearing up with the list of well measured NS masses, 
and DM clump simulations \citep{Bonnivard:2015pia,Charbonnier:2012gf}.

Therefore, we are particularly interested in the possibility of probing the existence and properties of dense dark matter clumps (DMCs) through the detected NSs. Indeed, dense DMCs could form binaries with stars or other dense DMCs within globular cluster (via three-body interaction or dynamical scattering), they could be captured by other binary stars \citep{Bonaca:2018fek, Wang:2019ftp}, or they may pass in the vicinity of supermassive black holes and give rise to tidal disruption events \citep{Ali-Haimoud:2015bfg}. 
Dense DMCs with larger densities are likely to survive from tidal interactions with the galactic environment \citep[e.g., see discussion in][]{Aslanyan:2015hmi, Berezinsky:2005py, Berezinsky:2013fxa},
	
but the viable spectra of their mass, density, and radius distribution are still highly uncertain. 
Along this line, the second aim of this work is to describe an approximate mapping between NSs configurations and DM clumps. The mapping we discuss relies on the similarities between NSs masses and DM clumps scales for the phase-space density distribution describing DM particles. 

\texttt{CLUMPY} helps us to simulate any Galactic or extragalactic DM halo including substructures: halo-to-halo concentration scatter, with the several levels of substructures, and triaxiality of the DM halos \citep{Bonnivard:2015pia,Charbonnier:2012gf}. This also includes clumps of sub-parsec scales, which are allowed by many DM models as a result of primordial adiabatic isocurvature fluctuations \citep{Kolb:1994fi}, cosmological phase transitions \citep{Schmid:1998mx}, topological defects \citep{Silk:1992bh}, accretion on primordial black holes 
\citep[PBHs,][]{Bertschinger:1985pd}, 
etc. Previously \texttt{CLUMPY} has been used to study DM annihilation and/or decay in dSph galaxies \citep{Walker:2011fs,Charbonnier:2011ft,Bonnivard:2015xpq, Bonnivard:2015vua, Walker:2015twz} and galaxy clusters \citep{Nezri:2012tu,Combet:2012tt,Maurin:2012tv}.

The paper is organized as follows. 
In Sec.\ref{sec:Data}, we discuss the NS data used in this study. In Sec.\ref{sec:NSs_in_DM_env}, discussed the effects of DM environment on NSs. 
In Sec.\ref{sec:Results}, we present our main results on the change of the NSs mass with respect to the DM clumps scale. In Sec. \ref{sec:Conclusions} we conclude.

\begin{figure}
	\centering
	\includegraphics[scale=0.42]{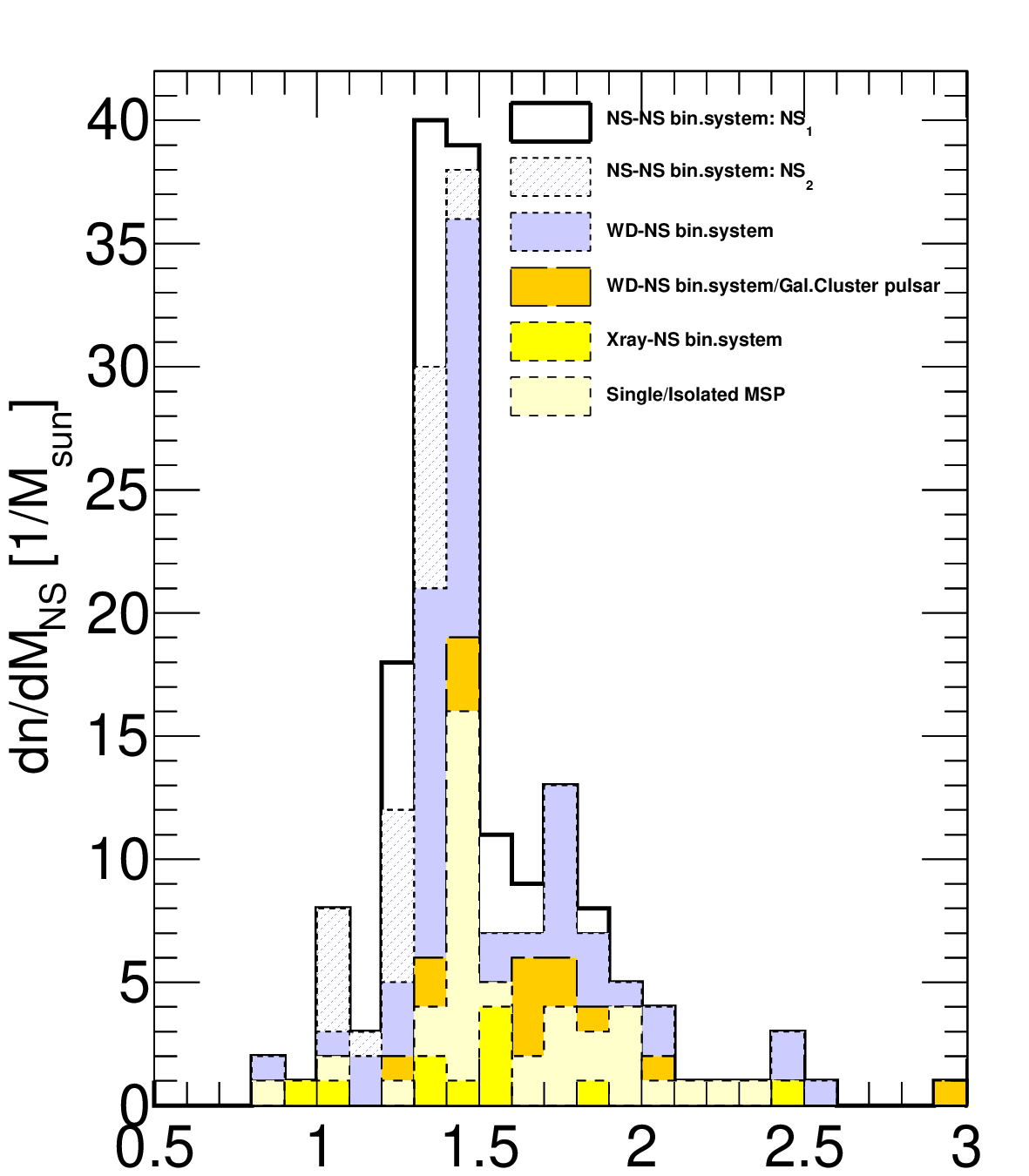}	
	\includegraphics[scale=0.42]{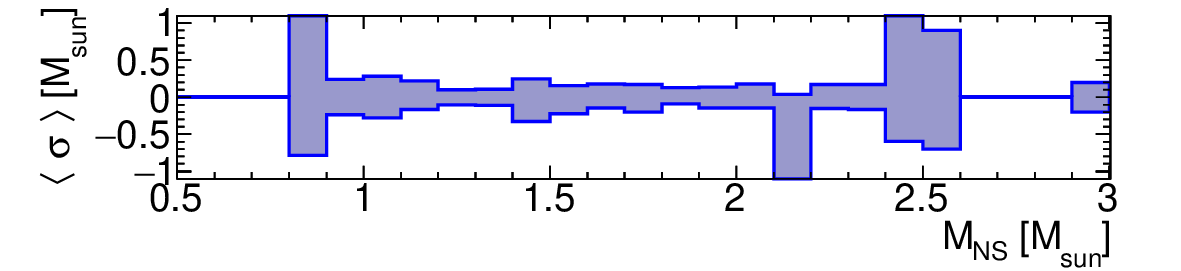} 
	\caption{
		Mass distribution of the examined NSs. Full set includes binary systems such as NS-NS, WD-NS, Xray-NS, and single Milisecond Pulsars (MSPs). Bottom pad shows the average uncertainty in the NS mass measurements.
	}
	\label{fig:NSs_massDistr}		
\end{figure}

\section{Data used}
\label{sec:Data}

In this paper, 
we use all currently available NS mass measurements summarized in Tables \ref{tab:NSlist_part1} - \ref{tab:NSlist_part2}. 
To find the associated DM clumps 
we use the Galactic latitude $b$, longitude $l$, and the distance estimates for the NS-NS binary systems, NSs in X-ray binaries, radio millisecond pulsars (MSPs), White Dwarfs-Neutron stars (WDs-NSs) systems, WDs-NSs Galactic Cluster pulsar. The mass distribution of the examined NSs is shown in  
Fig.~\ref{fig:NSs_massDistr}, where different color denote contributions from each of the considered star systems. 
The distribution includes NSs-NSs binary systems (black and dashed histogram), WDs-NSss binary systems (blue histogram), WDs-NSs binary systems/Galactic cluster pulsars (orange histogram), Xray-NSs binary systems (yellow histogram), single/isolated MSPs (light yellow histogram). The bottom panel in Fig.~\ref{fig:NSs_massDistr} shows the average uncertainty in the NSs mass measurements in the given bin.

\begin{figure}
\centering
\includegraphics[scale=0.25]{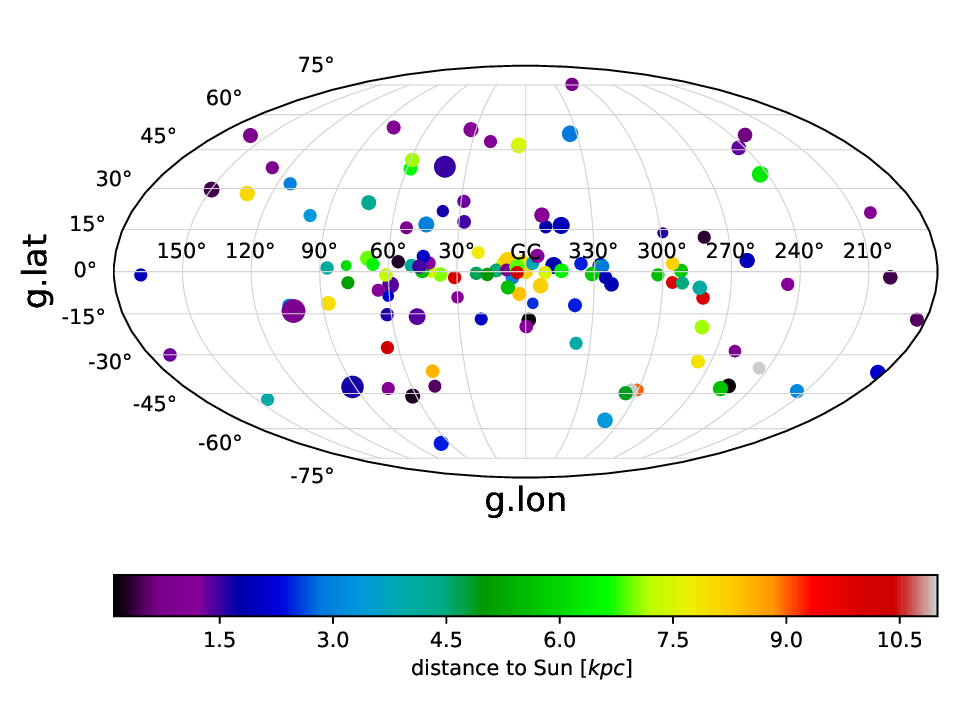}
\includegraphics[scale=0.25]{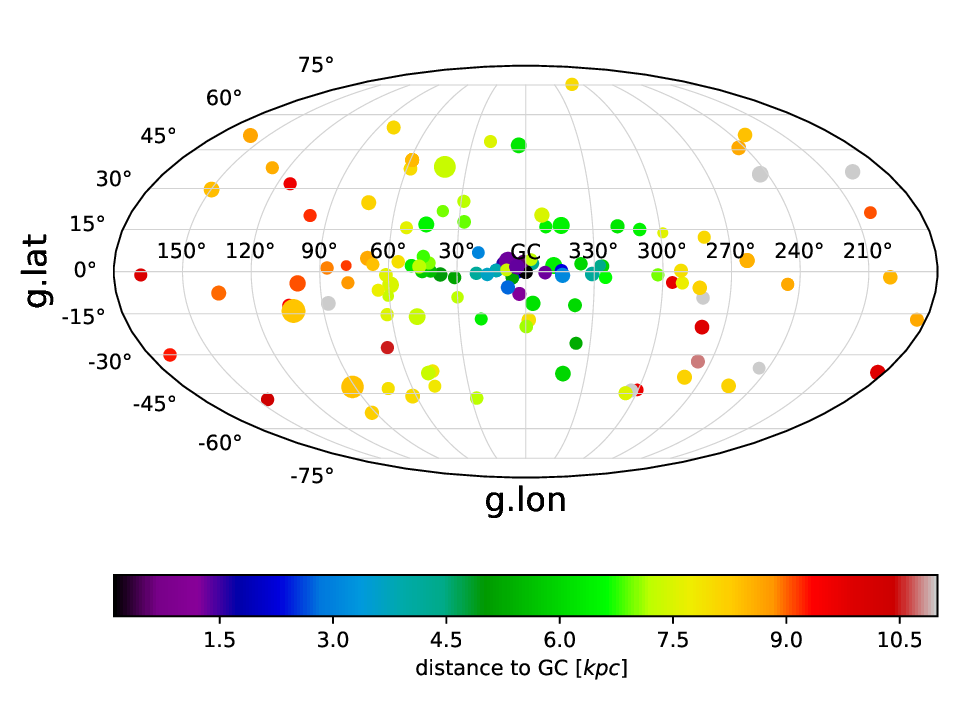} 

\includegraphics[scale=0.21]{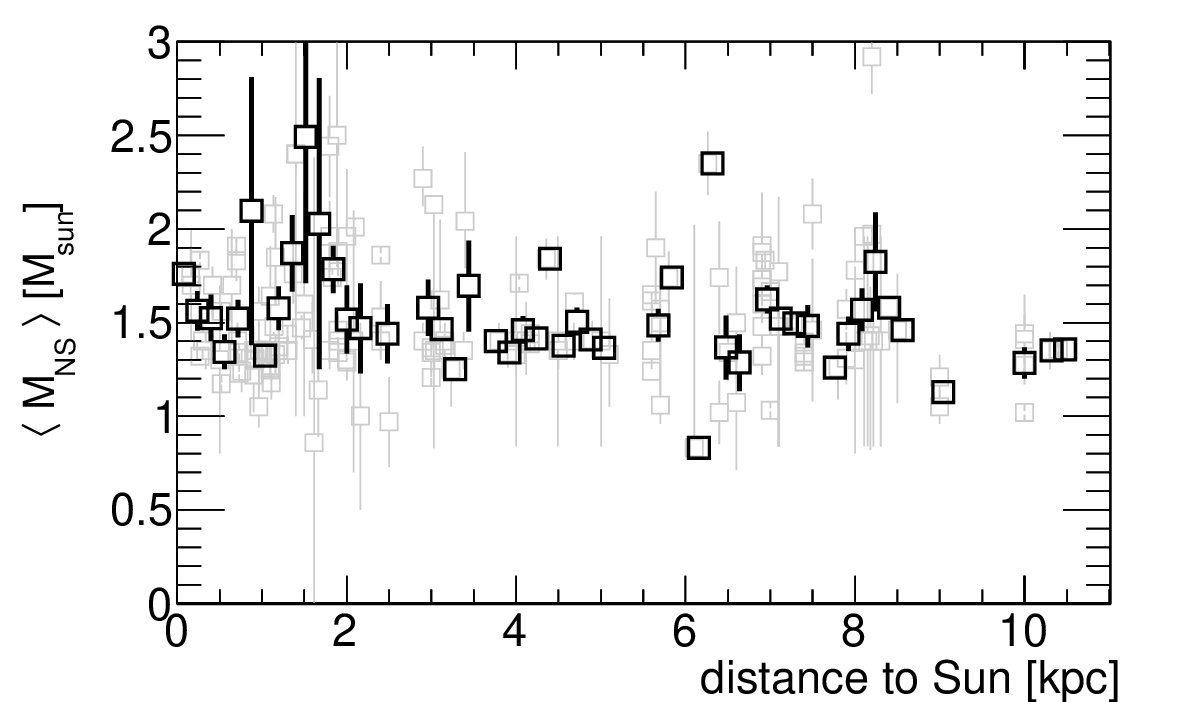} 
\includegraphics[scale=0.21]{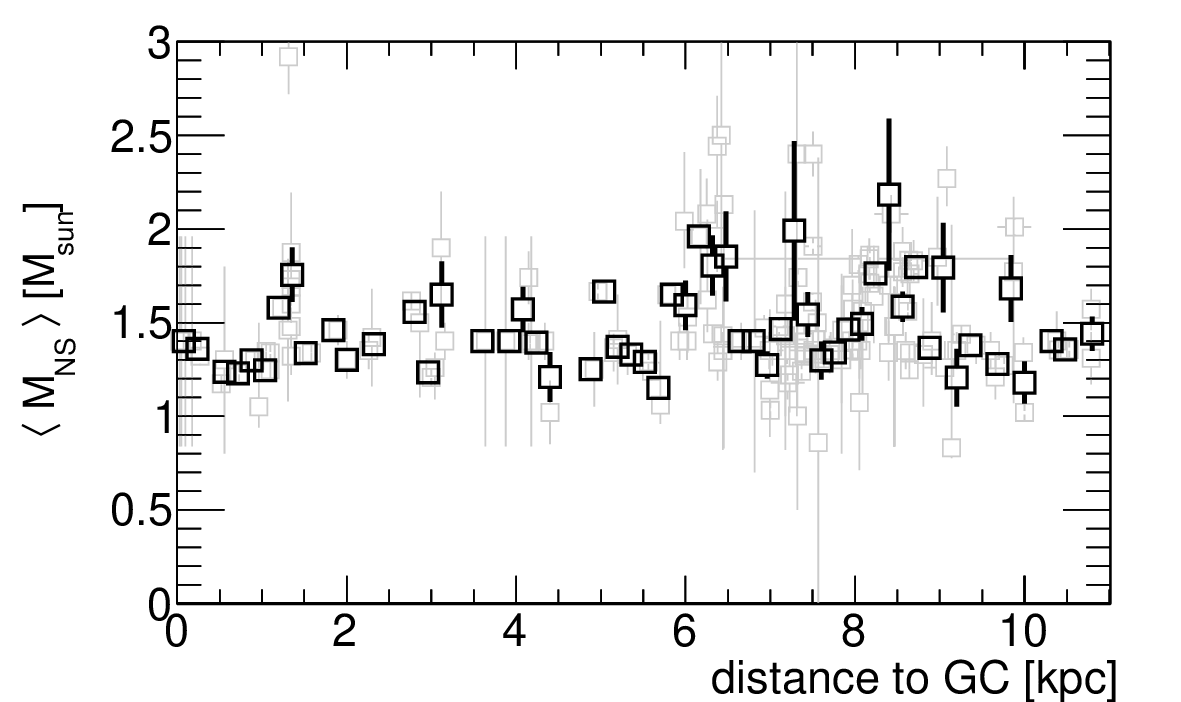}

\includegraphics[scale=0.21]{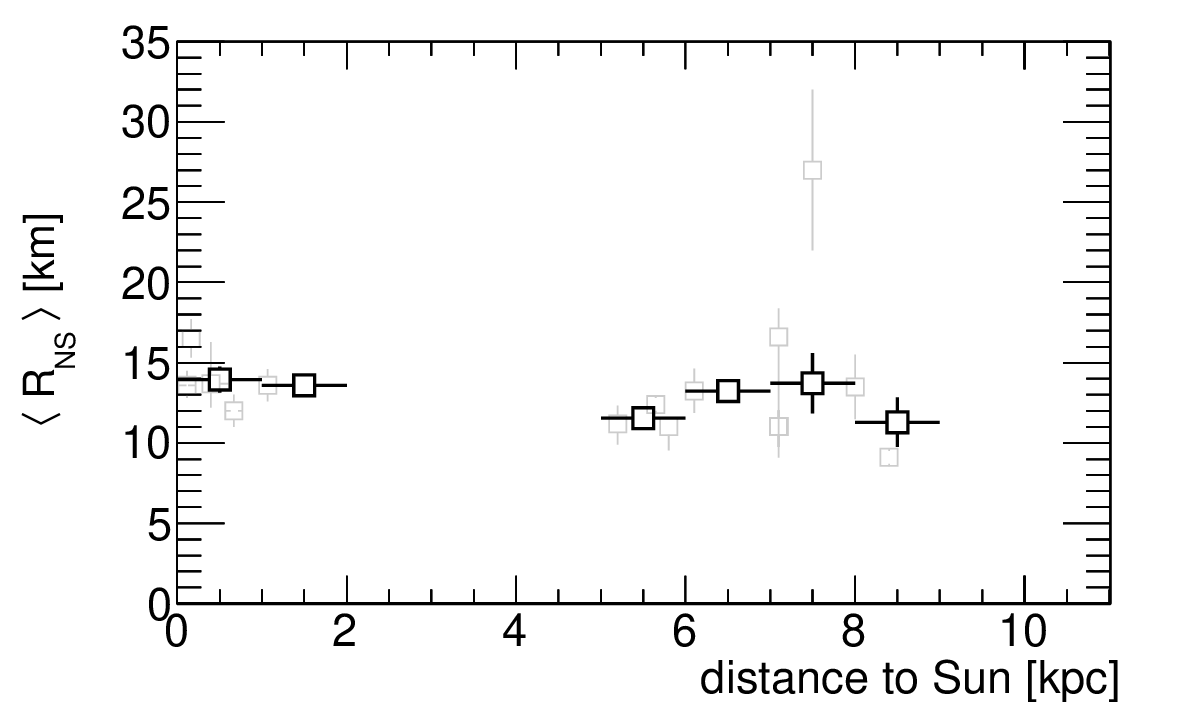}
\includegraphics[scale=0.21]{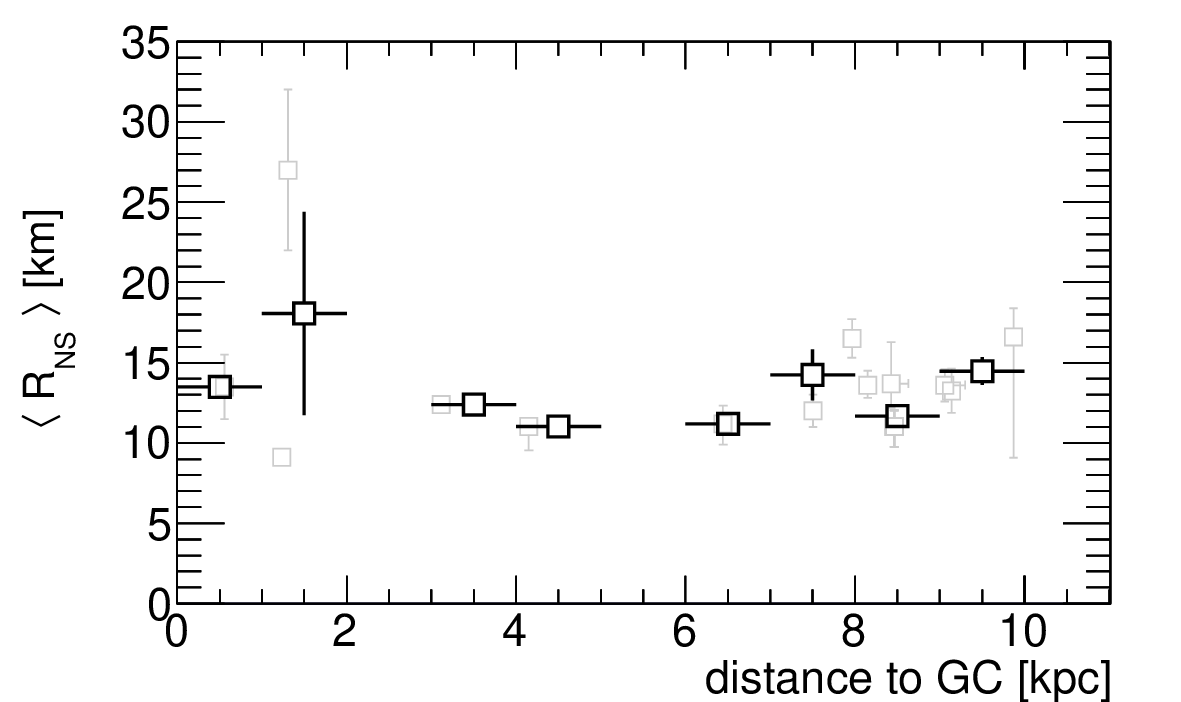} 
\caption{
	Top: Sky map of the examined NS with the color coded distances, while the NS masses are encoded in the scale of the circles. 
	Center: Mass distribution of the examined NSs as a function of the distance to Galactic centre.	Bottom: Radius distribution of the examined NSs as a function of the distance to Galactic centre. 
	Black boxes - the average mass and radius, the vertical error bars simply reflect the statistical uncertainty at the given bin; light grey boxes - the measured masses and radii of the NSs with the measured uncertainties, see Tables \ref{tab:NSlist_part1}-\ref{tab:NSlist_part2}. 
	Please compare with the original paper's Fig.~2	\citep{Deliyergiyev:2023uer}.
	Left: All distributions are plotted with respect to the distance to the Sun (parallax distance); 
	Right: All distributions are plotted with respect to the distance to the GC. 
}
\label{fig:GalaxyMap_NS}		
\end{figure}

Fig.~\ref{fig:GalaxyMap_NS} (top) is a sky map of the examined NS with the color coded distances, 
while the size of the circle markers represents the NSs masses. 
The bottom panel describes the average mass distribution of the NSs as a function of the distance from the Galactic center\footnote{
	The distance to GC was computed assuming that $Sgr A^{\star}$ is located in a nearly stationary position at the dynamical center of the Galaxy with the Galactic longtidute $l = 359.944\degree$, latitude
	$b = -0.046\degree$ \citep{Reid:2004rd}. Furthermore, we adopted the distance from the Sun to the Galactic center $8.122 \pm 0.031$ kpc \citep{GRAVITY:2018ofz}.
    }.

\section{Dark Matter clumps simulation} 
\label{sec:DarkMatter_clumps}

Many studies \citep{Berezinsky2003,Ricotti2009,Scott2009,Berezinsky:2013fxa,Bringmann2012,Berezinsky:2014wya}
pointed out that the DM distribution into a halo is not homogeneous, and superdense dark matter clumps (SDMC)
(i.e. bounded DM objects are virialized at the radiation dominated era), and ultra compact minihaloes (UCMH) formed from secondary accretion on SDCMs \citep{Berezinsky:2013fxa}, are present in the halo. 
\begin{figure*}
	\centering
	\includegraphics[scale=0.35]{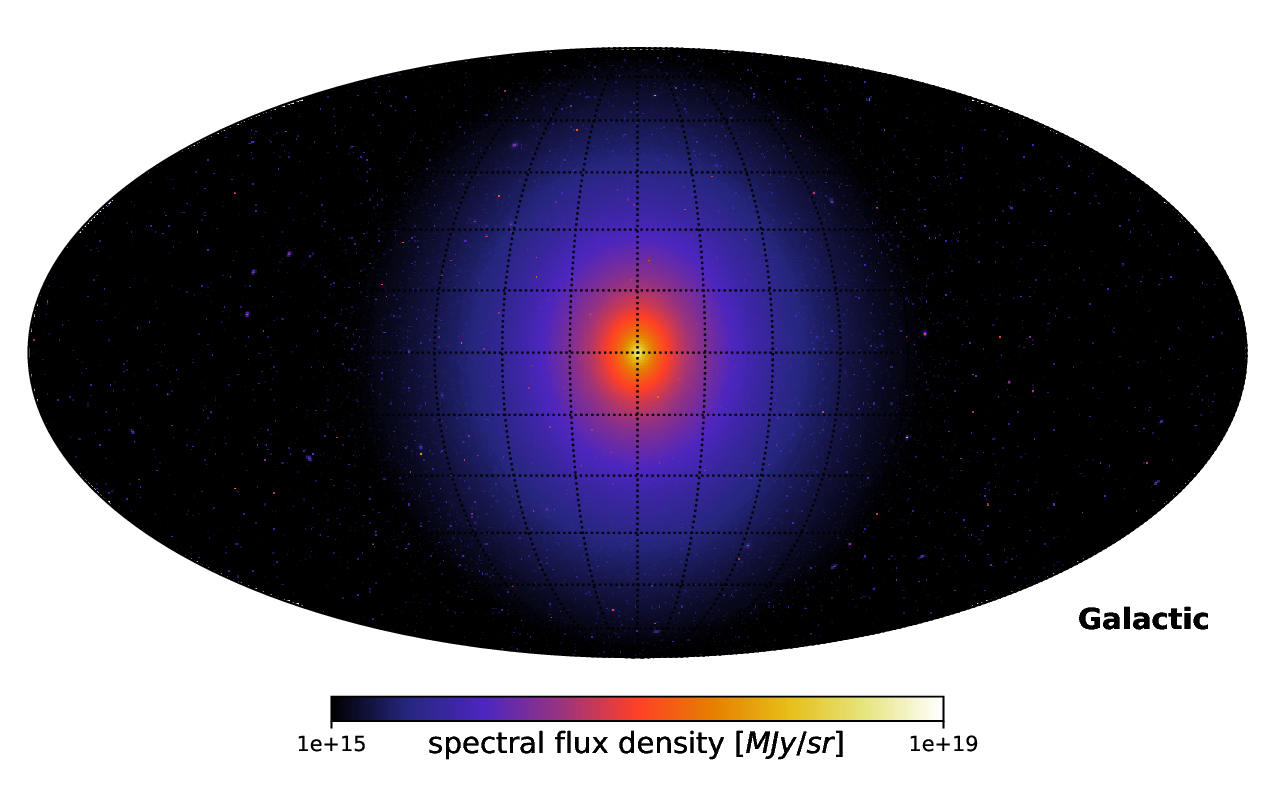}
	\includegraphics[scale=0.35]{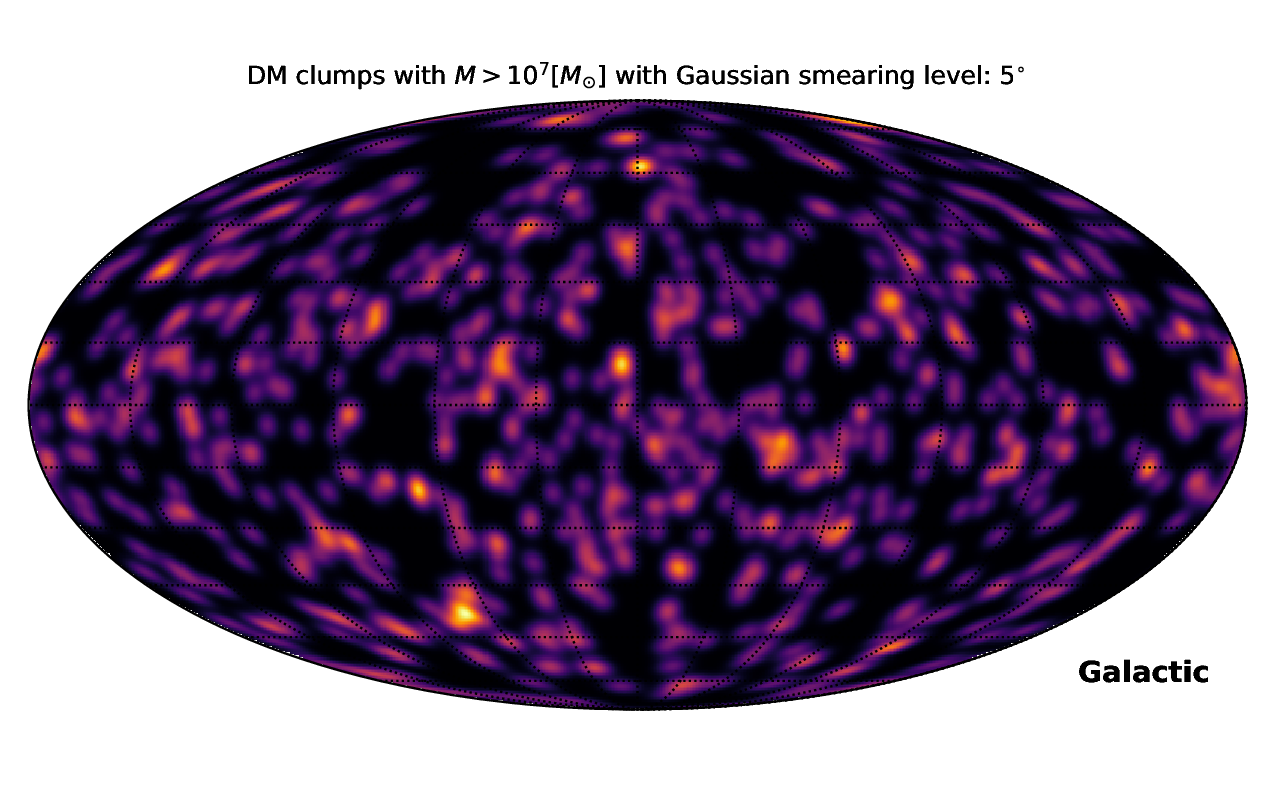}
	\caption{
		Left: Differential intensity skymap for an observer located at $R = 8.0$ kpc looking towards the centre of the galactic halo 
		for $\gamma$-rays from annihilation at 4 GeV for $m_{\chi}=100$ GeV and $\chi\chi\rightarrow b\bar{b}$ channel. The skymap is drawn in the \texttt{-g8} mode of \texttt{CLUMPY} \citep{Bonnivard:2015pia,Charbonnier:2012gf} for a numeric resolution of $N_{\rm{side}}=2048$ (corresponding to a pixel diameter of $\theta_{\rm{pix}}=\Omega^{1/2}=1.^{\prime}72$) with the parameters for a spherically symmetric halo specified in the text. The initial random seed is 1332. 
		The colour scale gives the $J$-factor values per steradian in case of DM annihilation \citep{Bonnivard:2015pia}. 
		Right: Extracted from the map on the left DM clumps with masses greater than $10^{7}$ $M_{\odot}$. A total number of 746 clumps is plotted with the Gaussian smearing of 5\%.				
	}
	\label{fig:GalaxyMap_NS_GaussianSmearing10}		
\end{figure*}

In literature are often encountered terms like "minihaloes", and "microhaloes", stressing the similitudes, when neglecting gas dynamics, between these DM objects and the DM haloes in which galaxies are found. Superdense clumps are formed by self-gravitating perturbed regions at the equivalence time, $t_{\rm eq}$, while objects formed at $t> t_{\rm eq}$ have lower densities, and are dubbed "clumps" \citep{Berezinsky:2014wya}. 
The clumps have a mass spectrum \citep{Berezinsky:2014wya} characterized by a minimum mass, $M_{\rm min}$ depending on the DM particle considered, and produced by the diffusion of DM out of the perturbation, and the following free streaming. In the case of neutralino, $10^{-11} {\rm M}_{\odot} < {\rm M}_{\rm min}< 10^{-3} {\rm M}_{\odot}$ (e.g., \citep{Bringmann:2009vf}). 

The characteristics of SDMC, and UCMH have been studied by means of simulations e.g. \citep{Ricotti2009}, or analytical models \citep{Berezinsky:2013fxa,Berezinsky:2014wya}. In this paper, we rely on the \texttt{CLUMPY} simulations which were used to quantify 
DM clump masses and scales in the vicinity of known NSs, assuming 100 and 500 GeV DM particle mass. 
We refer the reader to \citep{Charbonnier:2012gf,Bonnivard:2015pia} for an extensive description of the \texttt{CLUMPY} code features and validation, while here we briefly summarize the most important features for our analysis.

\texttt{CLUMPY} mostly has been used to calculate the $J$-factor 
\begin{equation}
	\begin{split}
		J\left( \vec{k},\Delta\Omega\right)&= \int_{\Delta\Omega}\int_{\rm{l.o.s.}}\rho^{2}_{\rm{dm}}dl~d\Omega\\
		&=\int_{0}^{2\pi}\int_{0}^{\theta_{\rm{int}}}\int_{\rm{l.o.s.}} \rho^{2}_{\rm{dm}}(\vec{k};l,\theta,\phi) {\rm{sin}}\theta~dl d\theta d\phi.
		\label{eq:Jfactor}
	\end{split}
\end{equation}
Here, we used the latest \texttt{CLUMPY} release, which includes additional outputs, that provides a better description of several quantities related to DM haloes and their substructures, which is our main interest. For this work, we mainly use \texttt{CLUMPY} in the so-called `skymap mode'\footnote{Simulation for the whole Galactic latitude and longitude ranges.} which allows the fast computation of full-sky maps of DM annihilation or decay signals. 
In \texttt{CLUMPY}, the DM density $\rho_{\rm{dm}}$ is integrated along the line of sight (l.o.s.), and up to a maximum angular distance $\theta_{\rm{int}}$. The overall DM density can be written $\rho_{\rm{tot}}=\rho_{\rm{sm}}+\rho_{\rm{subs}}$, where $\rho_{\rm{sm}}$ corresponds to the smooth component, and $\rho_{\rm{subs}}$ corresponds to the substructures of the Galactic DM halo. The galaxies with different DM clumps configurations which were used in this analysis were simulated using smoothing of the output $J$-factor skymaps with a Gaussian beam and calculating the angular power spectrum of the maps. The full width at half maximum of Gaussian smoothing was set to 0.8.

We analyze galaxies from $z = 0$, which should correspond to the characteristic overdensity value of $\Delta=200$. At this period of time, the galaxy should already change from being a very clumpy, gas-rich system that forms stars vigorously to a more regular and moderately star-forming galaxy.

As an illustrative example, 
we provide  2D-skymap plots for the Milky-Way DM haloes in Fig.~\ref{fig:GalaxyMap_NS_GaussianSmearing10} (left). In this example, the total density profile of the halo is parameterized by an Einasto profile with $r_{-2} = 15.14$ kpc, $\alpha_{E} = 0.17$ \citep{Fornasa:2012gu} and a local DM density of $\rho=0.4$ GeV cm$^{-3}$ at $R = 8.0$ kpc. All our Galaxy simulations use a virial radius of DM halo $r_{vir}=260$ kpc, yielding a total MW mass $M_{MW}=1.1\times 10^{12}$ $M_{\odot}$, in agreement with \citep{Nesti:2013uwa}, while DM clumps mass range is $10^{6}-10^{10}$ $M_{\odot}$. 
Fig.~\ref{fig:GalaxyMap_NS_GaussianSmearing10} (left) shows the color scale for the $J$-factor values per steradian in case of DM annihilation \citep{Bonnivard:2015pia}. In Fig.~\ref{fig:GalaxyMap_NS_GaussianSmearing10}(right) we show 746 extracted clumps from the sky map on the left with masses greater than $10^{7}$ $M_{\odot}$, where the clumps location has 5\% Gaussian smearing, the color indicates their spatial concentration. 

\texttt{CLUMPY} provides us with a multidimensional criterion for the definition of clumps that helps us more easily to select them in association with the NSs. This criterion includes parameters, such as, the clump $\Delta$-scale,  $R_{\Delta}$, the tidal and mass scales $R_{\rm{tidal}}$, and $M_{\rm{tidal}}$ respectively, and the clump masses $M_{\Delta}$, their galactic coordinates, distances, the critical DM density $\rho_{s}$, and the scale radius $r_{s}$ at that point. 

Whether substructures in subhalos are scaled-down versions of substructure in main halos remains an open question \citep{Springel:2008cc}. Using \texttt{CLUMPY} we only consider one level of substructure within the halo under scrutiny (Galactic halo or individual halo such as a dwarf spheroidal galaxy), and the properties of these sub-halos can be independently chosen from that of the parent halo.

To date, there is no consensus as to what the Galactic DM profile, $\rho_{\rm{dm}}(r)$, should be. There is some dynamical evidence that the Galactic DM halo might be triaxial \citep{Law:2009yq} and numerical simulations, such as the Aquarius \citep{Springel:2008cc} or the Via Lactea runs \citep{Diemand:2008in} also find non-spherical halos. All clumps have the same inner DM distribution, which in our case is an Einasto profile. The mass of the clump generally suffices to determine all its properties, i.e., its size $R^{\rm{cl}}_{\rm{vir}}$, and once an inner density profile is chosen, its scale radius $r_{s}$ and scale density $\rho_{s}$. 
Parametrizations of the mass-concentration relation have been established from numerical simulations for isolated field halos \citep{Bullock:1999he,Eke:2000av} but the concentration generally presents a significant scatter \citep{Wechsler:2001cs}. 
We used Moline's \citep{Moline:2016pbm} mass-concentration model of subclump.

A critical distance $l_{\rm{crit}}$ for the clumps in the Galaxy gives $\langle n_{\rm{cl}}\rangle$, 
the average number of clumps, obtained from a Poisson distribution of mean value $\langle n_{\rm{cl}}\rangle$.
	
Assuming that $N_{\rm{tot}}$ is the total number of clumps in a DM host halo, then the overall distribution of clumps is written as:
\begin{equation} 
	\frac{d^{2}N}{dVdM}=N_{\rm{tot}}\frac{d\mathcal{P}_{V}(r)}{dV}\frac{d\mathcal{P}_{M}(M)}{dM},
	\label{eq:DMclump_distr}
\end{equation}
where the spatial and mass distribution are probabilities, respectively normalised as:
\begin{equation} 
	\int_{V}\frac{d\mathcal{P}_{V}(r)}{dV} dV=1, ~{\rm{and}}~\int_{M_{\rm{min}}}^{M_{\rm{max}}}\frac{d\mathcal{P}_{M}(M)}{dM} dM=1.
	\label{eq:DMclump_norm}
\end{equation}
Analysis of N-body simulations have shown the mass distribution to vary as a simple power law
\begin{equation} 
	\frac{d\mathcal{P}_{M}(M)}{dM} \propto M^{-\alpha_{M}}
	\label{eq:MassDist_powerlaw}
\end{equation}
with $\alpha_{M}\approx 1.9$.

Given the mass and spatial distribution, the total number of clumps $N_{\rm{tot}}$ can be determined as following:
\begin{equation} 
	N_{\rm{tot}}=\frac{f_{\rm{cl}} {M}_{\rm{tot}}}{\langle M_{\rm{1cl}}\rangle}
	\label{eq:CLUMPY_Ntot_clumps}
\end{equation}
with $\langle M_{\rm{1cl}}\rangle$ the average mass of one DM clump, and the mass fraction $f_{\rm{cl}}$ of clumps.

From the spatial distribution $d\mathcal{P}_{V}(r)/dV$ one can define the average clump density
$\rho_{\rm{subs}}$ as
\begin{equation} 
	\langle \rho_{\rm{subs}(r)} \rangle = f_{\rm{cl}} {M}_{\rm{tot}} \frac{d\mathcal{P}_{V}(r)}{dV}
	\label{eq:CLUMPY_clumps_dens}
\end{equation}

We assume the standard $\Lambda$ cold dark matter ($\Lambda$CDM) cosmology with the $WMAP5$ cosmological parameters, namely $\Omega_{m}=0.27$, $\Omega_{\Lambda}=0.73$, $\Omega_{b}=0.0.045$,
$h=0.7$ and $\sigma_{8}=0.82$ \citep{Ullio:2002pj,Komatsu:2003kv, WMAP:2008lyn, Moline:2016pbm}. Each halo was selected to have a given virial mass at $z = 1$ and no ongoing major merger at that time. This latter criterion eliminates less than 10\% of the haloes, which tend to be in dense protocluster environments at $z\sim1$. 
The target virial masses at $z = 0$ were selected in the range $M_{\rm{vir}} = 10^{6} - 10^{10} M_{\odot}$. 

Generating skymaps with \texttt{CLUMPY} starts from setting DM properties: smooth DM profile, spatial and mass distribution of Galactic substructures, halo mass-concentration relation, DM particle mass, and annihilation/decay channels. The computation has been optimized to draw only subhalos that outshine the mean DM signal (set by a user-defined precision), leading to a decomposition of the substructure signal
$J_{\rm{subs}}^{\rm{tot}} = J_{\rm{drawn}} + \langle J_{\rm{subs}}\rangle$ into two components: $J_{\rm{drawn}}$ is the signal from the substructures drawn in a realization of the skymaps, and $\langle J_{\rm{subs}}\rangle$ is the average signal from all ‘unresolved’ halos\footnote{For legibility purpose, we define $\langle J_{\rm{subs}}\rangle$ to be the sum of $\langle J_{\rm{subs}}\rangle$ and $J_{\rm{cross-prod}}$ as defined in \citep{Bonnivard:2015pia}}, i.e., faint subhalos whose intrinsic $J$-factors do not pass the threshold defined from the precision level required by the user. Additional levels of clustering within subhalos are also considered using this average description.

To assess sensitivity prospects between NSs and DM clumps, we explore various parameter sets for the substructure density, while the average total Galactic halo density is left unchanged. We build sets of models by varying seven important properties of Galactic substructures such as the mass distribution, Eq.\eqref{eq:MassDist_powerlaw}, the power law index $\alpha_{M}$, the width of the mass-concentration distribution $\sigma_{c}$, the number of halos $N_{\rm{calib}}$ between $10^{6}-10^{10}$ $M_{\odot}$ (this number is used as a calibration for the total number of subhalos). We choose $N_{\rm{calib}}=150$ as our default value. For more subhalo-rich configurations, we used $N_{\rm{calib}}=200-300$ as motivated by the results of DM-only simulations \citep{Springel:2008cc}.

\section{Neutron Stars in the Dark Matter environment}
\label{sec:NSs_in_DM_env}

The accumulation of DM, studied 
by several authors \citep{Kouvaris:2010vv,deLavallaz:2010wp,Yang2011,Kouvaris:2010jy,Zheng2016}, happens in several phases \citep{Guver2012}. In the first one, the ambient DM is captured by the NS, when DM scatters with nucleons. In the second phase, scattering of DM with neutrons produces a decrease in the DM particle orbit radius. 
In the third phase, DM interacts with the already captured DM. DM thermalization with neutrons gives rise to the possibility to form a Bose-Einstein condensate, and DM becomes self-gravitating collapsing and forming a black hole \citep{Guver2012,Kouvaris2013}. 
The evolution of the DM number, $N_{\rm dm}$, is given by
\begin{equation} 
	\label{eq:capt}
	\frac{dN_{\rm dm}}{dt}=C_{\rm c}+C_{\rm s} N_{\rm dm}
\end{equation}
\citep{Guver2012}, where $C_{\rm c}$ is the capture rate due 
to DM-nucleon interaction, and $C_{\rm s}$ is the capture rate due to DM self-interactions \citep{Guver2012}, given by 
\begin{equation} 
	\label{eq:capt_SI}
	C_{s}=\sqrt{\frac{3}{2}}\frac{\rho_{\rm{dm}}}{m_{\rm{dm}}} (\sigma_{\rm{dm}}^{\rm{elas}} v_{\rm{esc}}(R))\frac{v_{\rm{esc}}(R)}{\vec{v}} \langle \hat{\phi}_{\rm{dm}} \rangle \frac{\rm{erf}(\eta)}{\eta}\frac{1}{1-\frac{2GM}{R}},
\end{equation}
where $v_{\rm{esc}}(R)$ is the escape velocity from the surface of the NS, $\sigma_{\rm{dm}}^{\rm{elas}}$ is the DM elastic scattering cross-section, $m_{\rm{dm}}$ is the mass of the DM particles,
$\hat{\phi}_{\rm{dm}}$ is a dimensionless potential, 
which embodies the compactness of the star, the parameter 
$\eta^{2}\equiv 3/2(v_{NS}/\vec{v})^{2}$ depends on the velocity, $v_{NS}$, of the NS in the Galaxy.

\begin{figure}
\centering
\includegraphics[scale=0.44]{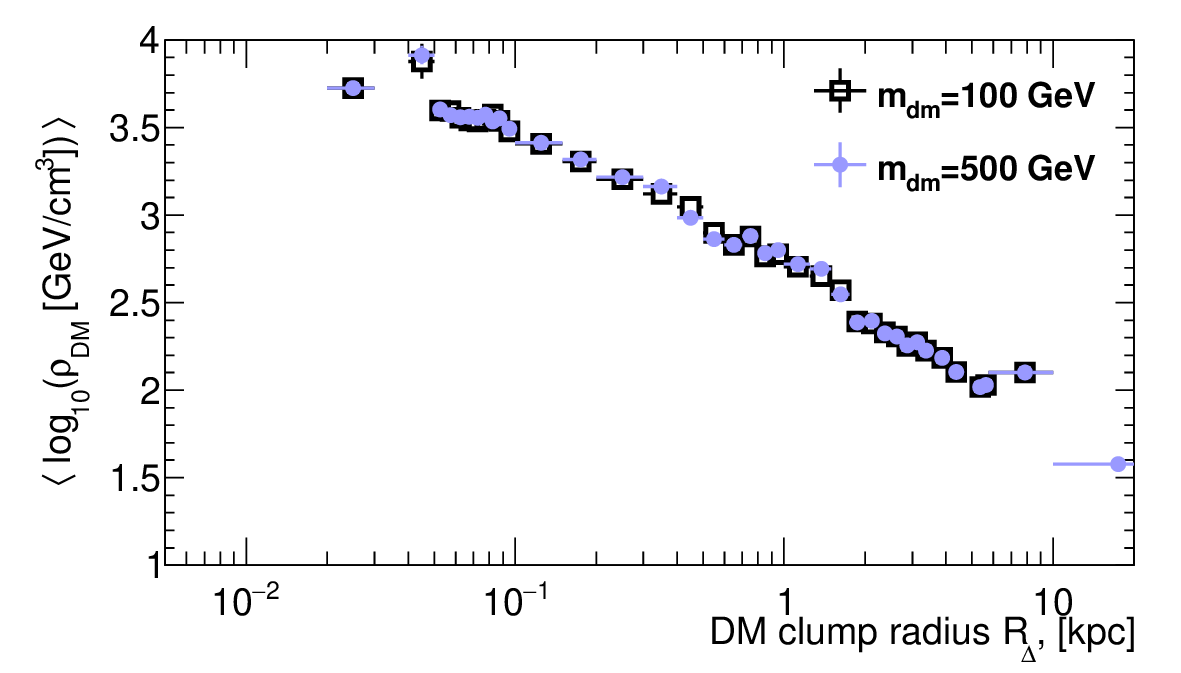}	

\includegraphics[scale=0.44]{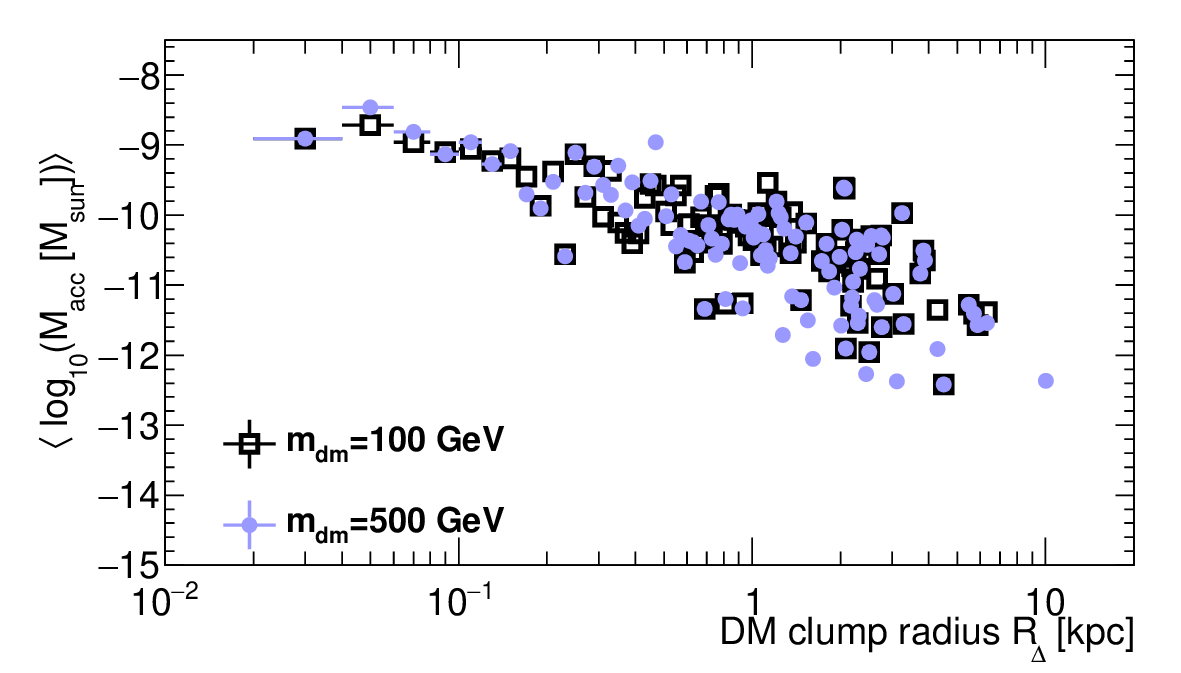}
\caption{
	Top: DM density distribution in a minihalo/clump. 
	Bottom: Accreted DM mass in a minihalo/clump, computed with the help of Eq.\eqref{eq:Kouv2013}, where DM density distribution in a minihalo/clump, was computed from the Einasto profile using the $\rho_{s}$ and $r_{s}$ values for the selected clump associated to the NS taken from \texttt{CLUMPY} simulation. 
	The results are shown for galaxy models assuming $m_{\rm{dm}}=100$ and 500 GeV. We used only those clumps that are associated with the NS.	
}
\label{fig:minihalo_dens}
\end{figure}

\begin{figure}
\centering
\includegraphics[scale=0.44]{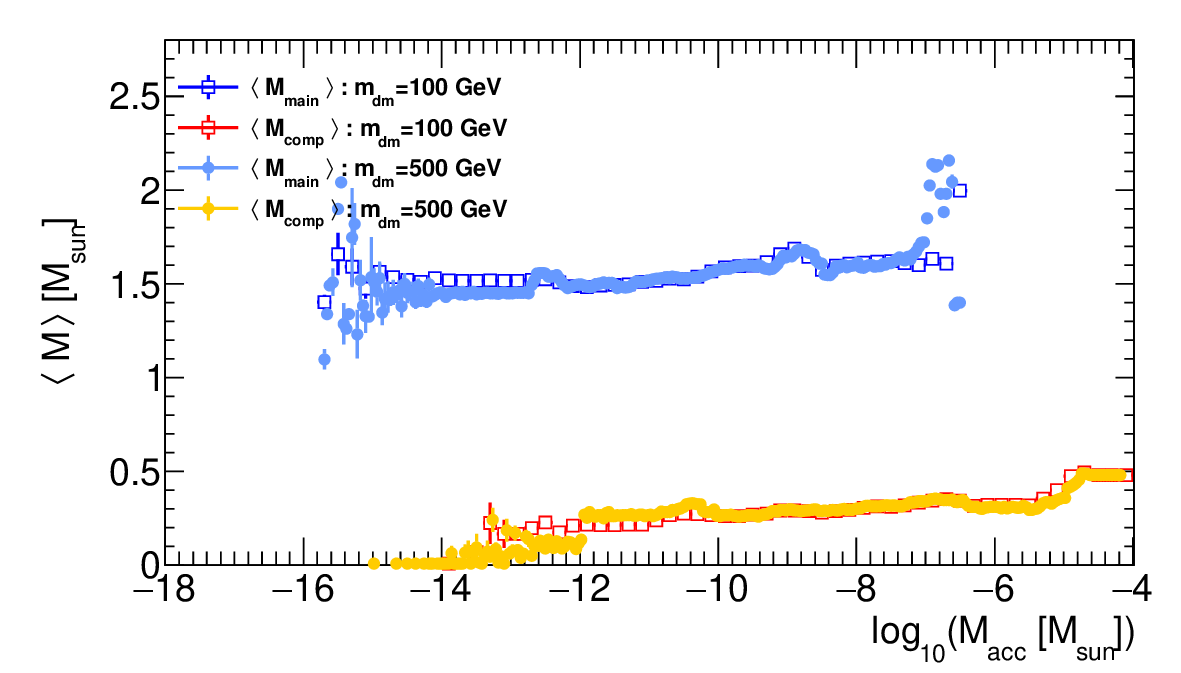}
\caption{	
 Change of the average star (NS/WD) mass with respect to the DM mass accreted inside the selected clumps, where the accreted mass was computed with the help of Eq.\eqref{eq:Kouv2013}. Results for the clumps obtained assuming particle mass $m_{\rm{md}}=100$ GeV (box) and 500 GeV (circles). 
 The profiles are shown for the main star (azure,blue) and companion star separately (red, orange).   
}
\label{fig:DM_mass_accretion_and_NS}		
\end{figure}

Following \citep{Kouvaris:2010vv,Kouvaris:2010jy,deLavallaz:2010wp,Zheng2011,Zhong2012}, we will neglect
the accretion due to self-interaction.

To calculate $C_{\rm c}$ \citep{Kouvaris2008}, we assumed a Maxwellian DM distribution. 
Then, were calculated the DM orbits that intersect the NS's, 
and then the subsample of those losing enough energy to be captured. The accretion rate can be written as
\begin{eqnarray}
	\label{eq:cc}
	C_{\rm c}&=&\frac{8 \pi^2}{3} \frac{\rho_{\rm dm}}{m_{\rm{dm}}} \left( \frac{3}{2 \pi \overline{v}^2} \right)^{3/2}
	G M R \overline{v}^2 \left( 1-e^{-3 E_0/\overline{v}^2} \right) \xi f \nonumber \\
	& &
	=1.1 \times 10^{27} s^{-1}  \left( \frac{\rho_{\rm{dm}}}{ 0.3 {{\rm{\frac{GeV}{cm^{3}}}}} } \right) \left(\frac{220 
		{{\rm{\frac{km}{s}}}}}{v}\right) \left(\frac{{{\rm{TeV}}}}{m_{\rm{dm}}} \right)\nonumber \\
	&& \left(\frac{M}{M_{\odot}}\right) \left(\frac{R}{R_{\odot}}\right) 
	\left( 1-e^{-3 E_0/\overline{v}^2} \right) f
\end{eqnarray}
\citep{Kouvaris:2010jy,Guver2012}, where $\rho_{\rm dm}$ is the local DM density, 
$\overline{v}$ is the average DM velocity in the Galactic halo \citep{Kouvaris:2010vv}, 
$\xi$ is the Pauli blocking factor defined in \citep{Guver2012, Bell:2020jou}.
$M$, and $R$ the mass and radius of the star, $E_0$ is the DM maximum energy per DM mass, which can give rise to capture, and $E_0 \gg 1/3 \overline{v}^2$, implying $e^{-3E_0/\overline{v}^2} \simeq 0$. $f$ is the fraction of particles undergoing scatterings in the star, and $f=1$ for $\sigma_{\rm dm} >10^{-45}$ $\rm cm^2$, or $f=0.45 \sigma_{\rm dm}/\sigma_{\rm crit}$, and $\sigma_{\rm crit} \simeq 6 \times 10^{-46}$ $\rm cm^2$.

The capture rate of fermionic DM scattering from neutrons within a NS in the zero temperature approximation can be written as following
\begin{eqnarray}
C_{c} &\approx &\frac{4\pi}{v_{NS}} \frac{\rho_{\rm{dm}}}{m_{\rm{dm}}} {\rm{Erf}} \left(\sqrt{\frac{3}{2}}\frac{v_{NS}}{v_{d}} \right) \nonumber \\
     && \times \int^{R_{NS}}_{0}r^{2}dr ~n_{n}(r) \frac{1-B(r)}{B(r)} \langle \sigma_{\rm{dm}}(r)\rangle, 
\label{eq:fermCaptRate}	
\end{eqnarray}
where $(1-B(r))$ plays the role of the escape velocity $v_{esc}^{2}(r)$ and $1/B(r)$ provides a relativistic correction, 
$v_{NS}$ is the NS velocity and $v_{d}$ is the DM velocity dispersion, while $n_{n}(r)$ is the neutron number density \citep{Bell:2020jou}. This approximation is valid for the intermediate DM mass range.

The capture rate has been estimated by several other authors \citep{Kouvaris2013,Zhong2012,Zheng2016,Guver2012} and the results are more or less in agreement with that of \citep{Kouvaris:2010jy}.

In the case of a typical NS, with a simple transformation of Eq.\eqref{eq:cc} one may get the following expression for the accreted total mass:
\begin{eqnarray} 
\label{eq:Kouv2013}
	M_{\rm acc} & =  9.29\times 10^{41} \left(\frac{M}{M_{\odot}}\right) \left(\frac{R}{R_{\odot}}\right) 
	\left( \frac{\rho_{\rm dm}}{0.3 \rm{\frac{GeV}{cm^3}}} \right) \left(\frac{\rm t}{\rm Gyr} \right) f \rm GeV,
\end{eqnarray}
which gives $\simeq 10^{-14} (10^{-11}) M_{\odot}$, for the DM trapped in a NS (WD). 
In Fig. \ref{fig:minihalo_dens}, we present two key aspects of our analysis. First, we display the average clump density, equation \eqref{eq:CLUMPY_clumps_dens}, as a function of clump radius. Second, we show the accreted DM mass within clumps. 
This mass is calculated using equation \eqref{eq:Kouv2013} and is based on the DM density distribution within these structures, determined from the Einasto profile with $\rho_{s}$ and $r_{s}$ values specific to the selected clumps associated with NS, derived from the CLUMPY simulation.

The previous equation is an underestimate of the accreted mass since it does not take into account the accretion by the NS progenitor, which is of the same order as that acquired in the NS phase \citep{Kouvaris:2010vv}, 
or the accretion due to DM self-interaction \citep{Guver2012}.

\begin{figure*}
\centering
\includegraphics[scale=0.4]{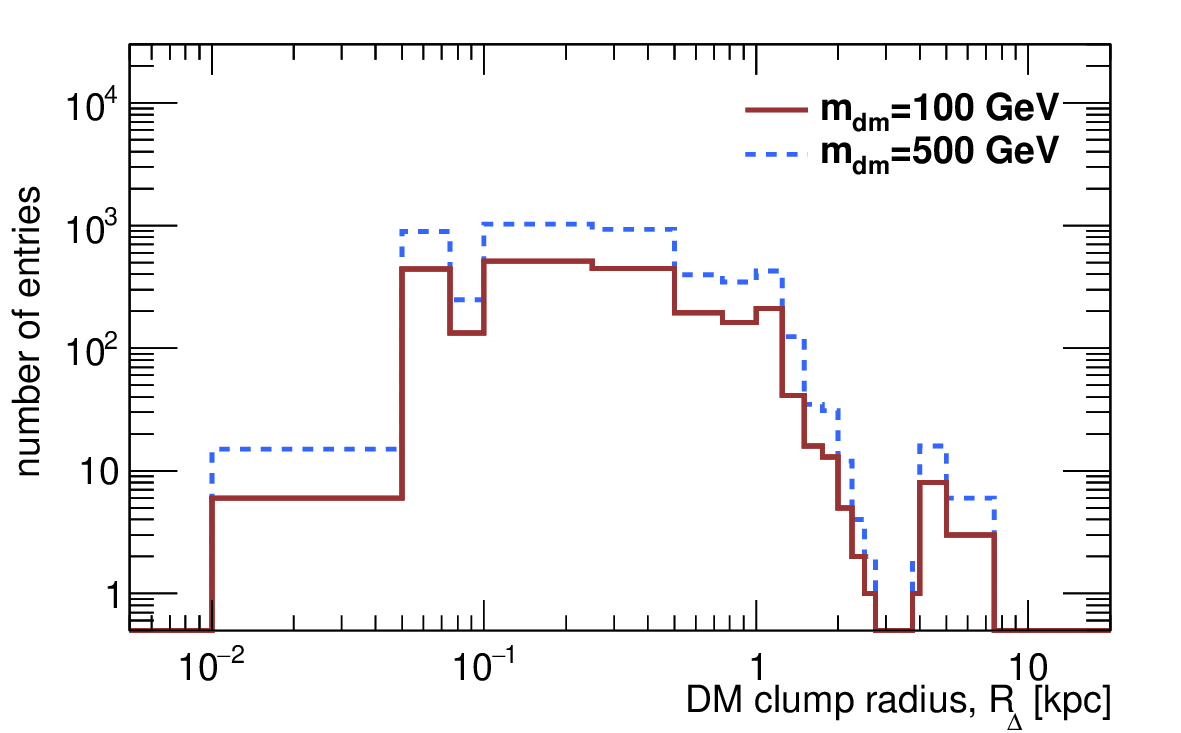}
\includegraphics[scale=0.4]{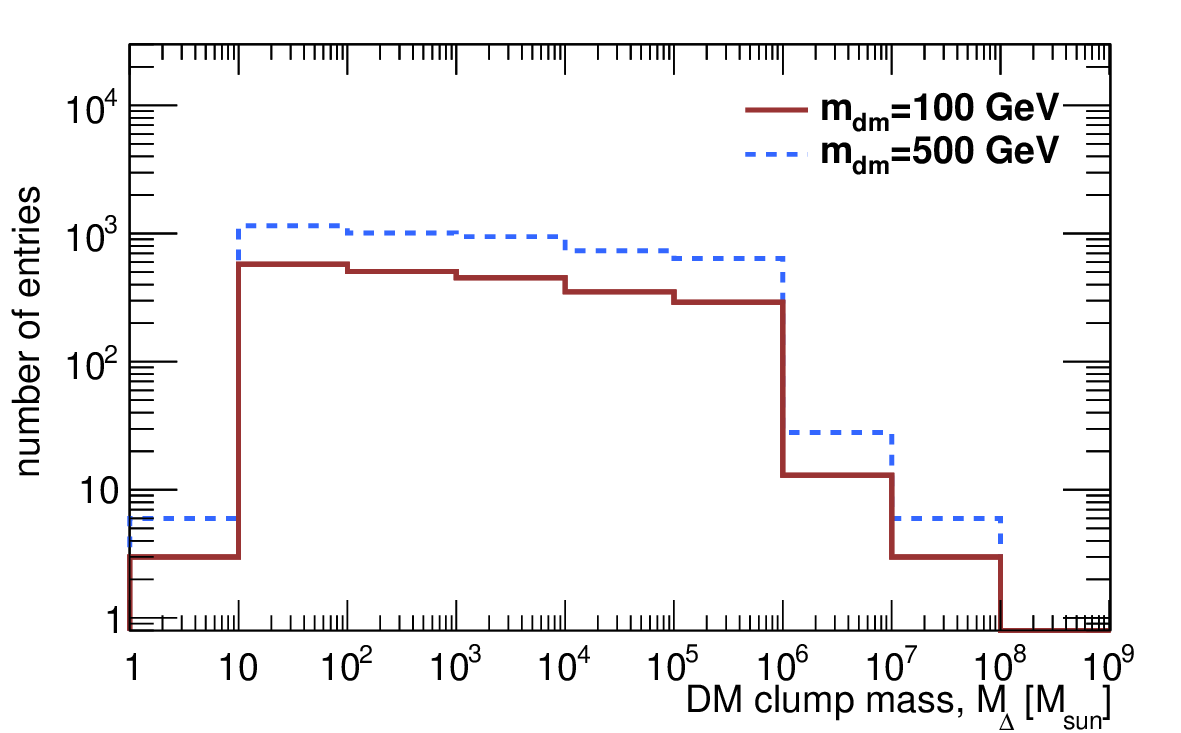}
\caption{
	(left): The outer bound of the DM-clumps in the vicinity of the examined. 
	(right): The mass of the DM clumps in the vicinity of the examined NSs in absolute units. 
	The real-space distance between NSs and DM clumps coordinates should be less than 0.01, 0.05, 0.1, 0.25, 0.5, 0.75, 1, 2, and 5 kpc, while the scale of the selected clumps should be greater than the distance. 
}
\label{fig:DMclump_Rtidal}		
\end{figure*}

The maximum density in the center of clumps can be estimated by means of the annihilation criterion, and gives
\begin{equation} 
	\rho(r_{\rm min}) \simeq \frac{m_{\rm{dm}}}{\langle\sigma v\rangle (t_0-t_f)},
	\label{eq:dmax}
\end{equation}
where $t_0$ is the present time (13.7 Gyr),
$t_f$ the formation time (59 Myr (0.49 Gyr) for non-contracted (contracted) UCMHs \citep{Scott2009}), 
$\langle\sigma v\rangle \simeq 3 \times 10^{-26}$ cm$^{3}$/s the thermal cross section, 
and $m_{\rm{dm}}$ the DM particle mass. 
For a 100 GeV particle, Eq.~\eqref{eq:dmax} gives $\rho(r_{\rm min})=7.7 \times 10^9$ GeV/cm$^3$, 
namely $\simeq 2.6 \times 10^{10}$ larger than the local DM density, implying that a NS would acquire a DM mass equal to $\simeq 7.5 \times 10^{-4} {\rm M}_{\odot}$.

\section{Results}
\label{sec:Results}

For this analysis we simulated 
19 
galaxies\footnote{For such high resolutions, the smoothing via spherical harmonics 
	\textendash ~ as done by the implementation in \texttt{CLUMPY} \textendash ~ is very computationally expensive. This is why the number of galaxies in our analysis is small.} with the help of \texttt{CLUMPY} \citep{Bonnivard:2015pia,Charbonnier:2012gf} with the numerical 
resolution of $N_{\rm{side}}=2048$ (corresponding to a pixel diameter of $\theta_{\rm{pix}}=\Omega^{1/2}=1.^{\prime}72$) with the parameters for both spherically symmetric and triaxial rotated halos. To compute fluxes for annihilation and decay, the tabulated DM $\gamma$-ray and $\nu$ spectra from \citep{Cirelli:2010xx} were used for $m_{\rm{dm}}=100$ 
and 
500 GeV, forced to decay only via $\chi\chi\rightarrow b\bar{b}$ channel\footnote{Default decay channel in \texttt{CLUMPY}.}. Since we are interested only in the location of the DM clumps, their mass and scale characteristics, the particular decay channel selection or their mixture was not relevant 
to our analysis.

The spatial distribution of subhaloes $d\mathcal{P}_{V}/dV$ is described by the Gao profile \citep{Gao:2004au,Madau:2008fr}. The mass distribution $d\mathcal{P}_{M} /dM$ follows a power-law with index $\alpha_{M}=1.9$, 
normalized 
by an abundance of 150-300 subhaloes in the mass range $10^{6}-10^{10}$ $M_{\odot}$. 

To address the main aim of our research we examined all currently measured NSs in single and binary systems\footnote{We collect 146 neutron stars with the measured masses, coordinates, and distances. We are aware that some NSs masses in our list were not exactly measured but were assumed to compute the companion star mass, such as \citep{Ng:2015zza, Cameron:2020pin}. }, which are listed in the Tables \ref{tab:NSlist_part1}-\ref{tab:NSlist_part2}.

Our analysis is based on the simplest approach, we select any massive object - DM clump in the vicinity of the NS. The real separation distance between NS and clump should be less than 0.1kpc (0.05 kpc for a finer, more computationally expensive check run, however yielding insufficient statistics), while the delta radius, $R_{\Delta}$, and tidal radius, $R_{\rm{tidal}}$, of such a clump, 
should be greater than this distance cut. For the spherical over density we use $M_{\Delta}$ and $R_{\Delta}$, for the tidal over density - $M_{\rm{tidal}}$ and $R_{\rm{tidal}}$. 
Once the clump matches our selection criteria, we explore dependencies between the parameters of this clump and NSs localized in this clump.

By using Eq.\eqref{eq:Kouv2013}, we computed the accreted DM mass for the examined NSs using the local density $\rho_{\rm{dm}}$, the scale density, $\rho_{s}$, and radius, $r_{s}$, for the selected clumps. 
In addition, we have to assume the formation age of 13.7 Gyr, a typical radius of 10 km, for NSs, and 7000 km, for the companion WDs, respectively, for the stars whose age is not measured yet, see Tables \ref{tab:NSlist_part1} - \ref{tab:NSlist_part2}. 
In Fig.~\ref{fig:DM_mass_accretion_and_NS} we show the results of such computation for the main star (NS) and companion WD.

We see that at the range of $10^{-14}-10^{-8}$ $M_{\odot}$ of $M_{\rm{acc}}$ there is no mass change for both types of stars. However, once the accreted mass goes beyond $10^{-8}$ $M_{\odot}$ the NSs mass tends to exponentially increase, while the WDs follow the same tendency but in a linear manner. This is in line with the results presented in \citep{Deliyergiyev:2019vti} if one assumes that DM interaction strength with ordinary matter is strong and of order $10^{3}$.

\begin{figure*}
	\centering
	\includegraphics[scale=0.4]{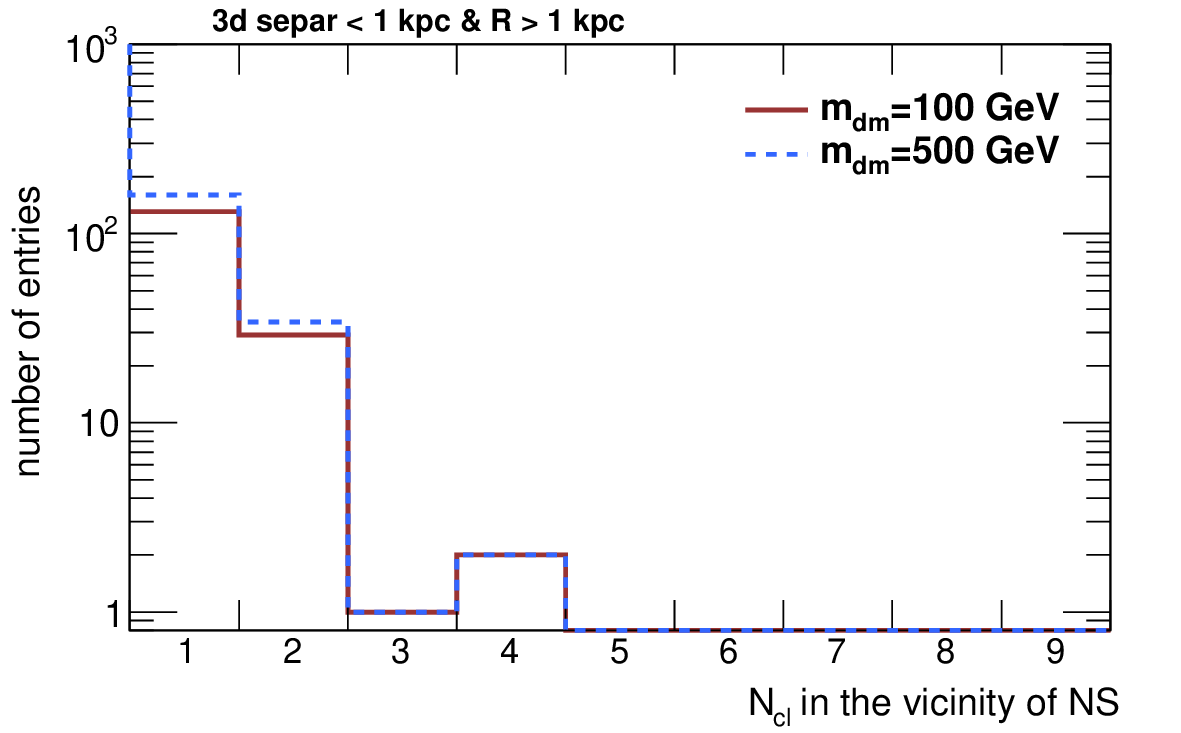}	 
	\includegraphics[scale=0.4]{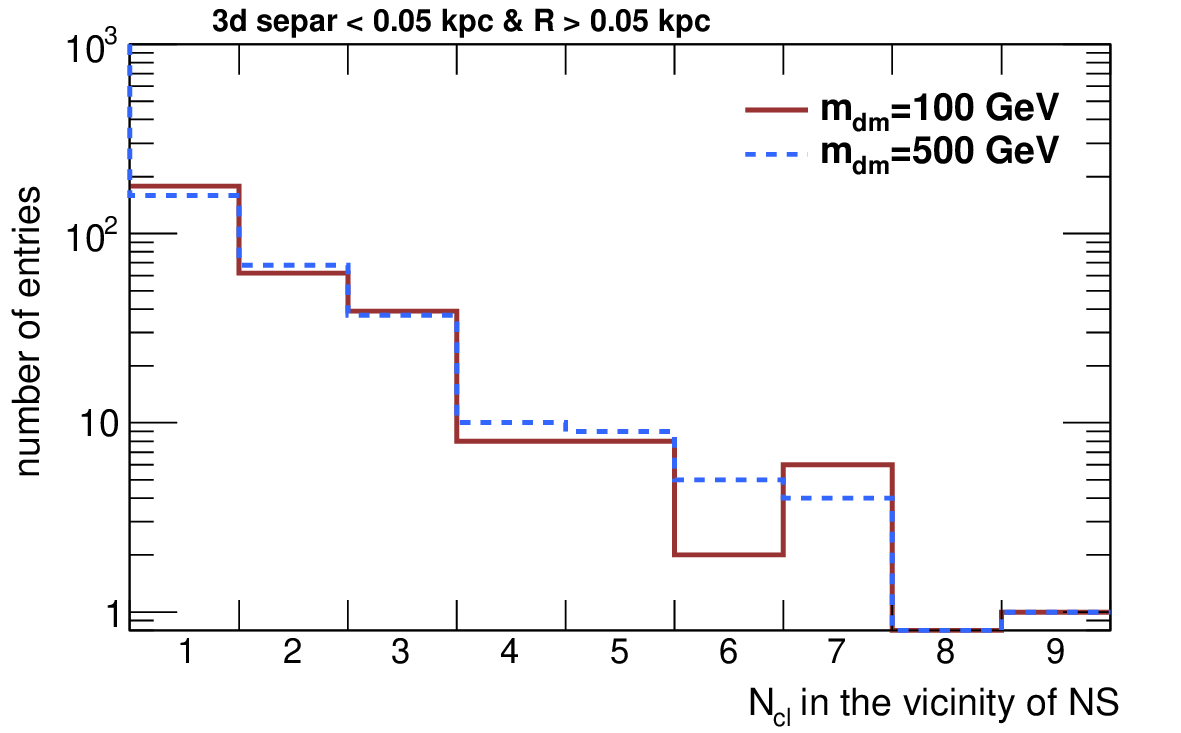}	
	\caption{
		The number of DM clumps in the vicinity of the examined NSs in absolute units. 
		(left): The multiplicity of DM clumps in the vicinity of the NSs when the real space separation distance between DM clumps and NSs coordinates should be less than 1 kpc, while the scale of the selected clumps should be greater than 1 kpc. (right): The distance between NSs and DM clumps coordinates should be less than 0.05 kpc, while the scale should be greater than 0.05 kpc. 
}
\label{fig:DMclump_multiplicity}		
\end{figure*}
\begin{figure*}
	\centering		
	\includegraphics[scale=0.5]{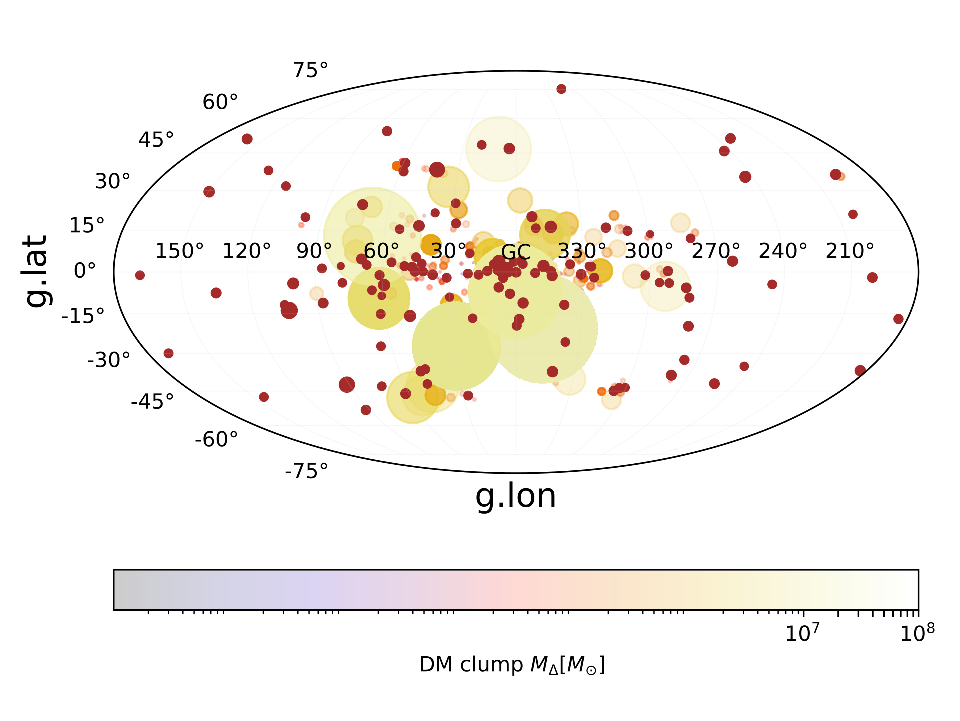}	
	\includegraphics[scale=0.5]{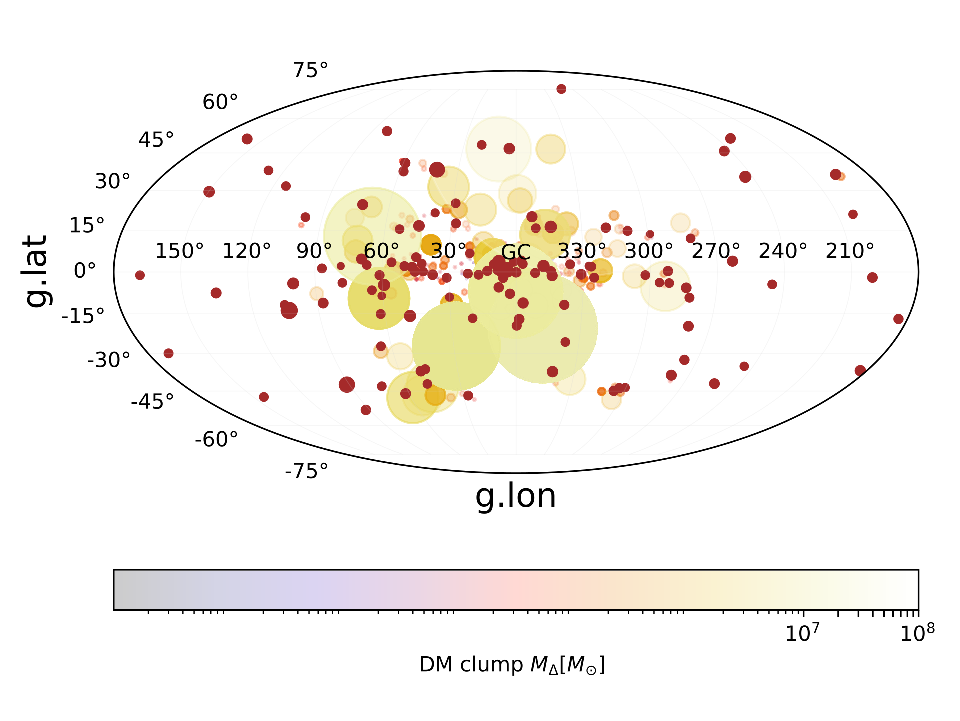}
	
	\includegraphics[scale=0.5]{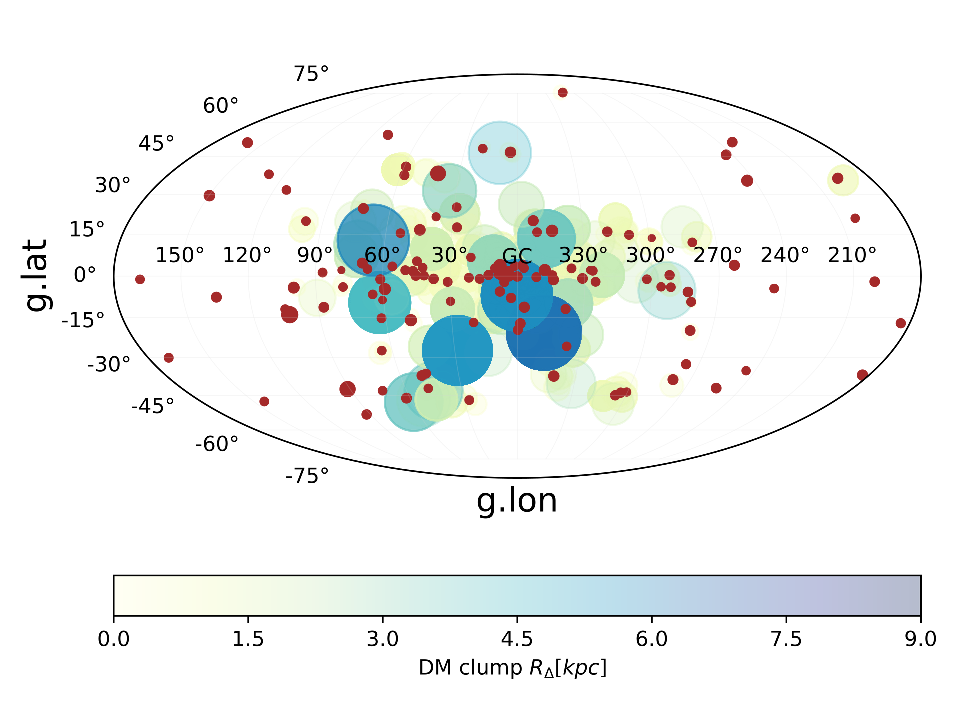}
	\includegraphics[scale=0.5]{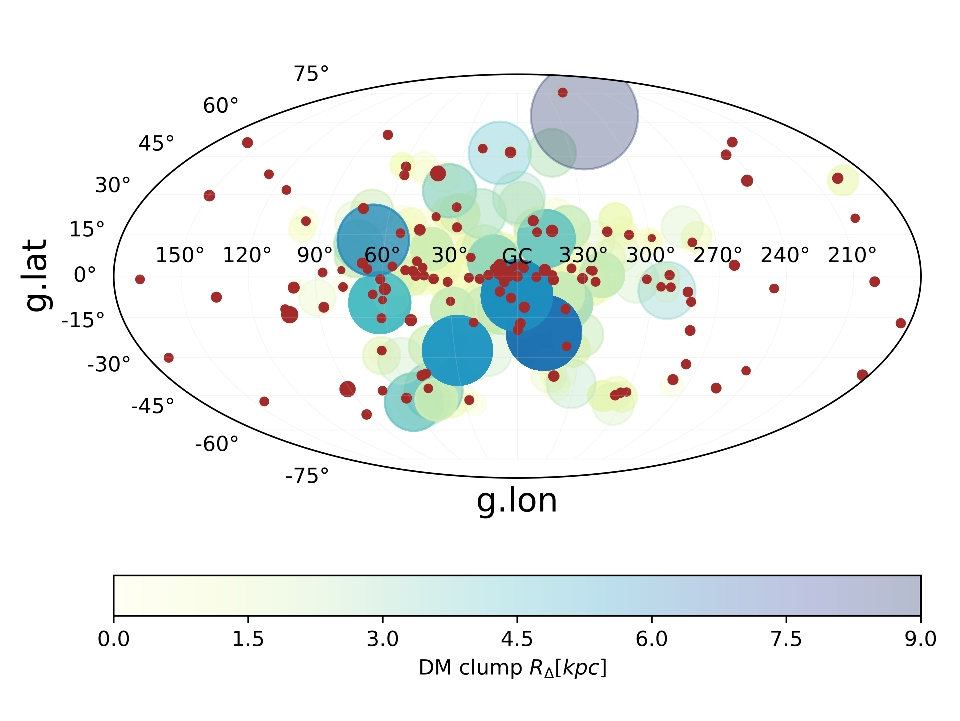}
	\caption{
		The distribution of DM clumps mass (top) and outer bound radius (bottom), as seen projected on the sky in Galactic coordinates. 
		DM clumps were selected with respect to the 3d-distance and scale. Namely, the 3d-separation distance should be less than 0.05 kpc, and the radius of the selected clump should be greater than 0.05 kpc.  
		The circle size indicates the mass of the DM clump (top), and the scale of the DM clumps (bottom). Red data points indicate known NSs. 
		(left) \texttt{CLUMPY} simulation using $m_{\rm{md}}=100$ GeV; (right) $m_{\rm{md}}=500$ GeV. 
  }
\label{fig:GalaxyMap_DMclumps}		
\end{figure*}

\begin{figure*}
	\centering	
	\includegraphics[scale=0.42]{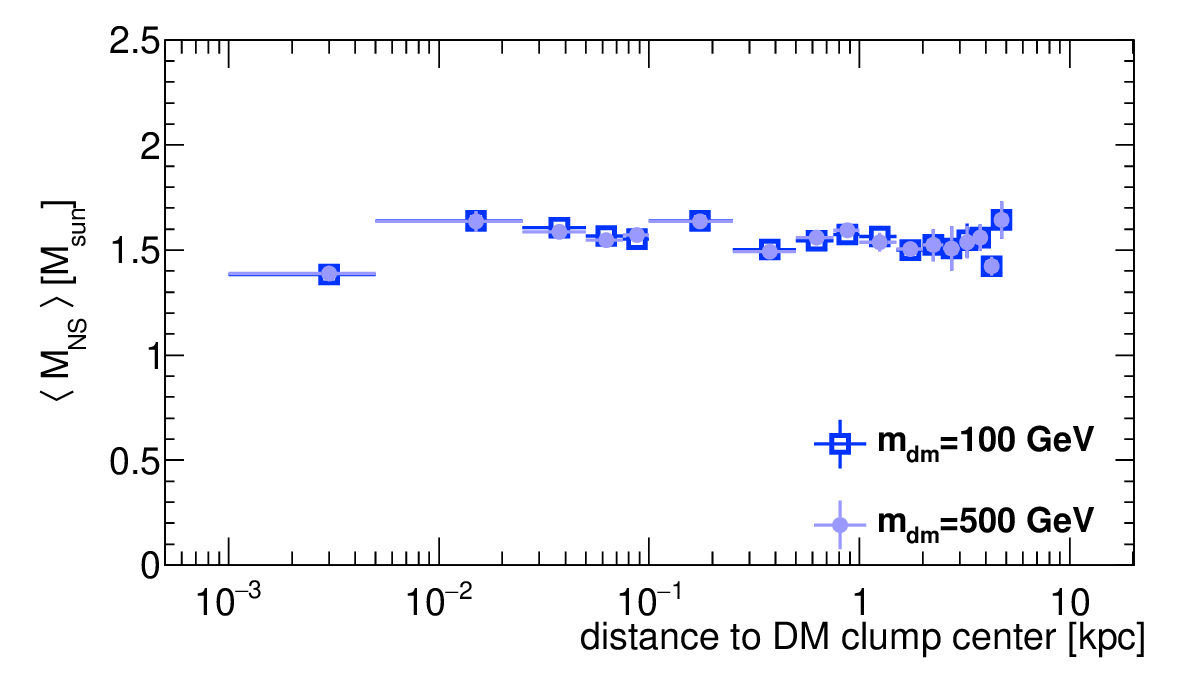}
	\includegraphics[scale=0.42]{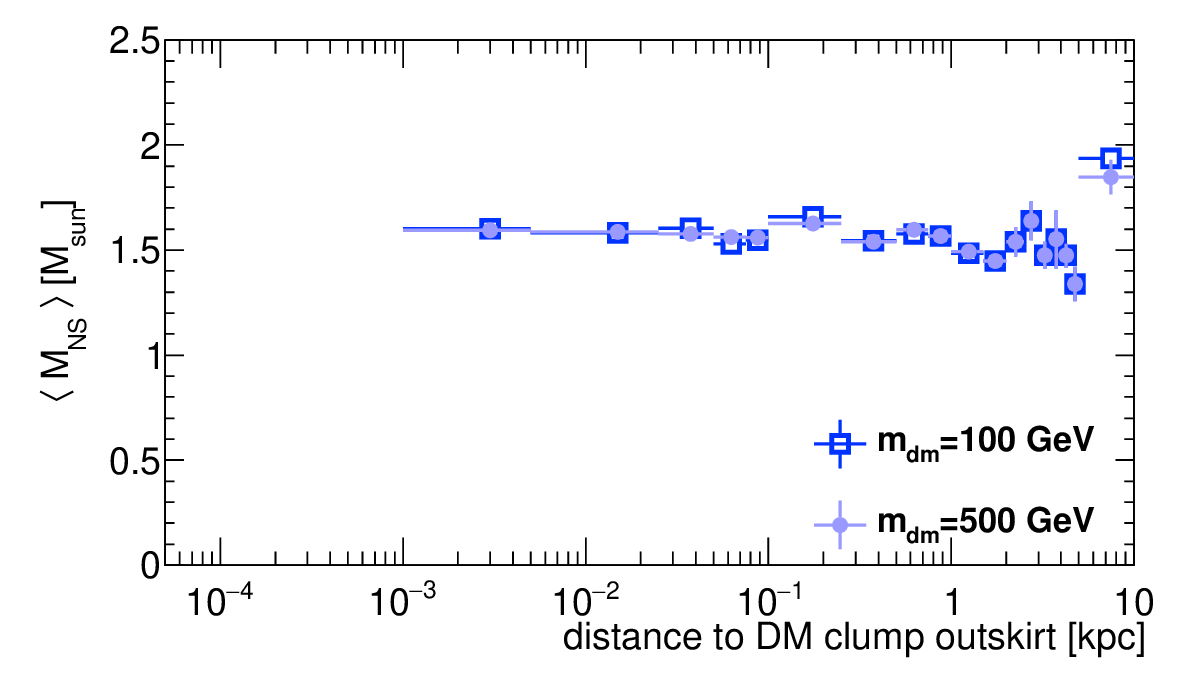}
	\caption{
		Left: The distribution of NS mass inside the DM clumps as function of distance to the clump center.
		Right: The distribution of NS mass inside the DM clumps as function of distance to the clump outskirt.
		Results are shown for the DM clumps produced with the help of \texttt{CLUMPY} assuming $m_{\rm{md}}=100$ GeV (boxes); $m_{\rm{md}}=500$ GeV (circles).
}
\label{fig:NSmass_vs_dist2ClumpCenter}
\end{figure*}

To proceed we need to understand the morphology of the selected clumps. In Fig.~\ref{fig:DMclump_Rtidal}, we plot the scale (left panel) and mass (right panel) distributions for the clumps associated with the NSs, 
namely the candidates' clumps that enclose one or more of the NSs,
requiring that the real-space distance between the NSs and clumps coordinates should be less than 0.01, 0.05, 0.1, 0.25, 0.5, 0.75, 1, 2 and 5 kpc, 
respectively, while the radius of the selected clump must be greater than this distance. This assures us that at least one NS is located inside the clump. 
We see that, after analyzing all modeled galaxies, the bulk of the clump scale distribution is localized in the range of 0.1 to 3 kpc, while the mass is distributed in the range of 10 to $10^{7}$ $M_{\odot}$.

The numbers of the clumps associated to one given NS are shown in Fig.~\ref{fig:DMclump_multiplicity}. We show these clump multiplicities for two different selection categories, when the real space separation distance between DM clumps and NSs coordinates should be less than 1 kpc, while
the clump radius is greater than 1 kpc (left panel), and when the real space separation distance between DM clumps and NSs coordinates
should be less than 0.05 kpc, while the clump radius is greater than 0.05 kpc (right panel). 
We see that the number of the big ($R_{\Delta}>1$ kpc) associated clumps at distances greater than 1 kpc is significantly reduced.

The results of this selection analysis are interpreted by mapping DM clumps on the sky map. Inclusion contours of the DM clumps are presented in Fig.~\ref{fig:GalaxyMap_DMclumps}, where we show the distributions of DM clumps into different categories based on their mass (top) and outer bound radius (bottom), as seen projected on the sky in Galactic coordinates. 
The scale of circles denotes the distance of these clumps to the Galactic center, while color denotes the mass (top) and radius (bottom). The red circles indicate NSs that were examined in this paper. The results of the analysis can then be used to infer properties of the underlying DM distribution, such as its spatial distribution, density, and velocity. 
The selected clumps are mostly located in the region of sky within Galactic longitude $60^{\circ} < l < 300^{\circ}$ and latitude $\vert b \vert < 45^{\circ}$, see Figs.\ref{fig:GalaxyMap_DMclumps}. 
On the one hand, we see that some NSs could theoretically interact or be wrapped up with/by more than one clump, see Fig.\ref{fig:DMclump_multiplicity}. The simulations with \texttt{CLUMPY} show us that 2-3 clumps with a radius of 0.1 kpc may wrap a single NS  within the separation distance of 0.01 kpc. Also, we note, that one clump may wrap up more than one NS, as seen from  Fig.~\ref{fig:GalaxyMap_DMclumps}. 
On the other hand, there is a number of NSs for which we did not manage to find an associated DM clump among simulated galaxies even in the vicinity of 6 kpc, that will have a radius greater than 6 kpc.
		
\begin{figure*}
\centering	
\includegraphics[scale=0.4]{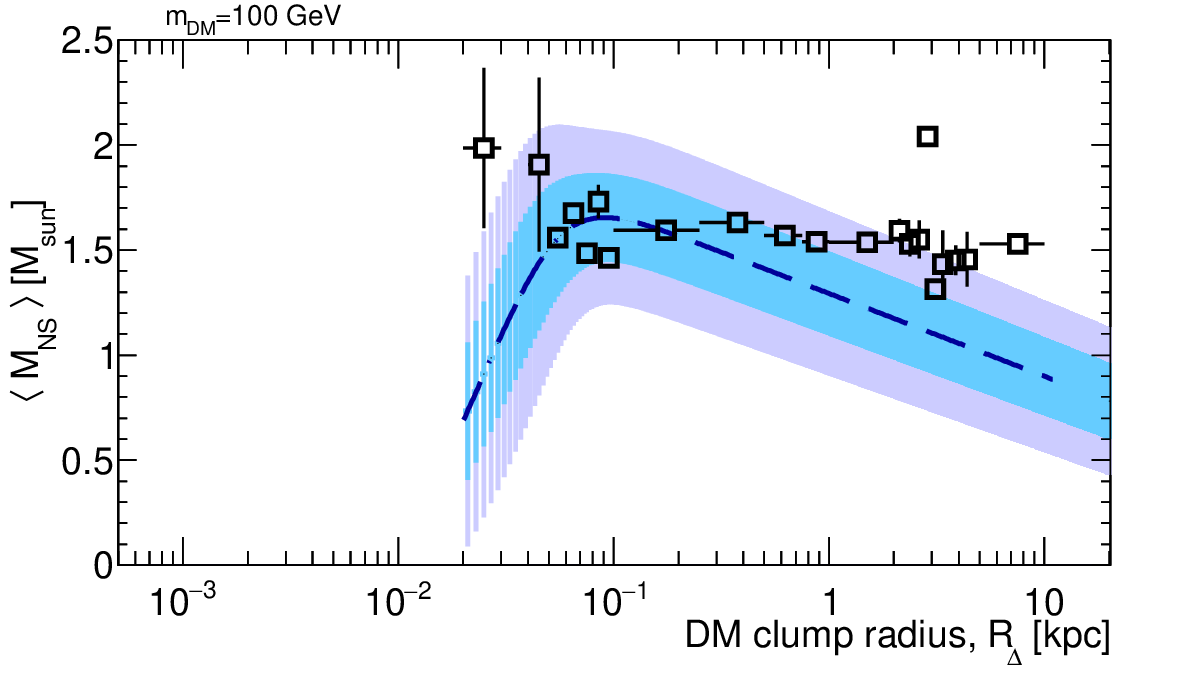}
\includegraphics[scale=0.4]{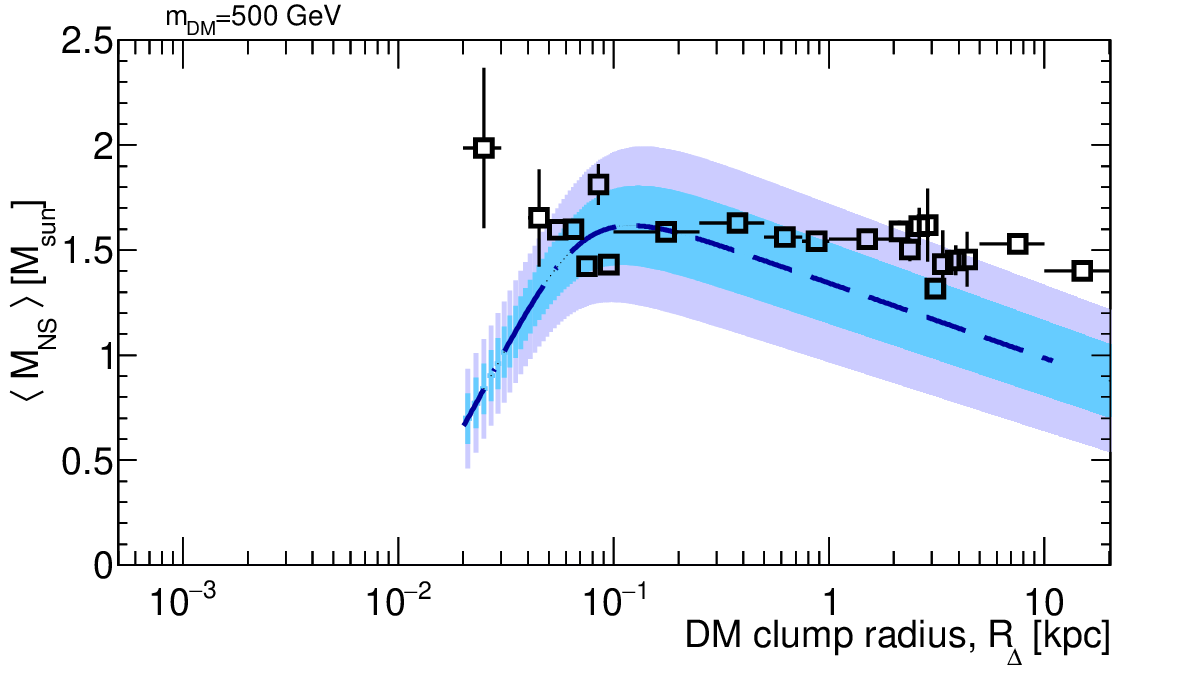}

\includegraphics[scale=0.4]{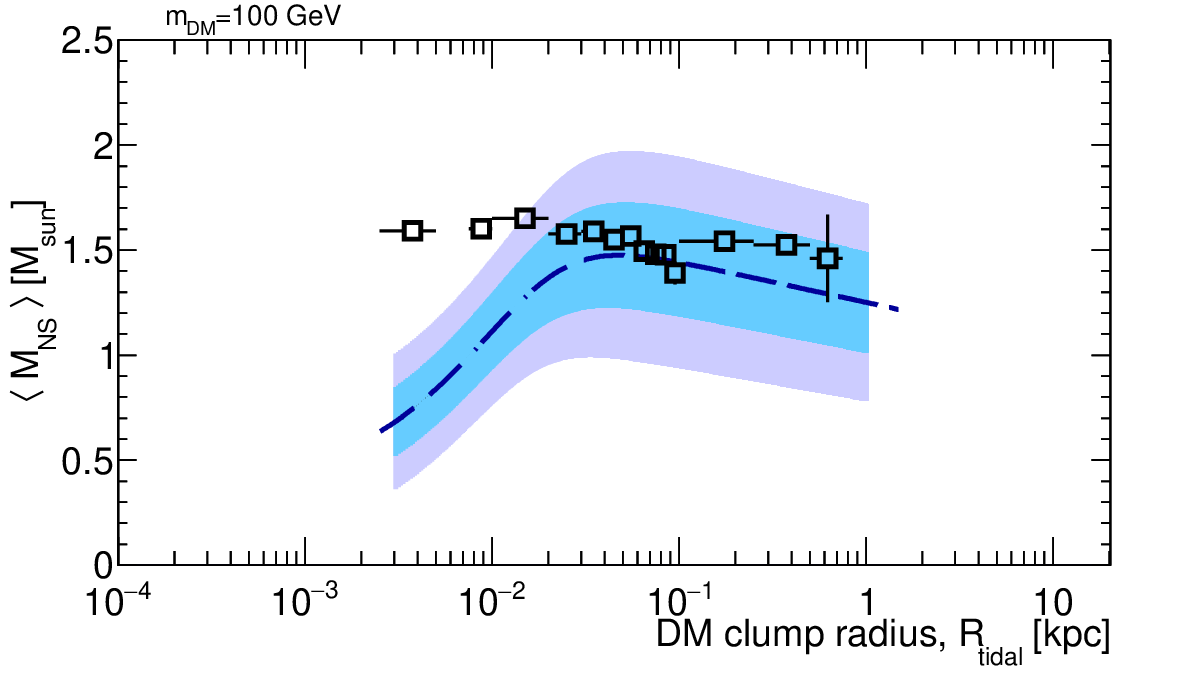}
\includegraphics[scale=0.4]{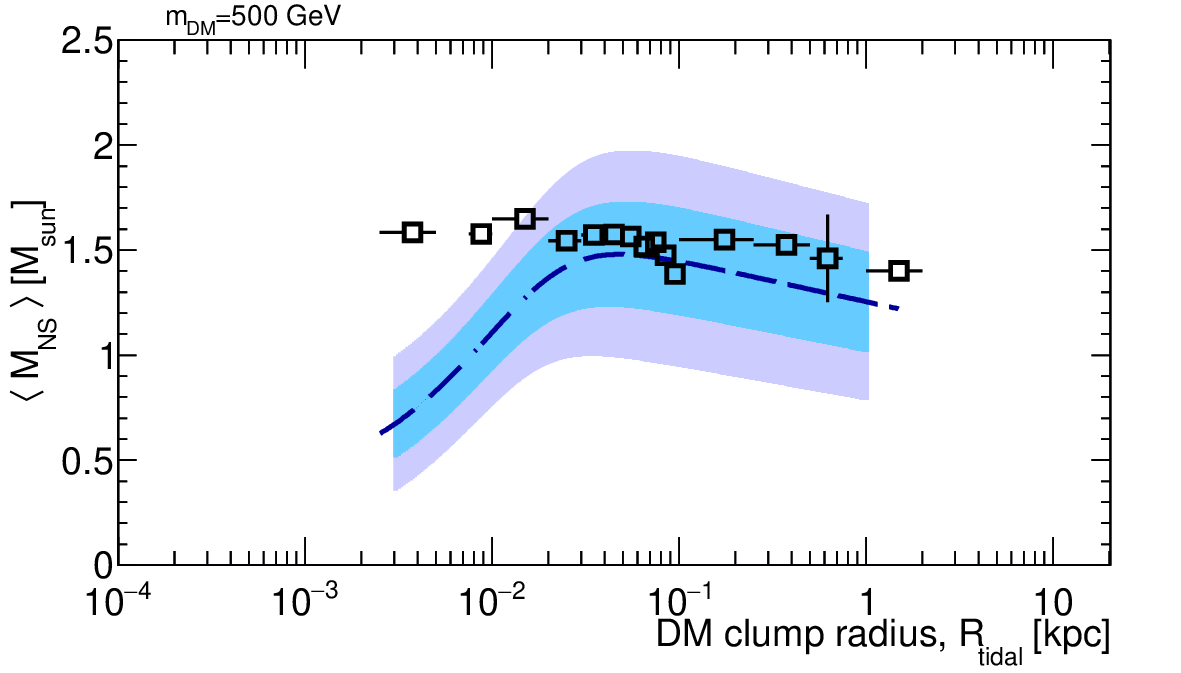}
\caption{
Fits to the NSs mass as a function of the clump radius (top) and tidal clump radius (bottom). Clumps were selected in the vicinity of 0.01, 0.05, 0.1, 0.25, 0.5, 0.75, 1, 2, and 5 kpc to the NSs, clumps scale was requested to be greater than this separation distance. Left: using the DM mass $m_{\rm{dm}}=100$ GeV. 
Right: using the DM mass $m_{\rm{dm}}=500$ GeV. 
}
\label{fig:fits2_NSmass_evolution}		
\end{figure*}
	
\begin{figure*}
\centering
\includegraphics[scale=0.4]{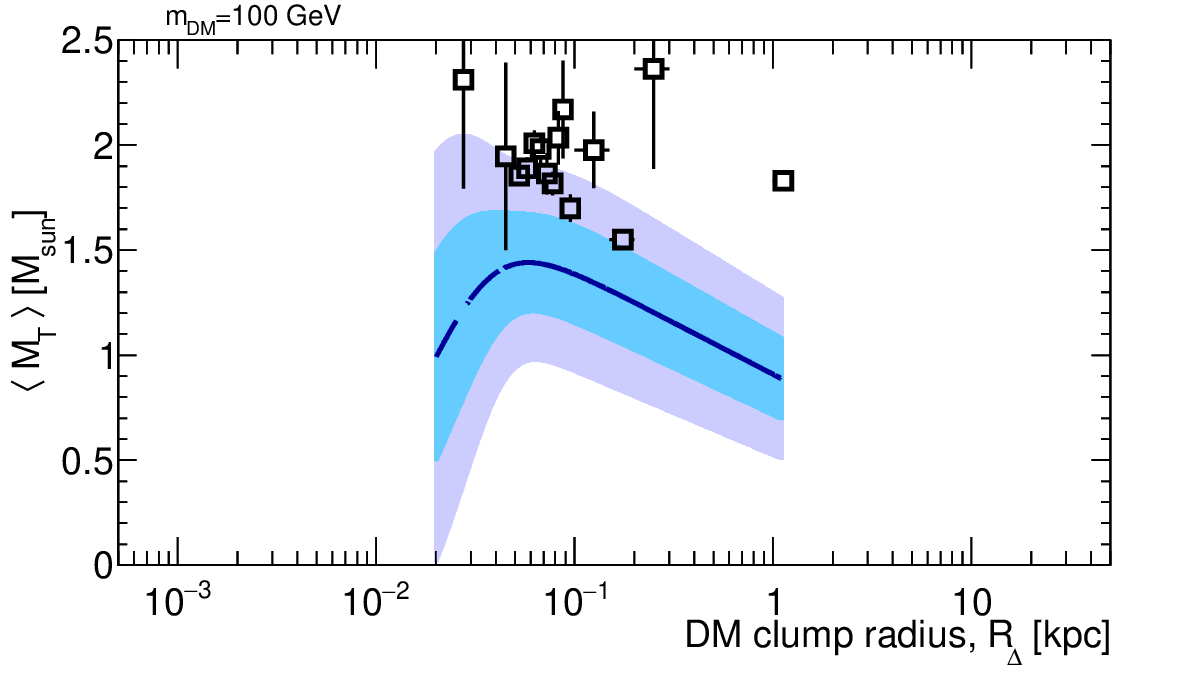}
\includegraphics[scale=0.4]{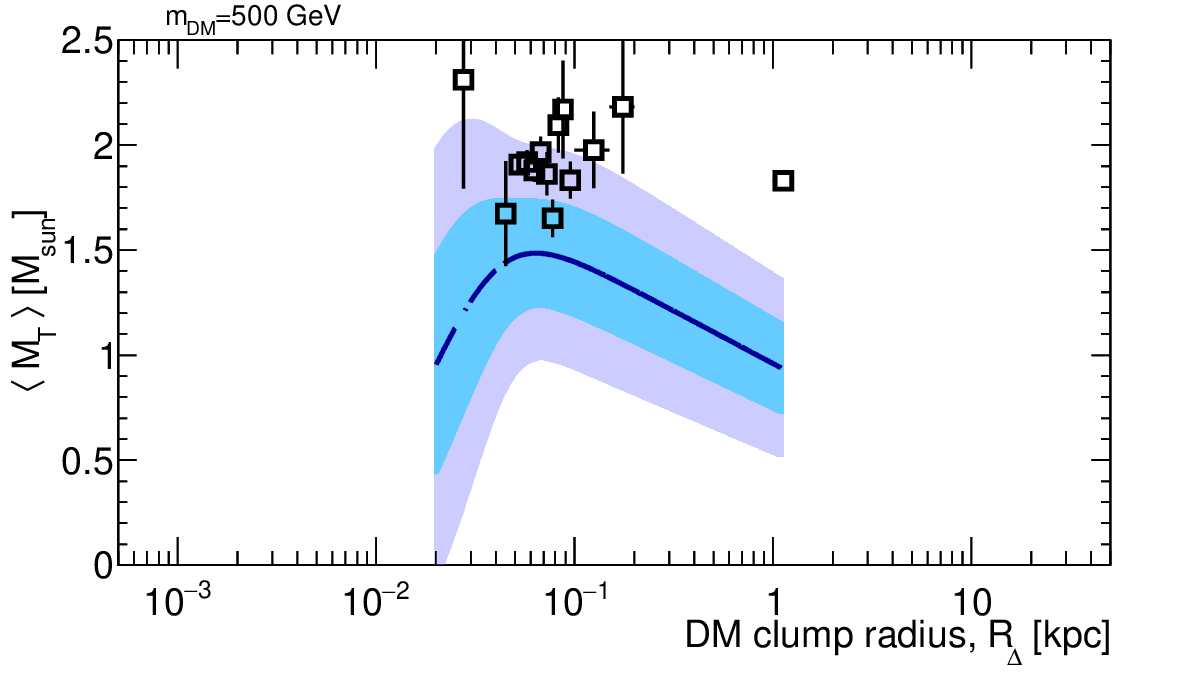}

\includegraphics[scale=0.4]{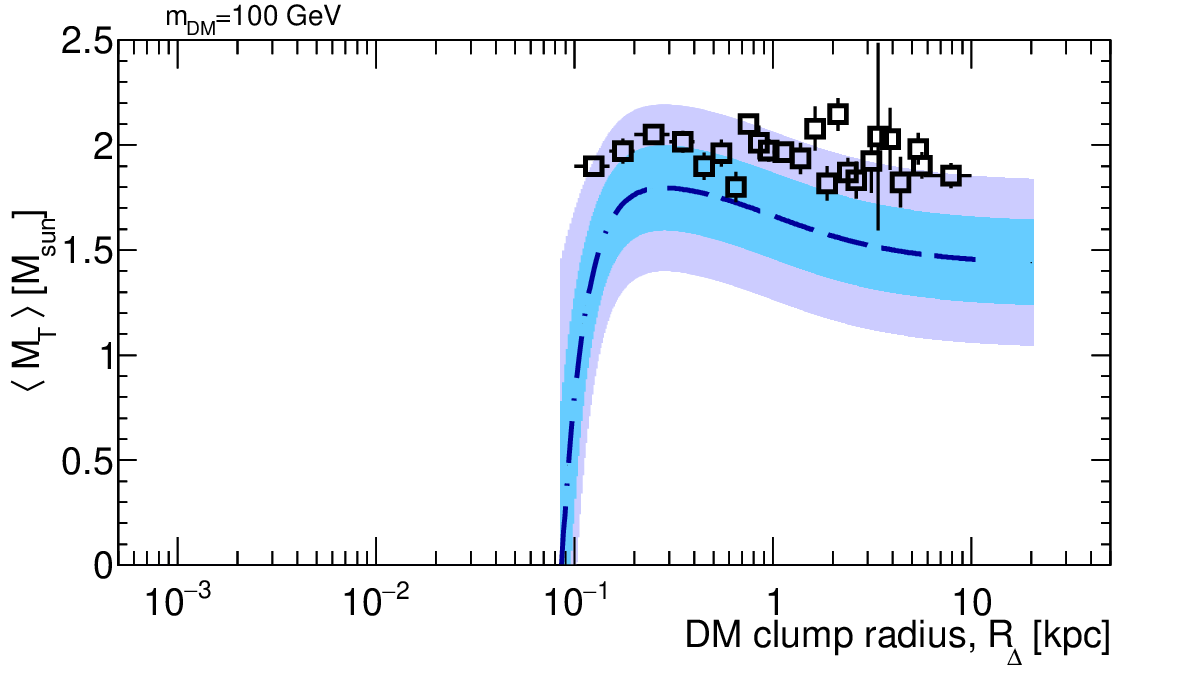}
\includegraphics[scale=0.4]{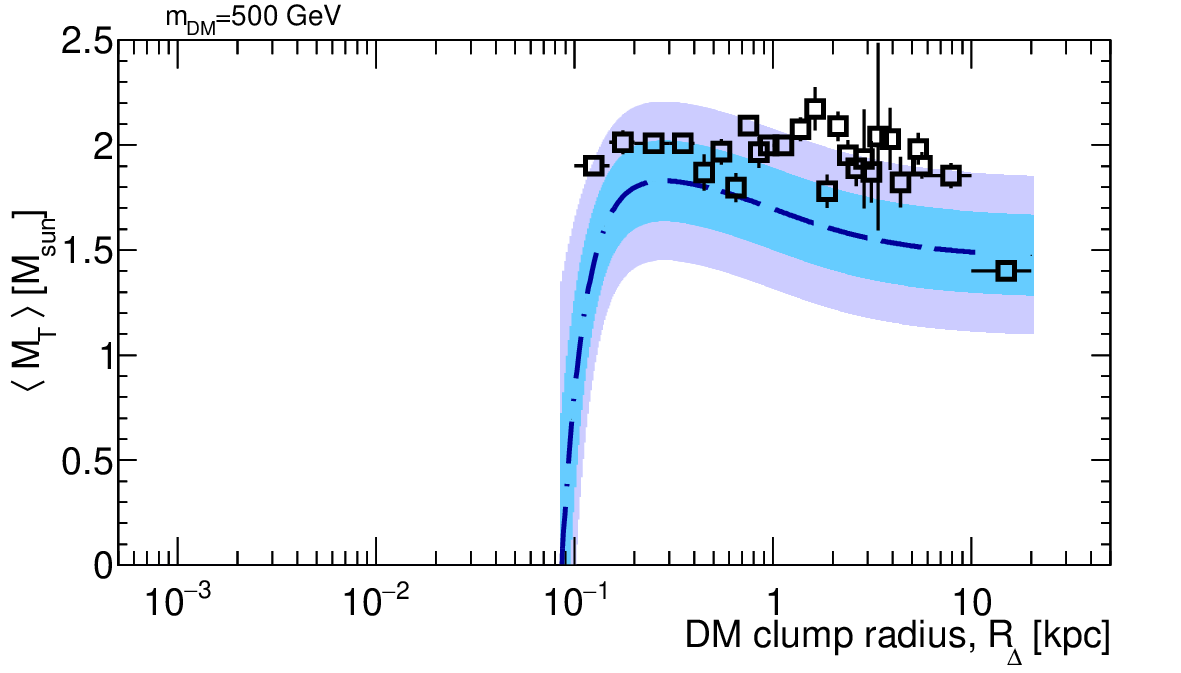}
\caption{
	Fits to the total mass of the binary systems as a function of the clump radius. Top: Clumps were selected in the vicinity of 0.05 kpc to the NSs, clumps scale was requested to be greater than this separation distance ($> 0.05$ kpc). Bottom: Clumps were selected in the vicinity of 0.1 kpc to the NSs, clumps scale was requested to be greater than this separation distance ($> 0.1$ kpc). 
	Left: using the DM mass $m_{\rm{dm}}=100$ GeV. 
	Right: using the DM mass $m_{\rm{dm}}=500$ GeV. Fits performed with the help of the power law function.	
}
\label{fig:fits2_Totalmass_evolution}		
\end{figure*}
	
Due to the inhomogeneity of the DM distribution in the galactic halo, one should expect that NSs and WDs forming in DM clumps have a different structure than those forming outside of them. This is because, in clumps, the DM density is larger than in the environment, and increases going toward the center of the clump. This implies a larger accretion of DM by NSs, and consequently, since larger content of DM in those objects implies a smaller mass \citep{Deliyergiyev:2019vti,DelPopolo:2020pzh}.
	
To determine how the NS mass change inside the clump obtained with the help of \texttt{CLUMPY}, 
the data analysis was performed in several steps. 
First, we plot the NSs mass distribution as a function of the distance to the center of the selected clump, Fig.~\ref{fig:NSmass_vs_dist2ClumpCenter}(left). 
We note that it decreases close to the DM clump center, then flattens towards the outskirt, in the range of distances from 0.01 to 10 kpc. While in the close distance range to the DM clump outskirt, Fig.~\ref{fig:NSmass_vs_dist2ClumpCenter}(right) the NS mass is almost flat.

Second, a fit to the star mass is performed with the main NS mass and total mass of the binary systems, split into the merged and differential selection categories with respect to the DM clump radius. 
The resulting observed tendency in the merged categories is shown in Fig.~\ref{fig:fits2_NSmass_evolution} for $R_{\Delta}$ and $R_{\rm tidal}$ separately, while the results for the differential categories are shown in Fig.~\ref{fig:fits2_Totalmass_evolution}.

More in detail, Fig.~\ref{fig:fits2_NSmass_evolution} shows the mass of NSs change with the clump scale in the case of DM particle mass 100 GeV and 500 GeV. The top panel of Fig.\ref{fig:fits2_NSmass_evolution} shows the NSs mass as a function of the clump radius, 
$R_\Delta$, with the 95\%, and 68\% confidence bands, and for $100$ GeV (left panel), and 500 GeV, (right panel). 
The bottom panels represent the same quantity for the tidal radius, $R_{\rm tidal}$. 
Clumps were selected in the vicinity of 0.01, 0.05, 0.1, 0.25, 0.5, 0.75, 1, 2, and 5 kpc to the NSs, 
and the clump scale was requested to be larger than this separation distance. 
We merged statistics from all clumps we found in the vicinity of 0.01, 0.05, 0.1, 0.25, 0.5, 0.75, 1, 2, and 5 kpc. 
In Fig.~\ref{fig:fits2_NSmass_evolution} (top), one may notice some transition region between very small clumps ($10^{-2}-10^{-1}$ kpc) and clumps with a radius greater than $10^{-1}$ kpc. We assume that such artifacts might be driven by the \texttt{CLUMPY} clump modeling.
	
Finally, in Fig.\ref{fig:fits2_Totalmass_evolution}, we plot the total mass of the binary systems as a function of the clump radius. The plot was built in a similar way to Fig.~\ref{fig:fits2_NSmass_evolution}, and again, we observe a decrease of the mass towards the outskirts of the clump. In other terms, the plots show that the NSs have a mass decreasing towards the outskirts of the clump, namely the opposite of what is predicted by \citep{DelPopolo:2019nng}, assuming that DM interaction strength with the ordinary matter is low.

However, we do not see this as a discrepancy \citep{DelPopolo:2019nng} and our results. In \citep{DelPopolo:2019nng} 
NS mass was shown as a function of distance, while Figs.~\ref{fig:fits2_NSmass_evolution} -\ref{fig:fits2_Totalmass_evolution} show the NS mass as a function of the clump scale.

\section{Conclusions and discussions}
\label{sec:Conclusions}

In this paper, we checked the prediction of \citep{DelPopolo:2019nng}, namely that the neutron stars (NSs) mass decreases with the increasing accumulation of dark matter (DM), and consequently, in our galaxy, the NSs mass decreases going toward the galactic center. 
The predictions made in \citep{DelPopolo:2019nng} were based on the distribution of DM in our galaxy, characterized by an increase going towards the galactic center. 
Changes in the NSs structure and decrease in their mass when getting closer to the galactic center was only produced in the very vicinity to that galactic center. 
Unfortunately, nowadays, we do not have enough information on NSs close to the galactic center. 
To verify our prediction, we then turned to DM clumps contained in the galaxy. These objects are characterized by an increase of the density towards their center, their density reaching similar values to that in the vicinity of the galactic center.

We then used the known NSs located in DM clumps. 
To check the quoted predictions, and to generate the clumps, we used \texttt{CLUMPY} simulations of any Galactic or extragalactic DM halo including substructures: halo-to-halo concentration scatter, with several levels of substructures, and triaxiality of the DM halos \citep{Bonnivard:2015pia,Charbonnier:2012gf}. The mass range of the DM clumps that we found in the vicinity of the neutron stars is $10< M_{\rm{clump}} < 10^{8} M_{\odot}$, while the scale of these clumps is $10^{-3}< R_{\rm{clump}} < 10$ kpc. 
We studied how the mass of NSs changes in the quoted clumps, showing that the NSs mass is almost stable going towards the outskirts of the DM clumps. 
This result is in agreement with \citep{DelPopolo:2019nng} taking into account that at distances greater than 0.01 kpc from the clump center, the NSs are much closer to the outskirt of selected DM clump. Namely, following \citep{DelPopolo:2019nng}, the NSs mass is already on the plateau region. However, if we take a close look on how the NS mass changes close to the clump outskirt, we start to see some structure in the NS mass distribution.

The obtained results are based on the techniques described. The clumpy morphology of galaxies has been produced by the idealized cosmological numerical simulations of isolated galaxies with the help of \texttt{CLUMPY}. 
Using simulations with feedback models as realistic as possible and in a full cosmological context is the most promising way to improve our understanding of the nature of star-forming clumps and the relations between clumps and NSs.
		
The reliability of the EoS constraints depends on the degree to which systematic errors can be controlled \citep[in many current analyses, these errors are significantly larger than the formal statistical uncertainties; see][]{Miller:2013tca, Miller:2016pom}, as well as on the precision of the astrophysical models that are applied to the data.
			
Because neither model atmosphere spectra  \citep[e.g.,][]{Suleimanov:2012sq}) nor the most accurately measured observed spectra \citep{Miller:2011wt} are exactly Planckian, using Planck fits as proxies throws away information and could even introduce biases. 
			
All current methods for determining the radii of NSs using their X-ray fluxes and spectra are subject to astrophysical effects that can confuse or bias the radius measurement. Furthermore, in most cases the data are not yet precise enough to determine whether the model being used correctly describes the data. It is therefore possible that a model may provide a statistically good fit and an apparently tight radius constraint but a value for the radius that is strongly biased relative to the true value.
			
There are recent developments on this subject \citep{Nattila:2017wtj} which aimed to assess the accuracy of future measurement methods and to explore possible biases in the results. Due to these experimental challenges, the radius measurements for the observed NSs are reported less frequently than their mass measurements, see Tables~\ref{tab:NSlist_part1}-\ref{tab:NSlist_part2}. 
			
Therefore, 
our analysis has benefited from recent advancements in technology, which provide 
unprecedented mass and distance measurements of the known NSs. Significant effort has been spent attempting to simulate the DM clumps distribution in the Galaxy, as well as to link this distribution to other aspects of DM clumps phenomenology.

The results from this research will provide valuable knowledge of the Galactic pulsar population for the planning of future DM survey strategies. 
		
However, our efforts are intrinsically constrained by those pulsars which are currently known and available for study. The properties of the NSs that were used, are collected in Tables \ref{tab:NSlist_part1}-\ref{tab:NSlist_part2}. 
Further characterizing the population statistics of NSs will allow for a greater understanding of the pulsar population and DM distribution as a whole and will be vital in the planning of future pulsars and DM surveys to be undertaken with next-generation radio telescopes such as MeerKAT \citep{Bailes:2020qai}, the Square Kilometre Array (SKA) \citep{Dutta:2022wuc} and the Five-hundred-meter Aperture Spherical Telescope (FAST) \citep{Hu:2019okh}. 
A more mature judgment can be formed when the number of the measured NS masses will increase by a factor of 5-10 from those discovered thus far. 
Apart from that, a major caveat in the estimation of DM clump mapping/detection is that our understanding of the underlying NSs distribution is still inadequate and hampered by the small number of known NSs, and the measured uncertainties of the NS masses. 
		
NSs formed in the DMCs could be detectable by next-generation GW detectors, such as Advanced LIGO \citep{LIGOScientific:2014pky}, Advanced Virgo \citep{VIRGO:2014yos}, KAGRA \citep{Somiya:2011np}, and the proposed space-based detector LISA \citep{LISA:2017pwj}. If the NS merger is accompanied by an electromagnetic counterpart, the GW source can be localized and a redshift can be determined. Knowing the redshift of the source allows us to infer its luminosity distance, which in turn constrains the distance between the source and the Earth. The distance to the source can then be used to infer NS properties, such as its mass, spin, and EoS. 
The NS distances inferred through dispersion measures exhibit appreciable variety and are susceptible to biases \citep{Verbiest:2012kh}, a typical measurement uncertainty of $\sim 20\%$ is not unreasonable \citep{Taylor:1993my,Yao_2017}, and indeed is expected to be readily achievable with next-generation radio telescopes \citep{Smits:2011zh}.  
		
The most promising GW sources from DMCs are expected to be low-mass binary NSs (LMBNSs). Low-mass binary systems are expected to form in DMCs due to the high stellar densities, which can lead to dynamical exchanges and three-body interactions \citep{Berezinsky:2014wya}. 

If the merging NSs have masses of $\sim 1.2 M_{\odot}$, the corresponding GW signal might be detectable.
			
Another option is to detect dark clumps via gravitational perturbations of visible objects \citep{Penarrubia:2010pa}. In this regard very wide binaries, i.e. those with separations $a \gtrsim 100$ AU, may enable simple experiments to test for the presence of dark substructures (clumps, halos), for even extremely weak tidal perturbations can disrupt them. In the stellar halo of the Milky Way, perturbers can be identified as inhomogeneities in the Galactic potential. There are huge list of different pertubers on such binaries, therefore isolation of such an effect is not straightforward. See the recent report on this subject in \cite{Inoue:2021dkv}.
				
Ref.~\citep{Penarrubia:2010pa} studied the disruption of wide ($\gtrsim 100$ AU) binaries as a result of interactions with dark substructures orbiting in dwarf galaxies, as well as exploring whether the effects can be detected with our present observational capabilities.
				
If there exists in nature dark matter clumps with compactness $GM/(Rc^{2}) \gtrsim  0.1$, i.e. comparable with neutron stars, GWs emitted during tidal encounters could be measured by ground-based detectors, such as LIGO \citep{Mendes:2016vdr}.

\section*{Acknowledgements}
We wish to acknowledge the support of the UniGe HPC center, which helps us to simulate galaxies. 
ADP thanks the Institute of Astronomy of the Academy of Science in Sofia for the hospitality. 
MLeD acknowledges the financial support by the Lanzhou University starting fund, 
the Fundamental Research Funds for the Central Universities (Grant No. lzujbky-2019-25), National Science Foundation of China (grant No. 12047501) and the 111 Project under Grant No. B20063.

\section*{Data Availability}

The CLUMPY synthetic data underlying this paper are publicly available. In case of the data loss, it can be reproduced with the help of the configuration parameters provided by authors. The code that supports the findings of this study will be shared upon a reasonable request to the corresponding author.



\bibliographystyle{mnras}
\bibliography{biblioNS,old_MasterBib2_1}

\begin{thebibliography}{}
\makeatletter
\relax
\def\mn@urlcharsother{\let\do\@makeother \do\$\do\&\do\#\do\^\do\_\do\%\do\~}
\def\mn@doi{\begingroup\mn@urlcharsother \@ifnextchar [ {\mn@doi@}
  {\mn@doi@[]}}
\def\mn@doi@[#1]#2{\def\@tempa{#1}\ifx\@tempa\@empty \href
  {http://dx.doi.org/#2} {doi:#2}\else \href {http://dx.doi.org/#2} {#1}\fi
  \endgroup}
\def\mn@eprint#1#2{\mn@eprint@#1:#2::\@nil}
\def\mn@eprint@arXiv#1{\href {http://arxiv.org/abs/#1} {{\tt arXiv:#1}}}
\def\mn@eprint@dblp#1{\href {http://dblp.uni-trier.de/rec/bibtex/#1.xml}
  {dblp:#1}}
\def\mn@eprint@#1:#2:#3:#4\@nil{\def\@tempa {#1}\def\@tempb {#2}\def\@tempc
  {#3}\ifx \@tempc \@empty \let \@tempc \@tempb \let \@tempb \@tempa \fi \ifx
  \@tempb \@empty \def\@tempb {arXiv}\fi \@ifundefined
  {mn@eprint@\@tempb}{\@tempb:\@tempc}{\expandafter \expandafter \csname
  mn@eprint@\@tempb\endcsname \expandafter{\@tempc}}}

\bibitem[\protect\citeauthoryear{Aad et~al.}{Aad et~al.}{2023}]{ATLAS:2022bzt}
Aad G.,  et~al., 2023, \mn@doi [JHEP] {10.1007/JHEP07(2023)116}, 07, 116

\bibitem[\protect\citeauthoryear{Aalseth et~al.}{Aalseth
  et~al.}{2011}]{CoGeNT:2010ols}
Aalseth C.~E.,  et~al., 2011, \mn@doi [Phys. Rev. Lett.]
  {10.1103/PhysRevLett.106.131301}, 106, 131301

\bibitem[\protect\citeauthoryear{Aasi et~al.}{Aasi
  et~al.}{2015}]{LIGOScientific:2014pky}
Aasi J.,  et~al., 2015, \mn@doi [Class. Quant. Grav.]
  {10.1088/0264-9381/32/7/074001}, 32, 074001

\bibitem[\protect\citeauthoryear{{Abbott, T. M. C. and others}}{{Abbott, T. M.
  C. and others}}{2023}]{PhysRevD.107.023531}
{Abbott, T. M. C. and others} 2023, \mn@doi [Phys. Rev. D]
  {10.1103/PhysRevD.107.023531}, 107, 023531

\bibitem[\protect\citeauthoryear{Abbott et~al.}{Abbott
  et~al.}{2022}]{DES:2021wwk}
Abbott T. M.~C.,  et~al., 2022, \mn@doi [Phys. Rev. D]
  {10.1103/PhysRevD.105.023520}, 105, 023520

\bibitem[\protect\citeauthoryear{Abdalla, Abramo, Sodre  \& Wang}{Abdalla
  et~al.}{2009}]{Abdalla:2007rd}
Abdalla E.,  Abramo L. R.~W.,  Sodre Jr. L.,   Wang B.,  2009, \mn@doi [Phys.
  Lett.] {10.1016/j.physletb.2009.02.008}, B673, 107

\bibitem[\protect\citeauthoryear{Abdalla, Abramo  \& de Souza}{Abdalla
  et~al.}{2010}]{Abdalla:2009mt}
Abdalla E.,  Abramo L.~R.,   de Souza J. C.~C.,  2010, \mn@doi [Phys. Rev.]
  {10.1103/PhysRevD.82.023508}, D82, 023508

\bibitem[\protect\citeauthoryear{Abuter, R.}{Abuter
et~al.}{2018}]{GRAVITY:2018ofz}
Abuter R., et~al., 2018, \mn@doi [Astron. Astrophys.]
{10.1051/0004-6361/201833718}, 615, L15

\bibitem[\protect\citeauthoryear{Acernese et~al.}{Acernese
  et~al.}{2015}]{VIRGO:2014yos}
Acernese F.,  et~al., 2015, \mn@doi [Class. Quant. Grav.]
  {10.1088/0264-9381/32/2/024001}, 32, 024001

\bibitem[\protect\citeauthoryear{Ackermann et~al.}{Ackermann
  et~al.}{2015}]{Fermi-LAT:2015att}
Ackermann M.,  et~al., 2015, \mn@doi [Phys. Rev. Lett.]
  {10.1103/PhysRevLett.115.231301}, 115, 231301

\bibitem[\protect\citeauthoryear{Ade et~al.}{Ade et~al.}{2014}]{Planck:2013pxb}
Ade P. A.~R.,  et~al., 2014, \mn@doi [Astron. Astrophys.]
  {10.1051/0004-6361/201321591}, 571, A16

\bibitem[\protect\citeauthoryear{Agnese et~al.}{Agnese
  et~al.}{2014}]{SuperCDMS:2013eoh}
Agnese R.,  et~al., 2014, \mn@doi [Phys. Rev. Lett.]
  {10.1103/PhysRevLett.112.041302}, 112, 041302

\bibitem[\protect\citeauthoryear{Akerib et~al.}{Akerib
  et~al.}{2017}]{LUX:2016ggv}
Akerib D.~S.,  et~al., 2017, \mn@doi [Phys. Rev. Lett.]
  {10.1103/PhysRevLett.118.021303}, 118, 021303

\bibitem[\protect\citeauthoryear{Ali-Ha\"\i{}moud, Kovetz  \&
  Silk}{Ali-Ha\"\i{}moud et~al.}{2016}]{Ali-Haimoud:2015bfg}
Ali-Ha\"\i{}moud Y.,  Kovetz E.~D.,   Silk J.,  2016, \mn@doi [Phys. Rev. D]
  {10.1103/PhysRevD.93.043508}, 93, 043508

\bibitem[\protect\citeauthoryear{Alonso-\'Alvarez \& Jaeckel}{Alonso-\'Alvarez
  \& Jaeckel}{2018}]{Alonso-Alvarez:2018tus}
Alonso-\'Alvarez G.,  Jaeckel J.,  2018, \mn@doi [JCAP]
  {10.1088/1475-7516/2018/10/022}, 10, 022

\bibitem[\protect\citeauthoryear{Alsing, Silva  \& Berti}{Alsing
  et~al.}{2018}]{Alsing:2017bbc}
Alsing J.,  Silva H.~O.,   Berti E.,  2018, \mn@doi [Mon. Not. Roy. Astron.
  Soc.] {10.1093/mnras/sty1065}, 478, 1377

\bibitem[\protect\citeauthoryear{Amaro-Seoane et~al.}{Amaro-Seoane
  et~al.}{2017}]{LISA:2017pwj}
Amaro-Seoane P.,  et~al., 2017, arXiv:1702.00786[astro-ph.IM]

\bibitem[\protect\citeauthoryear{Andersen \& Ransom}{Andersen \&
  Ransom}{2018}]{Andersen:2018nsx}
Andersen B.~C.,  Ransom S.~M.,  2018, \mn@doi [Astrophys. J. Lett.]
  {10.3847/2041-8213/aad59f}, 863, L13

\bibitem[\protect\citeauthoryear{Antoniadis, van Kerkwijk, Koester, Freire,
  Wex, Tauris, Kramer  \& Bassa}{Antoniadis et~al.}{2012}]{Antoniadis:2012vy}
Antoniadis J.,  van Kerkwijk M.~H.,  Koester D.,  Freire P. C.~C.,  Wex N.,
  Tauris T.~M.,  Kramer M.,   Bassa C.~G.,  2012, \mn@doi [Mon. Not. Roy.
  Astron. Soc.] {10.1111/j.1365-2966.2012.21124.x}, 423, 3316

\bibitem[\protect\citeauthoryear{Antoniadis et~al.}{Antoniadis
  et~al.}{2013}]{Antoniadis:2013pzd}
Antoniadis J.,  et~al., 2013, \mn@doi [Science] {10.1126/science.1233232}, 340,
  6131

\bibitem[\protect\citeauthoryear{Aprile et~al.}{Aprile
  et~al.}{2012}]{XENON100:2012itz}
Aprile E.,  et~al., 2012, \mn@doi [Phys. Rev. Lett.]
  {10.1103/PhysRevLett.109.181301}, 109, 181301

\bibitem[\protect\citeauthoryear{Aprile et~al.}{Aprile
  et~al.}{2016}]{XENON100:2016sjq}
Aprile E.,  et~al., 2016, \mn@doi [Phys. Rev. D] {10.1103/PhysRevD.94.122001},
  94, 122001

\bibitem[\protect\citeauthoryear{Aprile et~al.}{Aprile
  et~al.}{2022}]{XENONCollaboration:2022kmb}
Aprile E.,  et~al., 2022, \mn@doi [Phys. Rev. Lett.]
  {10.1103/PhysRevLett.129.161805}, 129, 161805

\bibitem[\protect\citeauthoryear{Archibald et~al.,}{Archibald
  et~al.}{2018}]{Archibald:2018oxs}
Archibald A.~M.,  et~al., 2018, \mn@doi [Nature] {10.1038/s41586-018-0265-1},
  559, 73

\bibitem[\protect\citeauthoryear{Armand \& Herrmann}{Armand \&
  Herrmann}{2022}]{Armand:2022sjf}
Armand C.,  Herrmann B.,  2022, \mn@doi [JCAP] {10.1088/1475-7516/2022/11/055},
  11, 055

\bibitem[\protect\citeauthoryear{Arzoumanian et~al.}{Arzoumanian
  et~al.}{2018}]{NANOGrav:2017wvv}
Arzoumanian Z.,  et~al., 2018, \mn@doi [Astrophys. J. Suppl.]
  {10.3847/1538-4365/aab5b0}, 235, 37

\bibitem[\protect\citeauthoryear{Aslanyan, Price, Adams, Bringmann, Clark,
  Easther, Lewis  \& Scott}{Aslanyan et~al.}{2016}]{Aslanyan:2015hmi}
Aslanyan G.,  Price L.~C.,  Adams J.,  Bringmann T.,  Clark H.~A.,  Easther R.,
   Lewis G.~F.,   Scott P.,  2016, \mn@doi [Phys. Rev. Lett.]
  {10.1103/PhysRevLett.117.141102}, 117, 141102

\bibitem[\protect\citeauthoryear{Bailes et~al.}{Bailes
  et~al.}{2020}]{Bailes:2020qai}
Bailes M.,  et~al., 2020, \mn@doi [Publ. Astron. Soc. Austral.]
  {10.1017/pasa.2020.19}, 37, e028

\bibitem[\protect\citeauthoryear{Bak~Nielsen, Patruno  \& D'Angelo}{Bak~Nielsen
  et~al.}{2017}]{BakNielsen:2017ssx}
Bak~Nielsen A.-S.,  Patruno A.,   D'Angelo C.,  2017, \mn@doi [Mon. Not. Roy.
  Astron. Soc.] {10.1093/mnras/stx491}, 468, 824

\bibitem[\protect\citeauthoryear{Balakrishnan et~al.}{Balakrishnan
  et~al.}{2023}]{Balakrishnan:2023hdi}
Balakrishnan V.,  et~al., 2023, \mn@doi [Astrophys. J. Lett.]
  {10.3847/2041-8213/acae99}, 942, L35

\bibitem[\protect\citeauthoryear{Barr, Freire, Kramer, Champion, Berezina,
  Bassa, Lyne  \& Stappers}{Barr et~al.}{2017}]{Barr:2016vxv}
Barr E.~D.,  Freire P. C.~C.,  Kramer M.,  Champion D.~J.,  Berezina M.,  Bassa
  C.~G.,  Lyne A.~G.,   Stappers B.~W.,  2017, \mn@doi [Mon. Not. Roy. Astron.
  Soc.] {10.1093/mnras/stw2947}, 465, 1711

\bibitem[\protect\citeauthoryear{Bassa, van Kerkwijk, Koester  \&
  Verbunt}{Bassa et~al.}{2006}]{Bassa:2006mj}
Bassa C.~G.,  van Kerkwijk M.~H.,  Koester D.,   Verbunt F.,  2006, \mn@doi
  [Astron. Astrophys.] {10.1051/0004-6361:20065181}, 456, 295

\bibitem[\protect\citeauthoryear{Bell, Bessell, Stappers, Bailes  \&
  Kaspi}{Bell et~al.}{1995}]{Bell:1995hr}
Bell J.~F.,  Bessell M.~S.,  Stappers B.~W.,  Bailes M.,   Kaspi V.~M.,  1995,
  \mn@doi [Astrophys. J. Lett.] {10.1086/309565}, 447, L117

\bibitem[\protect\citeauthoryear{Bell, Busoni, Robles  \& Virgato}{Bell
  et~al.}{2020}]{Bell:2020jou}
Bell N.~F.,  Busoni G.,  Robles S.,   Virgato M.,  2020, \mn@doi [JCAP]
  {10.1088/1475-7516/2020/09/028}, 09, 028

\bibitem[\protect\citeauthoryear{Bellm et~al.}{Bellm
  et~al.}{2016}]{Bellm:2015dfa}
Bellm E.~C.,  et~al., 2016, \mn@doi [Astrophys. J.]
  {10.3847/0004-637X/816/2/74}, 816, 74

\bibitem[\protect\citeauthoryear{Berezina et~al.}{Berezina
  et~al.}{2017}]{Berezina:2017vts}
Berezina M.,  et~al., 2017, \mn@doi [Mon. Not. Roy. Astron. Soc.]
  {10.1093/mnras/stx1518}, 470, 4421

\bibitem[\protect\citeauthoryear{Berezinsky, Dokuchaev  \&
  Eroshenko}{Berezinsky et~al.}{2003}]{Berezinsky2003}
Berezinsky V.,  Dokuchaev V.,   Eroshenko Y.,  2003, \mn@doi [Phys. Rev. D]
  {10.1103/PhysRevD.68.103003}, 68, 103003

\bibitem[\protect\citeauthoryear{Berezinsky, Dokuchaev  \&
  Eroshenko}{Berezinsky et~al.}{2005}]{10.1142:97898127018480012}
Berezinsky V.~S.,  Dokuchaev V.~I.,   Eroshenko Y.~N.,  2005, \mn@doi [The
  Identification of Dark Matter] {{10.1142/9789812701848$\_$0012}}, pp 81--86

\bibitem[\protect\citeauthoryear{Berezinsky, Dokuchaev  \&
  Eroshenko}{Berezinsky et~al.}{2006}]{Berezinsky:2005py}
Berezinsky V.,  Dokuchaev V.,   Eroshenko Y.,  2006, \mn@doi [Phys. Rev. D]
  {10.1103/PhysRevD.73.063504}, 73, 063504

\bibitem[\protect\citeauthoryear{Berezinsky, Dokuchaev  \&
  Eroshenko}{Berezinsky et~al.}{2013}]{Berezinsky:2013fxa}
Berezinsky V.~S.,  Dokuchaev V.~I.,   Eroshenko Y.~N.,  2013, \mn@doi [JCAP]
  {10.1088/1475-7516/2013/11/059}, 11, 059

\bibitem[\protect\citeauthoryear{Berezinsky, Dokuchaev  \&
  Eroshenko}{Berezinsky et~al.}{2014}]{Berezinsky:2014wya}
Berezinsky V.~S.,  Dokuchaev V.~I.,   Eroshenko Y.~N.,  2014, \mn@doi [Phys.
  Usp.] {10.3367/UFNe.0184.201401a.0003}, 57, 1

\bibitem[\protect\citeauthoryear{Bernabei et~al.}{Bernabei
  et~al.}{2010}]{DAMA:2010gpn}
Bernabei R.,  et~al., 2010, \mn@doi [Eur. Phys. J. C]
  {10.1140/epjc/s10052-010-1303-9}, 67, 39

\bibitem[\protect\citeauthoryear{Berthereau et~al.,}{Berthereau
  et~al.}{2023}]{Berthereau:2023aod}
Berthereau A.,  et~al., 2023, \mn@doi [Astron. Astrophys.]
  {10.1051/0004-6361/202346228}, 674, A71

\bibitem[\protect\citeauthoryear{Bertolami, Gil~Pedro  \& Le~Delliou}{Bertolami
  et~al.}{2007}]{Bertolami:2007zm}
Bertolami O.,  Gil~Pedro F.,   Le~Delliou M.,  2007, \mn@doi [Phys. Lett.]
  {10.1016/j.physletb.2007.08.046}, B654, 165

\bibitem[\protect\citeauthoryear{Bertolami, Gil~Pedro  \& Le~Delliou}{Bertolami
  et~al.}{2008}]{Bertolami:2008rz}
Bertolami O.,  Gil~Pedro F.,   Le~Delliou M.,  2008, \mn@doi [EAS Publ. Ser.]
  {10.1051/eas:0830019}, 30, 161

\bibitem[\protect\citeauthoryear{Bertolami, Pedro  \& Le~Delliou}{Bertolami
  et~al.}{2009}]{Bertolami:2007tq}
Bertolami O.,  Pedro F.~G.,   Le~Delliou M.,  2009, \mn@doi [Gen. Rel. Grav.]
  {10.1007/s10714-009-0810-1}, 41, 2839

\bibitem[\protect\citeauthoryear{Bertolami, Gil~Pedro  \& Le~Delliou}{Bertolami
  et~al.}{2012}]{Bertolami:2012yp}
Bertolami O.,  Gil~Pedro F.,   Le~Delliou M.,  2012, \mn@doi [Gen. Rel. Grav.]
  {10.1007/s10714-012-1327-6}, 44, 1073

\bibitem[\protect\citeauthoryear{Bertone \& Fairbairn}{Bertone \&
  Fairbairn}{2008}]{Bertone:2007ae}
Bertone G.,  Fairbairn M.,  2008, \mn@doi [Phys. Rev.]
  {10.1103/PhysRevD.77.043515}, D77, 043515

\bibitem[\protect\citeauthoryear{Bertschinger}{Bertschinger}{1985}]{Bertschinger:1985pd}
Bertschinger E.,  1985, \mn@doi [Astrophys. J. Suppl.] {10.1086/191028}, 58, 39

\bibitem[\protect\citeauthoryear{Betoule et~al.}{Betoule
  et~al.}{2014}]{SDSS:2014iwm}
Betoule M.,  et~al., 2014, \mn@doi [Astron. Astrophys.]
  {10.1051/0004-6361/201423413}, 568, A22

\bibitem[\protect\citeauthoryear{Bhalerao, van Kerkwijk  \& Harrison}{Bhalerao
  et~al.}{2012}]{Bhalerao_2012}
Bhalerao V.~B.,  van Kerkwijk M.~H.,   Harrison F.~A.,  2012, \mn@doi [The
  Astrophysical Journal] {10.1088/0004-637X/757/1/10}, 757, 10

\bibitem[\protect\citeauthoryear{Bhat \& Paul}{Bhat \&
  Paul}{2020}]{Bhat:2019tnz}
Bhat S.~A.,  Paul A.,  2020, \mn@doi [Eur. Phys. J. C]
  {10.1140/epjc/s10052-020-8072-x}, 80, 544

\bibitem[\protect\citeauthoryear{Bhat, Bailes  \& Verbiest}{Bhat
  et~al.}{2008}]{Bhat:2008ck}
Bhat N. D.~R.,  Bailes M.,   Verbiest J. P.~W.,  2008, \mn@doi [Phys. Rev. D]
  {10.1103/PhysRevD.77.124017}, 77, 124017

\bibitem[\protect\citeauthoryear{Bonaca, Hogg, Price-Whelan  \& Conroy}{Bonaca
  et~al.}{2018}]{Bonaca:2018fek}
Bonaca A.,  Hogg D.~W.,  Price-Whelan A.~M.,   Conroy C.,  2018, ]
  {10.3847/1538-4357/ab2873}

\bibitem[\protect\citeauthoryear{Bonnivard et~al.}{Bonnivard
  et~al.}{2015}]{Bonnivard:2015xpq}
Bonnivard V.,  et~al., 2015, \mn@doi [Mon. Not. Roy. Astron. Soc.]
  {10.1093/mnras/stv1601}, 453, 849

\bibitem[\protect\citeauthoryear{Bonnivard, H\"utten, Nezri, Charbonnier,
  Combet  \& Maurin}{Bonnivard et~al.}{2016a}]{Bonnivard:2015pia}
Bonnivard V.,  H\"utten M.,  Nezri E.,  Charbonnier A.,  Combet C.,   Maurin
  D.,  2016a, \mn@doi [Comput. Phys. Commun.] {10.1016/j.cpc.2015.11.012}, 200,
  336

\bibitem[\protect\citeauthoryear{Bonnivard, Maurin  \& Walker}{Bonnivard
  et~al.}{2016b}]{Bonnivard:2015vua}
Bonnivard V.,  Maurin D.,   Walker M.~G.,  2016b, \mn@doi [Mon. Not. Roy.
  Astron. Soc.] {10.1093/mnras/stw1691}, 462, 223

\bibitem[\protect\citeauthoryear{Bringmann}{Bringmann}{2009}]{Bringmann:2009vf}
Bringmann T.,  2009, \mn@doi [New J. Phys.] {10.1088/1367-2630/11/10/105027},
  11, 105027

\bibitem[\protect\citeauthoryear{{Bringmann}, {Scott}  \& {Akrami}}{{Bringmann}
  et~al.}{2012}]{Bringmann2012}
{Bringmann} T.,  {Scott} P.,   {Akrami} Y.,  2012, \mn@doi [\prd]
  {10.1103/PhysRevD.85.125027}, \href
  {http://adsabs.harvard.edu/abs/2012PhRvD..85l5027B} {85, 125027}

\bibitem[\protect\citeauthoryear{Bullock, Kolatt, Sigad, Somerville, Kravtsov,
  Klypin, Primack  \& Dekel}{Bullock et~al.}{2001}]{Bullock:1999he}
Bullock J.~S.,  Kolatt T.~S.,  Sigad Y.,  Somerville R.~S.,  Kravtsov A.~V.,
  Klypin A.~A.,  Primack J.~R.,   Dekel A.,  2001, \mn@doi [Mon. Not. Roy.
  Astron. Soc.] {10.1046/j.1365-8711.2001.04068.x}, 321, 559

\bibitem[\protect\citeauthoryear{Cameron et~al.}{Cameron
  et~al.}{2018}]{Cameron:2017ody}
Cameron A.~D.,  et~al., 2018, \mn@doi [Mon. Not. Roy. Astron. Soc.]
  {10.1093/mnrasl/sly003}, 475, L57

\bibitem[\protect\citeauthoryear{Cameron et~al.}{Cameron
  et~al.}{2020}]{Cameron:2020pin}
Cameron A.~D.,  et~al., 2020, \mn@doi [Mon. Not. Roy. Astron. Soc.]
  {10.1093/mnras/staa039}, 493, 1063

\bibitem[\protect\citeauthoryear{Cameron et~al.}{Cameron
  et~al.}{2022}]{Cameron:2022soa}
Cameron A.~D.,  et~al., 2022, in {16th Marcel Grossmann Meeting on~Recent
  Developments in Theoretical and Experimental General Relativity, Astrophysics
  and Relativistic Field Theories}.  (\mn@eprint {arXiv} {2203.15995}),
  \mn@doi{10.1142/9789811269776_0312}

\bibitem[\protect\citeauthoryear{Casares, Hernandez, Israelian  \&
  Rebolo}{Casares et~al.}{2010}]{Casares:2009vq}
Casares J.,  Hernandez J. I.~G.,  Israelian G.,   Rebolo R.,  2010, \mn@doi
  [Mon. Not. Roy. Astron. Soc.] {10.1111/j.1365-2966.2009.15828.x}, 401, 2517

\bibitem[\protect\citeauthoryear{Charbonnier et~al.}{Charbonnier
  et~al.}{2011}]{Charbonnier:2011ft}
Charbonnier A.,  et~al., 2011, \mn@doi [Mon. Not. Roy. Astron. Soc.]
  {10.1111/j.1365-2966.2011.19387.x}, 418, 1526

\bibitem[\protect\citeauthoryear{Charbonnier, Combet  \& Maurin}{Charbonnier
  et~al.}{2012}]{Charbonnier:2012gf}
Charbonnier A.,  Combet C.,   Maurin D.,  2012, \mn@doi [Comput. Phys. Commun.]
  {10.1016/j.cpc.2011.10.017}, 183, 656

\bibitem[\protect\citeauthoryear{Chatrchyan et~al.}{Chatrchyan
  et~al.}{2012}]{CMS:2012ucb}
Chatrchyan S.,  et~al., 2012, \mn@doi [JHEP] {10.1007/JHEP09(2012)094}, 09, 094

\bibitem[\protect\citeauthoryear{Chen et~al.}{Chen et~al.}{2023}]{Chen:2023lzp}
Chen W.,  et~al., 2023, \mn@doi [Mon. Not. Roy. Astron. Soc.]
  {10.1093/mnras/stad029}, 520, 3847

\bibitem[\protect\citeauthoryear{Ciarcelluti \& Sandin}{Ciarcelluti \&
  Sandin}{2011}]{Ciarcelluti:2010ji}
Ciarcelluti P.,  Sandin F.,  2011, \mn@doi [Phys. Lett.]
  {10.1016/j.physletb.2010.11.021}, B695, 19

\bibitem[\protect\citeauthoryear{Cirelli et~al.,}{Cirelli
  et~al.}{2011}]{Cirelli:2010xx}
Cirelli M.,  et~al., 2011, \mn@doi [JCAP] {10.1088/1475-7516/2012/10/E01}, 03,
  051

\bibitem[\protect\citeauthoryear{Clark et~al.}{Clark
  et~al.}{2021}]{Clark:2020hbv}
Clark C.~J.,  et~al., 2021, \mn@doi [Mon. Not. Roy. Astron. Soc.]
  {10.1093/mnras/staa3484}, 502, 915

\bibitem[\protect\citeauthoryear{Cocozza, Ferraro, Possenti  \&
  D'Amico}{Cocozza et~al.}{2006}]{Cocozza:2006mm}
Cocozza G.,  Ferraro F.~R.,  Possenti A.,   D'Amico N.,  2006, \mn@doi
  [Astrophys. J. Lett.] {10.1086/504040}, 641, L129

\bibitem[\protect\citeauthoryear{Combet, Maurin, Nezri, Pointecouteau, Hinton
  \& White}{Combet et~al.}{2012}]{Combet:2012tt}
Combet C.,  Maurin D.,  Nezri E.,  Pointecouteau E.,  Hinton J.~A.,   White R.,
   2012, \mn@doi [Phys. Rev. D] {10.1103/PhysRevD.85.063517}, 85, 063517

\bibitem[\protect\citeauthoryear{Conrad}{Conrad}{2014}]{Conrad:2014tla}
Conrad J.,  2014, in {Interplay between Particle and Astroparticle physics
  (IPA2014) London, United Kingdom, August 18-22, 2014}.  (\mn@eprint {arXiv}
  {1411.1925}), \url
  {http://inspirehep.net/record/1326617/files/arXiv:1411.1925.pdf}

\bibitem[\protect\citeauthoryear{Corongiu et~al.}{Corongiu
  et~al.}{2012}]{Corongiu:2012as}
Corongiu A.,  et~al., 2012, \mn@doi [Astrophys. J.]
  {10.1088/0004-637X/760/2/100}, 760, 100

\bibitem[\protect\citeauthoryear{Dai \& Stojkovic}{Dai \&
  Stojkovic}{2009}]{Dai:2009ik}
Dai D.-C.,  Stojkovic D.,  2009, \mn@doi [JHEP]
  {10.1088/1126-6708/2009/08/052}, 08, 052

\bibitem[\protect\citeauthoryear{Das, Malik  \& Nayak}{Das
  et~al.}{2019}]{Das:2018frc}
Das A.,  Malik T.,   Nayak A.~C.,  2019, \mn@doi [Phys. Rev. D]
  {10.1103/PhysRevD.99.043016}, 99, 043016

\bibitem[\protect\citeauthoryear{Das, Kumar, Kumar, Kumar~Biswal, Nakatsukasa,
  Li  \& Patra}{Das et~al.}{2020}]{Das:2020vng}
Das H.~C.,  Kumar A.,  Kumar B.,  Kumar~Biswal S.,  Nakatsukasa T.,  Li A.,
  Patra S.~K.,  2020, \mn@doi [Mon. Not. Roy. Astron. Soc.]
  {10.1093/mnras/staa1435}, 495, 4893

\bibitem[\protect\citeauthoryear{Das, Malik  \& Nayak}{Das
  et~al.}{2022}]{Das:2020ecp}
Das A.,  Malik T.,   Nayak A.~C.,  2022, \mn@doi [Phys. Rev. D]
  {10.1103/PhysRevD.105.123034}, 105, 123034

\bibitem[\protect\citeauthoryear{Del~Popolo, Le~Delliou  \&
  Deliyergiyev}{Del~Popolo et~al.}{2020a}]{DelPopolo:2020hel}
Del~Popolo A.,  Le~Delliou M.,   Deliyergiyev M.,  2020a, \mn@doi [Universe]
  {10.3390/universe6120222}, 6, 222

\bibitem[\protect\citeauthoryear{Del~Popolo, Deliyergiyev, Le~Delliou, Tolos
  \& Burgio}{Del~Popolo et~al.}{2020b}]{DelPopolo:2019nng}
Del~Popolo A.,  Deliyergiyev M.,  Le~Delliou M.,  Tolos L.,   Burgio F.,
  2020b, \mn@doi [Phys. Dark Univ.] {10.1016/j.dark.2020.100484}, 28, 100484

\bibitem[\protect\citeauthoryear{Del~Popolo, Deliyergiyev  \&
  Le~Delliou}{Del~Popolo et~al.}{2020c}]{DelPopolo:2020pzh}
Del~Popolo A.,  Deliyergiyev M.,   Le~Delliou M.,  2020c, \mn@doi [Phys. Dark
  Univ.] {10.1016/j.dark.2020.100622}, 30, 100622

\bibitem[\protect\citeauthoryear{Del~Popolo, Deliyergiyev, Le~Delliou, Tolos
  \& Burgio}{Del~Popolo et~al.}{2020d}]{DelPopolo:2019rox}
Del~Popolo A.,  Deliyergiyev M.,  Le~Delliou M.,  Tolos L.,   Burgio F.,
  2020d, \mn@doi [PoS] {10.22323/1.364.0098}, EPS-HEP2019, 098


\bibitem[\protect\citeauthoryear{Deliyergiyev, Del~Popolo, Le~Delliou}{Deliyergiyev et~al.}{2023}]{Deliyergiyev:2023uer}
Deliyergiyev M.,  Del~Popolo A.,  Le~Delliou M.,  2023, \mn@doi [Mon. Not. Roy. Astron. Soc.] {10.1093/mnras/stad3311}, 527, 4483--4504


\bibitem[\protect\citeauthoryear{Deliyergiyev, Del~Popolo, Tolos, Le~Delliou,
  Lee  \& Burgio}{Deliyergiyev et~al.}{2019}]{Deliyergiyev:2019vti}
Deliyergiyev M.,  Del~Popolo A.,  Tolos L.,  Le~Delliou M.,  Lee X.,   Burgio
  F.,  2019, \mn@doi [Phys. Rev. D] {10.1103/PhysRevD.99.063015}, 99, 063015

\bibitem[\protect\citeauthoryear{Demorest, Pennucci, Ransom, Roberts  \&
  Hessels}{Demorest et~al.}{2010}]{Demorest:2010bx}
Demorest P.,  Pennucci T.,  Ransom S.,  Roberts M.,   Hessels J.,  2010,
  \mn@doi [Nature] {10.1038/nature09466}, 467, 1081

\bibitem[\protect\citeauthoryear{Desvignes et~al.}{Desvignes
  et~al.}{2016}]{EPTA:2016ndq}
Desvignes G.,  et~al., 2016, \mn@doi [Mon. Not. Roy. Astron. Soc.]
  {10.1093/mnras/stw483}, 458, 3341

\bibitem[\protect\citeauthoryear{Diemand, Kuhlen, Madau, Zemp, Moore, Potter
  \& Stadel}{Diemand et~al.}{2008}]{Diemand:2008in}
Diemand J.,  Kuhlen M.,  Madau P.,  Zemp M.,  Moore B.,  Potter D.,   Stadel
  J.,  2008, \mn@doi [Nature] {10.1038/nature07153}, 454, 735

\bibitem[\protect\citeauthoryear{Dutta, Ghosh, Kar  \& Mukhopadhyaya}{Dutta
  et~al.}{2022}]{Dutta:2022wuc}
Dutta K.,  Ghosh A.,  Kar A.,   Mukhopadhyaya B.,  2022, \mn@doi [JCAP]
  {10.1088/1475-7516/2022/09/005}, 09, 005

\bibitem[\protect\citeauthoryear{Eke, Navarro  \& Steinmetz}{Eke
  et~al.}{2001}]{Eke:2000av}
Eke V.~R.,  Navarro J.~F.,   Steinmetz M.,  2001, \mn@doi [Astrophys. J.]
  {10.1086/321345}, 554, 114

\bibitem[\protect\citeauthoryear{Ellis, Hütsi, Kannike, Marzola, Raidal  \&
  Vaskonen}{Ellis et~al.}{2018}]{Ellis:2018bkr}
Ellis J.,  Hütsi G.,  Kannike K.,  Marzola L.,  Raidal M.,   Vaskonen V.,
  2018, \mn@doi [Phys. Rev.] {10.1103/PhysRevD.97.123007}, D97, 123007

\bibitem[\protect\citeauthoryear{Falanga, Bozzo, Lutovinov, Bonnet-Bidaud,
  Fetisova  \& Puls}{Falanga et~al.}{2015}]{Falanga:2015mra}
Falanga M.,  Bozzo E.,  Lutovinov A.,  Bonnet-Bidaud J.~M.,  Fetisova Y.,
  Puls J.,  2015, \mn@doi [Astron. Astrophys.] {10.1051/0004-6361/201425191},
  577, A130

\bibitem[\protect\citeauthoryear{{Fan}, {Yang}  \& {Chang}}{{Fan}
  et~al.}{2012}]{Zhong2012}
{Fan} Y.-z.,  {Yang} R.-z.,   {Chang} J.,  2012, preprint, \href
  {http://adsabs.harvard.edu/abs/2012arXiv1204.2564F} {} (\mn@eprint {arXiv}
  {1204.2564})

\bibitem[\protect\citeauthoryear{Ferdman}{Ferdman}{2017}]{Ferdman:2017vkx}
Ferdman R.~D.,  2017, \mn@doi [IAU Symp.] {10.1017/S1743921317009139}, 337, 146

\bibitem[\protect\citeauthoryear{Ferdman et~al.}{Ferdman
  et~al.}{2014}]{Ferdman:2014rna}
Ferdman R.~D.,  et~al., 2014, \mn@doi [Mon. Not. Roy. Astron. Soc.]
  {10.1093/mnras/stu1223}, 443, 2183

\bibitem[\protect\citeauthoryear{Ferdman et~al.}{Ferdman
  et~al.}{2020}]{Ferdman:2020huz}
Ferdman R.~D.,  et~al., 2020, \mn@doi [Nature] {10.1038/s41586-020-2439-x},
  583, 211

\bibitem[\protect\citeauthoryear{Fonseca, Stairs  \& Thorsett}{Fonseca
  et~al.}{2014}]{Fonseca:2014qla}
Fonseca E.,  Stairs I.~H.,   Thorsett S.~E.,  2014, \mn@doi [Astrophys. J.]
  {10.1088/0004-637X/787/1/82}, 787, 82

\bibitem[\protect\citeauthoryear{Fonseca et~al.}{Fonseca
  et~al.}{2016}]{Fonseca:2016tux}
Fonseca E.,  et~al., 2016, \mn@doi [Astrophys. J.]
  {10.3847/0004-637X/832/2/167}, 832, 167

\bibitem[\protect\citeauthoryear{Fornasa et~al.,}{Fornasa
  et~al.}{2013}]{Fornasa:2012gu}
Fornasa M.,  et~al., 2013, \mn@doi [Mon. Not. Roy. Astron. Soc.]
  {10.1093/mnras/sts444}, 429, 1529

\bibitem[\protect\citeauthoryear{Fortin, Bejger, Haensel  \& Zdunik}{Fortin
  et~al.}{2016}]{Fortin:2014ufa}
Fortin M.,  Bejger M.,  Haensel P.,   Zdunik J.~L.,  2016, \mn@doi [Astron.
  Astrophys.] {10.1051/0004-6361/201424911}, 586, A109

\bibitem[\protect\citeauthoryear{Freire}{Freire}{2008}]{Freire:2007sg}
Freire P. C.~C.,  2008, \mn@doi [AIP Conf. Proc.] {10.1063/1.2900274}, 983, 459

\bibitem[\protect\citeauthoryear{Freire, Ransom, Begin, Stairs, Hessels, Frey
  \& Camilo}{Freire et~al.}{2008a}]{Freire:2007jd}
Freire P. C.~C.,  Ransom S.~M.,  Begin S.,  Stairs I.~H.,  Hessels J. W.~T.,
  Frey L.~H.,   Camilo F.,  2008a, \mn@doi [Astrophys. J.] {10.1086/526338},
  675, 670

\bibitem[\protect\citeauthoryear{Freire, Wolszczan, Berg  \& Hessels}{Freire
  et~al.}{2008b}]{Freire:2007xg}
Freire P. C.~C.,  Wolszczan A.,  Berg M. v.~d.,   Hessels J. W.~T.,  2008b,
  \mn@doi [Astrophys. J.] {10.1086/587832}, 679, 1433

\bibitem[\protect\citeauthoryear{Gao, White, Jenkins, Stoehr  \& Springel}{Gao
  et~al.}{2004}]{Gao:2004au}
Gao L.,  White S. D.~M.,  Jenkins A.,  Stoehr F.,   Springel V.,  2004, \mn@doi
  [Mon. Not. Roy. Astron. Soc.] {10.1111/j.1365-2966.2004.08360.x}, 355, 819

\bibitem[\protect\citeauthoryear{{Gelino}, {Tomsick}  \& {Heindl}}{{Gelino}
  et~al.}{2002}]{2002AAS...201.5405G}
{Gelino} D.~M.,  {Tomsick} J.~A.,   {Heindl} W.~A.,  2002, Bulletin of the
  American Astronomical Society, \href
  {https://ui.adsabs.harvard.edu/abs/2002AAS...201.5405G} {201, 54.05}

\bibitem[\protect\citeauthoryear{Goldman \& Nussinov}{Goldman \&
  Nussinov}{1989}]{Goldman:1989nd}
Goldman I.,  Nussinov S.,  1989, \mn@doi [Phys. Rev.]
  {10.1103/PhysRevD.40.3221}, D40, 3221

\bibitem[\protect\citeauthoryear{Goldman, Mohapatra, Nussinov, Rosenbaum  \&
  Teplitz}{Goldman et~al.}{2013}]{Goldman:2013qla}
Goldman I.,  Mohapatra R.~N.,  Nussinov S.,  Rosenbaum D.,   Teplitz V.,  2013,
  \mn@doi [Phys. Lett.] {10.1016/j.physletb.2013.07.017}, B725, 200

\bibitem[\protect\citeauthoryear{Gonzalez-Caniulef, Guillot  \&
  Reisenegger}{Gonzalez-Caniulef et~al.}{2019}]{Gonzalez-Caniulef:2019wzi}
Gonzalez-Caniulef D.,  Guillot S.,   Reisenegger A.,  2019, \mn@doi [Mon. Not.
  Roy. Astron. Soc.] {10.1093/mnras/stz2941}, 490, 5848

\bibitem[\protect\citeauthoryear{Green}{Green}{2017}]{Green:2017odb}
Green A.~M.,  2017, \mn@doi [J. Phys. G] {10.1088/1361-6471/aa7819}, 44, 084001

\bibitem[\protect\citeauthoryear{Guillot et~al.}{Guillot
  et~al.}{2019}]{Guillot:2019vqp}
Guillot S.,  et~al., 2019, \mn@doi [Astrophys. J. Lett.]
  {10.3847/2041-8213/ab511b}, 887, L27

\bibitem[\protect\citeauthoryear{Guo et~al.}{Guo et~al.}{2021}]{Guo:2021bqa}
Guo Y.~J.,  et~al., 2021, \mn@doi [Astron. Astrophys.]
  {10.1051/0004-6361/202141450}, 654, A16

\bibitem[\protect\citeauthoryear{Guver, Ozel, Cabrera-Lavers  \&
  Wroblewski}{Guver et~al.}{2010a}]{Guver:2008gc}
Guver T.,  Ozel F.,  Cabrera-Lavers A.,   Wroblewski P.,  2010a, \mn@doi
  [Astrophys. J.] {10.1088/0004-637X/712/2/964}, 712, 964

\bibitem[\protect\citeauthoryear{Guver, Wroblewski, Camarota  \& Ozel}{Guver
  et~al.}{2010b}]{Guver:2010td}
Guver T.,  Wroblewski P.,  Camarota L.,   Ozel F.,  2010b, \mn@doi [Astrophys.
  J.] {10.1088/0004-637X/719/2/1807}, 719, 1807

\bibitem[\protect\citeauthoryear{{G{\"u}ver}, {Emre Erkoca}, {Hall Reno}  \&
  {Sarcevic}}{{G{\"u}ver} et~al.}{2014}]{Guver2012}
{G{\"u}ver} T.,  {Emre Erkoca} A.,  {Hall Reno} M.,   {Sarcevic} I.,  2014,
  \mn@doi [\jcap] {10.1088/1475-7516/2014/05/013}, \href
  {http://adsabs.harvard.edu/abs/2014JCAP...05..013G} {5, 013}

\bibitem[\protect\citeauthoryear{Haniewicz, Ferdman, Freire, Champion, Bunting,
  Lorimer  \& McLaughlin}{Haniewicz et~al.}{2020}]{Haniewicz:2020jro}
Haniewicz H.~T.,  Ferdman R.~D.,  Freire P. C.~C.,  Champion D.~J.,  Bunting
  K.~A.,  Lorimer D.~R.,   McLaughlin M.~A.,  2020, \mn@doi [Mon. Not. Roy.
  Astron. Soc.] {10.1093/mnras/staa3466}, 500, 4620

\bibitem[\protect\citeauthoryear{Hildebrandt et~al.}{Hildebrandt
  et~al.}{2017}]{Hildebrandt:2016iqg}
Hildebrandt H.,  et~al., 2017, \mn@doi [Mon. Not. Roy. Astron. Soc.]
  {10.1093/mnras/stw2805}, 465, 1454

\bibitem[\protect\citeauthoryear{Hobbs et~al.}{Hobbs
  et~al.}{2004}]{Hobbs:2004gj}
Hobbs G.,  et~al., 2004, \mn@doi [Mon. Not. Roy. Astron. Soc.]
  {10.1111/j.1365-2966.2004.08042.x}, 352, 1439

\bibitem[\protect\citeauthoryear{Hu, Wang, Wu, Wang, Zhang  \& Chen}{Hu
  et~al.}{2020}]{Hu:2019okh}
Hu W.,  Wang X.,  Wu F.,  Wang Y.,  Zhang P.,   Chen X.,  2020, \mn@doi [Mon.
  Not. Roy. Astron. Soc.] {10.1093/mnras/staa650}, 493, 5854

\bibitem[\protect\citeauthoryear{Inoue, Minezaki, Matsushita  \&
  Nakanishi}{Inoue et~al.}{2023}]{Inoue:2021dkv}
Inoue K.~T.,  Minezaki T.,  Matsushita S.,   Nakanishi K.,  2023, \mn@doi
  [Astrophys. J.] {10.3847/1538-4357/aceb5f}, 954, 197

\bibitem[\protect\citeauthoryear{Jacoby, Cameron, Jenet, Anderson, Murty  \&
  Kulkarni}{Jacoby et~al.}{2006}]{Jacoby:2006dy}
Jacoby B.~A.,  Cameron P.~B.,  Jenet F.~A.,  Anderson S.~B.,  Murty R.~N.,
  Kulkarni S.~R.,  2006, \mn@doi [Astrophys. J. Lett.] {10.1086/505742}, 644,
  L113

\bibitem[\protect\citeauthoryear{Jennings, Kaplan, Chatterjee, Cordes  \&
  Deller}{Jennings et~al.}{2018}]{Jennings:2018psk}
Jennings R.~J.,  Kaplan D.~L.,  Chatterjee S.,  Cordes J.~M.,   Deller A.~T.,
  2018, \mn@doi [Astrophys. J.] {10.3847/1538-4357/aad084}, 864, 26

\bibitem[\protect\citeauthoryear{Jiang, Wang, Chen, Liu, Leng, Yuan  \&
  Qian}{Jiang et~al.}{2021}]{Jiang:2021ako}
Jiang L.,  Wang N.,  Chen W.-C.,  Liu W.-M.,  Leng C.-w.,  Yuan J.-P.,   Qian
  X.-L.,  2021, \mn@doi [Res. Astron. Astrophys.] {10.1088/1674-4527/21/9/231},
  21, 231

\bibitem[\protect\citeauthoryear{Kandel \& Romani}{Kandel \&
  Romani}{2020}]{Kandel_2020}
Kandel D.,  Romani R.~W.,  2020, \mn@doi [The Astrophysical Journal]
  {10.3847/1538-4357/ab7b62}, 892, 101

\bibitem[\protect\citeauthoryear{Kandel \& Romani}{Kandel \&
  Romani}{2023}]{Kandel:2022qor}
Kandel D.,  Romani R.~W.,  2023, \mn@doi [Astrophys. J.]
  {10.3847/1538-4357/aca524}, 942, 6

\bibitem[\protect\citeauthoryear{Kaplan, Bhalerao, van Kerkwijk, Koester,
  Kulkarni  \& Stovall}{Kaplan et~al.}{2013}]{Kaplan:2013hii}
Kaplan D.~L.,  Bhalerao V.~B.,  van Kerkwijk M.~H.,  Koester D.,  Kulkarni
  S.~R.,   Stovall K.,  2013, \mn@doi [Astrophys. J.]
  {10.1088/0004-637X/765/2/158}, 765, 158

\bibitem[\protect\citeauthoryear{Karkevandi, Shakeri, Sagun  \&
  Ivanytskyi}{Karkevandi et~al.}{2022}]{Karkevandi:2021ygv}
Karkevandi D.~R.,  Shakeri S.,  Sagun V.,   Ivanytskyi O.,  2022, \mn@doi
  [Phys. Rev. D] {10.1103/PhysRevD.105.023001}, 105, 023001

\bibitem[\protect\citeauthoryear{Keith, Kramer, Lyne, Eatough, Stairs,
  Possenti, Camilo  \& Manchester}{Keith et~al.}{2009}]{Keith:2008ga}
Keith M.~J.,  Kramer M.,  Lyne A.~G.,  Eatough R.~P.,  Stairs I.~H.,  Possenti
  A.,  Camilo F.,   Manchester R.~N.,  2009, \mn@doi [Mon. Not. Roy. Astron.
  Soc.] {10.1111/j.1365-2966.2008.14234.x}, 393, 623

\bibitem[\protect\citeauthoryear{Kennedy et~al.}{Kennedy
  et~al.}{2022}]{Kennedy:2022zml}
Kennedy M.~R.,  et~al., 2022, \mn@doi [Mon. Not. Roy. Astron. Soc.]
  {10.1093/mnras/stac379}, 512, 3001

\bibitem[\protect\citeauthoryear{Kiziltan, Kottas, De~Yoreo  \&
  Thorsett}{Kiziltan et~al.}{2013}]{Kiziltan:2013oja}
Kiziltan B.,  Kottas A.,  De~Yoreo M.,   Thorsett S.~E.,  2013, \mn@doi
  [Astrophys. J.] {10.1088/0004-637X/778/1/66}, 778, 66

\bibitem[\protect\citeauthoryear{Kolb \& Tkachev}{Kolb \&
  Tkachev}{1994}]{Kolb:1994fi}
Kolb E.~W.,  Tkachev I.~I.,  1994, \mn@doi [Phys. Rev. D]
  {10.1103/PhysRevD.50.769}, 50, 769

\bibitem[\protect\citeauthoryear{Komatsu}{Komatsu}{2003}]{Komatsu:2003kv}
Komatsu E.,  2003, in {4th International Conference on Physics Beyond the
  Standard Model: Beyond the Desert (BEYOND 03)}. pp 75--91

\bibitem[\protect\citeauthoryear{Komatsu et~al.}{Komatsu
  et~al.}{2009}]{WMAP:2008lyn}
Komatsu E.,  et~al., 2009, \mn@doi [Astrophys. J. Suppl.]
  {10.1088/0067-0049/180/2/330}, 180, 330

\bibitem[\protect\citeauthoryear{Konacki \& Wolszczan}{Konacki \&
  Wolszczan}{2003}]{Konacki:2003xa}
Konacki M.,  Wolszczan A.,  2003, \mn@doi [Astrophys. J. Lett.]
  {10.1086/377093}, 591, L147

\bibitem[\protect\citeauthoryear{{Kouvaris}}{{Kouvaris}}{2008}]{Kouvaris2008}
{Kouvaris} C.,  2008, \mn@doi [\prd] {10.1103/PhysRevD.77.023006}, \href
  {http://adsabs.harvard.edu/abs/2008PhRvD..77b3006K} {77, 023006}

\bibitem[\protect\citeauthoryear{{Kouvaris}}{{Kouvaris}}{2013}]{Kouvaris2013}
{Kouvaris} C.,  2013, preprint, \href
  {http://adsabs.harvard.edu/abs/2013arXiv1308.3222K} {} (\mn@eprint {arXiv}
  {1308.3222})

\bibitem[\protect\citeauthoryear{Kouvaris \& Nielsen}{Kouvaris \&
  Nielsen}{2015}]{Kouvaris:2015rea}
Kouvaris C.,  Nielsen N.~G.,  2015, \mn@doi [Phys. Rev.]
  {10.1103/PhysRevD.92.063526}, D92, 063526

\bibitem[\protect\citeauthoryear{Kouvaris \& Tinyakov}{Kouvaris \&
  Tinyakov}{2010}]{Kouvaris:2010vv}
Kouvaris C.,  Tinyakov P.,  2010, \mn@doi [Phys. Rev.]
  {10.1103/PhysRevD.82.063531}, D82, 063531

\bibitem[\protect\citeauthoryear{Kouvaris \& Tinyakov}{Kouvaris \&
  Tinyakov}{2011a}]{Kouvaris:2010jy}
Kouvaris C.,  Tinyakov P.,  2011a, \mn@doi [Phys. Rev. D]
  {10.1103/PhysRevD.83.083512}, 83, 083512

\bibitem[\protect\citeauthoryear{Kouvaris \& Tinyakov}{Kouvaris \&
  Tinyakov}{2011b}]{Kouvaris:2011fi}
Kouvaris C.,  Tinyakov P.,  2011b, \mn@doi [Phys. Rev. Lett.]
  {10.1103/PhysRevLett.107.091301}, 107, 091301

\bibitem[\protect\citeauthoryear{Kouvaris, Lang\ae{}ble  \& Nielsen}{Kouvaris
  et~al.}{2016}]{Kouvaris:2016ltf}
Kouvaris C.,  Lang\ae{}ble K.,   Nielsen N.~G.,  2016, \mn@doi [JCAP]
  {10.1088/1475-7516/2016/10/012}, 10, 012

\bibitem[\protect\citeauthoryear{Kramer et~al.}{Kramer
  et~al.}{2021}]{Kramer:2021jcw}
Kramer M.,  et~al., 2021, \mn@doi [Phys. Rev. X] {10.1103/PhysRevX.11.041050},
  11, 041050

\bibitem[\protect\citeauthoryear{Lattimer \& Prakash}{Lattimer \&
  Prakash}{2004}]{Lattimer2004}
Lattimer J.~M.,  Prakash M.,  2004, \mn@doi [Science]
  {10.1126/science.1090720}, 304, 536

\bibitem[\protect\citeauthoryear{Law, Majewski  \& Johnston}{Law
  et~al.}{2009}]{Law:2009yq}
Law D.~R.,  Majewski S.~R.,   Johnston K.~V.,  2009, \mn@doi [Astrophys. J.
  Lett.] {10.1088/0004-637X/703/1/L67}, 703, L67

\bibitem[\protect\citeauthoryear{Le~Delliou, Bertolami  \&
  Gil~Pedro}{Le~Delliou et~al.}{2007}]{LeDelliou:2007am}
Le~Delliou M.,  Bertolami O.,   Gil~Pedro F.,  2007, \mn@doi [AIP Conf. Proc.]
  {10.1063/1.2823818}, 957, 421

\bibitem[\protect\citeauthoryear{Le~Delliou, Marcondes, Lima~Neto  \&
  Abdalla}{Le~Delliou et~al.}{2015}]{LeDelliou:2014fto}
Le~Delliou M.,  Marcondes R. J.~F.,  Lima~Neto G.~B.,   Abdalla E.,  2015,
  \mn@doi [Mon. Not. Roy. Astron. Soc.] {10.1093/mnras/stv1561}, 453, 2

\bibitem[\protect\citeauthoryear{Le~Delliou, Marcondes  \& Neto}{Le~Delliou
  et~al.}{2019}]{LeDelliou:2018vua}
Le~Delliou M.,  Marcondes R. J.~F.,   Neto G.~a. B.~L.,  2019, \mn@doi [Mon.
  Not. Roy. Astron. Soc.] {10.1093/mnras/stz2757}, 490, 1944

\bibitem[\protect\citeauthoryear{Leung, Chu  \& Lin}{Leung
  et~al.}{2011}]{Leung:2011zz}
Leung S.~C.,  Chu M.~C.,   Lin L.~M.,  2011, \mn@doi [Phys. Rev.]
  {10.1103/PhysRevD.84.107301}, D84, 107301

\bibitem[\protect\citeauthoryear{Li, Huang  \& Xu}{Li et~al.}{2012}]{Li:2012ii}
Li A.,  Huang F.,   Xu R.-X.,  2012, \mn@doi [Astropart. Phys.]
  {10.1016/j.astropartphys.2012.07.006}, 37, 70

\bibitem[\protect\citeauthoryear{Linares, Shahbaz  \& Casares}{Linares
  et~al.}{2018}]{Linares:2018ppq}
Linares M.,  Shahbaz T.,   Casares J.,  2018, \mn@doi [Astrophys. J.]
  {10.3847/1538-4357/aabde6}, 859, 54

\bibitem[\protect\citeauthoryear{Liu et~al.}{Liu et~al.}{2020}]{Liu:2020hkx}
Liu K.,  et~al., 2020, \mn@doi [Mon. Not. Roy. Astron. Soc.]
  {10.1093/mnras/staa2993}, 499, 2276

\bibitem[\protect\citeauthoryear{Lohmer, Kramer, Driebe, Jessner, Mitra  \&
  Lyne}{Lohmer et~al.}{2004}]{Lohmer:2004aw}
Lohmer O.,  Kramer M.,  Driebe T.,  Jessner A.,  Mitra D.,   Lyne A.~G.,  2004,
  \mn@doi [Astron. Astrophys.] {10.1051/0004-6361:20041031}, 426, 631

\bibitem[\protect\citeauthoryear{Lynch, Freire, Ransom  \& Jacoby}{Lynch
  et~al.}{2012}]{Lynch:2011aa}
Lynch R.~S.,  Freire P. C.~C.,  Ransom S.~M.,   Jacoby B.~A.,  2012, \mn@doi
  [Astrophys. J.] {10.1088/0004-637X/745/2/109}, 745, 109

\bibitem[\protect\citeauthoryear{Lynch et~al.}{Lynch
  et~al.}{2018}]{Lynch:2018zxo}
Lynch R.~S.,  et~al., 2018, \mn@doi [Astrophys. J.] {10.3847/1538-4357/aabf8a},
  859, 93

\bibitem[\protect\citeauthoryear{Madau, Diemand  \& Kuhlen}{Madau
  et~al.}{2008}]{Madau:2008fr}
Madau P.,  Diemand J.,   Kuhlen M.,  2008, \mn@doi [Astrophys. J.]
  {10.1086/587545}, 679, 1260

\bibitem[\protect\citeauthoryear{Mason, Norton, Clark, Negueruela  \&
  Roche}{Mason et~al.}{2010}]{Mason:2009ia}
Mason A.~B.,  Norton A.~J.,  Clark J.~S.,  Negueruela I.,   Roche P.,  2010,
  \mn@doi [Astron. Astrophys.] {10.1051/0004-6361/200913394}, 509, A79

\bibitem[\protect\citeauthoryear{Mason, Clark, Norton, Crowther, Tauris,
  Langer, Negueruela  \& Roche}{Mason et~al.}{2012}]{Mason:2011yp}
Mason A.~B.,  Clark J.~S.,  Norton A.~J.,  Crowther P.~A.,  Tauris T.~M.,
  Langer N.,  Negueruela I.,   Roche P.,  2012, \mn@doi [Mon. Not. Roy. Astron.
  Soc.] {10.1111/j.1365-2966.2012.20596.x}, 422, 199

\bibitem[\protect\citeauthoryear{Mata~S\'anchez, Istrate, van Kerkwijk, Breton
  \& Kaplan}{Mata~S\'anchez et~al.}{2020}]{MataSanchez:2020pys}
Mata~S\'anchez D.,  Istrate A.~G.,  van Kerkwijk M.~H.,  Breton R.~P.,   Kaplan
  D.~L.,  2020, \mn@doi [Mon. Not. Roy. Astron. Soc.] {10.1093/mnras/staa983},
  494, 4031

\bibitem[\protect\citeauthoryear{Maurin, Combet, Nezri  \&
  Pointecouteau}{Maurin et~al.}{2012}]{Maurin:2012tv}
Maurin D.,  Combet C.,  Nezri E.,   Pointecouteau E.,  2012, \mn@doi [Astron.
  Astrophys.] {10.1051/0004-6361/201218986}, 547, A16

\bibitem[\protect\citeauthoryear{McCullough \& Fairbairn}{McCullough \&
  Fairbairn}{2010}]{McCullough:2010ai}
McCullough M.,  Fairbairn M.,  2010, \mn@doi [Phys. Rev.]
  {10.1103/PhysRevD.81.083520}, D81, 083520

\bibitem[\protect\citeauthoryear{McDermott, Yu  \& Zurek}{McDermott
  et~al.}{2012}]{McDermott:2011jp}
McDermott S.~D.,  Yu H.-B.,   Zurek K.~M.,  2012, \mn@doi [Phys. Rev. D]
  {10.1103/PhysRevD.85.023519}, 85, 023519

\bibitem[\protect\citeauthoryear{McKee et~al.}{McKee
  et~al.}{2020}]{McKee:2020pzp}
McKee J.~W.,  et~al., 2020, \mn@doi [Mon. Not. Roy. Astron. Soc.]
  {10.1093/mnras/staa2994}, 499, 4082

\bibitem[\protect\citeauthoryear{Mendes \& Yang}{Mendes \&
  Yang}{2017}]{Mendes:2016vdr}
Mendes R. F.~P.,  Yang H.,  2017, \mn@doi [Class. Quant. Grav.]
  {10.1088/1361-6382/aa842d}, 34, 185001

\bibitem[\protect\citeauthoryear{Mignani, Corongiu, Pallanca  \&
  Ferraro}{Mignani et~al.}{2013}]{Mignani:2012vg}
Mignani R.~P.,  Corongiu A.,  Pallanca C.,   Ferraro F.~R.,  2013, \mn@doi
  [Mon. Not. Roy. Astron. Soc.] {10.1093/mnras/sts671}, 430, 1008

\bibitem[\protect\citeauthoryear{Miller}{Miller}{2020}]{Miller:2013tca}
Miller M.~C.,  2020, \mn@doi [Astrophys. Space Sci. Libr.]
  {10.1007/978-3-662-62110-3_1}, 461, 1

\bibitem[\protect\citeauthoryear{Miller \& Lamb}{Miller \&
  Lamb}{2016}]{Miller:2016pom}
Miller M.~C.,  Lamb F.~K.,  2016, \mn@doi [Eur. Phys. J. A]
  {10.1140/epja/i2016-16063-8}, 52, 63

\bibitem[\protect\citeauthoryear{Miller, Boutloukos, Lo  \& Lamb}{Miller
  et~al.}{2011}]{Miller:2011wt}
Miller M.~C.,  Boutloukos S.,  Lo K.~H.,   Lamb F.~K.,  2011

\bibitem[\protect\citeauthoryear{Miller et~al.}{Miller
  et~al.}{2021}]{Miller:2021qha}
Miller M.~C.,  et~al., 2021, \mn@doi [Astrophys. J. Lett.]
  {10.3847/2041-8213/ac089b}, 918, L28

\bibitem[\protect\citeauthoryear{Moldon, Ribo, Paredes, Brisken, Dhawan,
  Kramer, Lyne  \& Stappers}{Moldon et~al.}{2012}]{Moldon:2012xk}
Moldon J.,  Ribo M.,  Paredes J.~M.,  Brisken W.,  Dhawan V.,  Kramer M.,  Lyne
  A.~G.,   Stappers B.~W.,  2012, \mn@doi [Astron. Astrophys.]
  {10.1051/0004-6361/201219205}, 543, A26

\bibitem[\protect\citeauthoryear{Molin\'e, S\'anchez-Conde, Palomares-Ruiz  \&
  Prada}{Molin\'e et~al.}{2017}]{Moline:2016pbm}
Molin\'e A.,  S\'anchez-Conde M.~A.,  Palomares-Ruiz S.,   Prada F.,  2017,
  \mn@doi [Mon. Not. Roy. Astron. Soc.] {10.1093/mnras/stx026}, 466, 4974

\bibitem[\protect\citeauthoryear{Mukhopadhyay \&
  Schaffner-Bielich}{Mukhopadhyay \&
  Schaffner-Bielich}{2016}]{Mukhopadhyay:2015xhs}
Mukhopadhyay P.,  Schaffner-Bielich J.,  2016, \mn@doi [Phys. Rev.]
  {10.1103/PhysRevD.93.083009}, D93, 083009

\bibitem[\protect\citeauthoryear{N\"attil\"a, Miller, Steiner, Kajava,
  Suleimanov  \& Poutanen}{N\"attil\"a et~al.}{2017}]{Nattila:2017wtj}
N\"attil\"a J.,  Miller M.~C.,  Steiner A.~W.,  Kajava J. J.~E.,  Suleimanov
  V.~F.,   Poutanen J.,  2017, \mn@doi [Astron. Astrophys.]
  {10.1051/0004-6361/201731082}, 608, A31

\bibitem[\protect\citeauthoryear{Nesti \& Salucci}{Nesti \&
  Salucci}{2013}]{Nesti:2013uwa}
Nesti F.,  Salucci P.,  2013, \mn@doi [JCAP] {10.1088/1475-7516/2013/07/016},
  07, 016

\bibitem[\protect\citeauthoryear{Nezri, White, Combet, Maurin, Pointecouteau
  \& Hinton}{Nezri et~al.}{2012}]{Nezri:2012tu}
Nezri E.,  White R.,  Combet C.,  Maurin D.,  Pointecouteau E.,   Hinton J.~A.,
   2012, \mn@doi [Mon. Not. Roy. Astron. Soc.]
  {10.1111/j.1365-2966.2012.21484.x}, 425, 477

\bibitem[\protect\citeauthoryear{Ng et~al.}{Ng et~al.}{2015}]{Ng:2015zza}
Ng C.,  et~al., 2015, \mn@doi [Mon. Not. Roy. Astron. Soc.]
  {10.1093/mnras/stv753}, 450, 2922

\bibitem[\protect\citeauthoryear{Ng, Guillemot, Freire, Kramer, Champion,
  Cognard, Theureau  \& Barr}{Ng et~al.}{2020}]{Ng:2020uck}
Ng C.,  Guillemot L.,  Freire P. C.~C.,  Kramer M.,  Champion D.~J.,  Cognard
  I.,  Theureau G.,   Barr E.~D.,  2020, \mn@doi [Mon. Not. Roy. Astron. Soc.]
  {10.1093/mnras/staa337}, 493, 1261

\bibitem[\protect\citeauthoryear{Nice, Splaver  \& Stairs}{Nice
  et~al.}{2001}]{Nice:2000fw}
Nice D.~J.,  Splaver E.~M.,   Stairs I.~H.,  2001, \mn@doi [Astrophys. J.]
  {10.1086/319079}, 549, 516

\bibitem[\protect\citeauthoryear{Oppenheimer \& Volkoff}{Oppenheimer \&
  Volkoff}{1939}]{Oppenheimer:1939ne}
Oppenheimer J.~R.,  Volkoff G.~M.,  1939, \mn@doi [Phys. Rev.]
  {10.1103/PhysRev.55.374}, 55, 374

\bibitem[\protect\citeauthoryear{Orosz \& van Kerkwijk}{Orosz \& van
  Kerkwijk}{2003}]{Orosz:2002ci}
Orosz J.~A.,  van Kerkwijk M.~H.,  2003, \mn@doi [Astron. Astrophys.]
  {10.1051/0004-6361:20021468}, 397, 237

\bibitem[\protect\citeauthoryear{Panotopoulos \& Lopes}{Panotopoulos \&
  Lopes}{2017}]{Panotopoulos:2017idn}
Panotopoulos G.,  Lopes I.,  2017, \mn@doi [Phys. Rev. D]
  {10.1103/PhysRevD.96.083004}, 96, 083004

\bibitem[\protect\citeauthoryear{Penarrubia, Koposov, Walker, Gilmore, Evans
  \& Mackay}{Penarrubia et~al.}{2010}]{Penarrubia:2010pa}
Penarrubia J.,  Koposov S.~E.,  Walker M.~G.,  Gilmore G.,  Evans N.~W.,
  Mackay C.~D.,  2010, arXiv:1005.5388[astro-ph.GA]

\bibitem[\protect\citeauthoryear{Perez-Garcia \& Silk}{Perez-Garcia \&
  Silk}{2012}]{PerezGarcia:2011hh}
Perez-Garcia M.~A.,  Silk J.,  2012, \mn@doi [Phys. Lett.]
  {10.1016/j.physletb.2012.03.065}, B711, 6

\bibitem[\protect\citeauthoryear{Perez-Garcia, Silk  \& Stone}{Perez-Garcia
  et~al.}{2010}]{PerezGarcia:2010ap}
Perez-Garcia M.~A.,  Silk J.,   Stone J.~R.,  2010, \mn@doi [Phys. Rev. Lett.]
  {10.1103/PhysRevLett.105.141101}, 105, 141101

\bibitem[\protect\citeauthoryear{P\'erez de~los Heros}{P\'erez de~los
  Heros}{2020}]{PerezdelosHeros:2020qyt}
P\'erez de~los Heros C.,  2020, \mn@doi [Symmetry] {10.3390/sym12101648}, 12,
  1648

\bibitem[\protect\citeauthoryear{Pieri, Bertone  \& Branchini}{Pieri
  et~al.}{2008}]{Pieri:2007ir}
Pieri L.,  Bertone G.,   Branchini E.,  2008, \mn@doi [Mon. Not. Roy. Astron.
  Soc.] {10.1111/j.1365-2966.2007.12828.x}, 384, 1627

\bibitem[\protect\citeauthoryear{Posselt, Schreyer, Perna, Sommer, Klein  \&
  Slane}{Posselt et~al.}{2010}]{Posselt:2010yg}
Posselt B.,  Schreyer K.,  Perna R.,  Sommer M.~W.,  Klein B.,   Slane P.,
  2010, \mn@doi [Mon. Not. Roy. Astron. Soc.]
  {10.1111/j.1365-2966.2010.16557.x}, 405, 1840

\bibitem[\protect\citeauthoryear{Ransom et~al.}{Ransom
  et~al.}{2014}]{Ransom:2014xla}
Ransom S.~M.,  et~al., 2014, \mn@doi [Nature] {10.1038/nature12917}, 505, 520

\bibitem[\protect\citeauthoryear{Rawls, Orosz, McClintock, Torres, Bailyn  \&
  Buxton}{Rawls et~al.}{2011}]{Rawls:2011jw}
Rawls M.~L.,  Orosz J.~A.,  McClintock J.~E.,  Torres M. A.~P.,  Bailyn C.~D.,
   Buxton M.~M.,  2011, \mn@doi [Astrophys. J.] {10.1088/0004-637X/730/1/25},
  730, 25

\bibitem[\protect\citeauthoryear{Reardon et~al.}{Reardon
  et~al.}{2016}]{Reardon:2015kba}
Reardon D.~J.,  et~al., 2016, \mn@doi [Mon. Not. Roy. Astron. Soc.]
  {10.1093/mnras/stv2395}, 455, 1751

\bibitem[\protect\citeauthoryear{Reardon et~al.}{Reardon
  et~al.}{2021}]{Reardon:2021gko}
Reardon D.~J.,  et~al., 2021, \mn@doi [Mon. Not. Roy. Astron. Soc.]
  {10.1093/mnras/stab1990}, 507, 2137

\bibitem[\protect\citeauthoryear{Reid, Mark J.  \& Brunthaler, A}{Reid
et~al.}{2004}]{Reid:2004rd}
Reid M.~J.,  Brunthaler A.,  2004, \mn@doi [The Astrophysical Journal]
{10.1086/424960}, 616, 872--884

\bibitem[\protect\citeauthoryear{Rezzolla, Most  \& Weih}{Rezzolla
  et~al.}{2018}]{Rezzolla2017}
Rezzolla L.,  Most E.~R.,   Weih L.~R.,  2018, \mn@doi [Astrophys. J.]
  {10.3847/2041-8213/aaa401}, 852, L25

\bibitem[\protect\citeauthoryear{{Ricotti} \& {Gould}}{{Ricotti} \&
  {Gould}}{2009}]{Ricotti2009}
{Ricotti} M.,  {Gould} A.,  2009, \mn@doi [\apj] {10.1088/0004-637X/707/2/979},
  \href {http://adsabs.harvard.edu/abs/2009ApJ...707..979R} {707, 979}

\bibitem[\protect\citeauthoryear{Ridolfi, Freire, Gupta  \& Ransom}{Ridolfi
  et~al.}{2019}]{Ridolfi:2019wgs}
Ridolfi A.,  Freire P. C.~C.,  Gupta Y.,   Ransom S.~M.,  2019, \mn@doi [Mon.
  Not. Roy. Astron. Soc.] {10.1093/mnras/stz2645}, 490, 3860

\bibitem[\protect\citeauthoryear{Ridolfi et~al.}{Ridolfi
  et~al.}{2021}]{Ridolfi:2021idl}
Ridolfi A.,  et~al., 2021, \mn@doi [Mon. Not. Roy. Astron. Soc.]
  {10.1093/mnras/stab790}, 504, 1407

\bibitem[\protect\citeauthoryear{Riley et~al.}{Riley
  et~al.}{2019}]{Riley:2019yda}
Riley T.~E.,  et~al., 2019, \mn@doi [Astrophys. J. Lett.]
  {10.3847/2041-8213/ab481c}, 887, L21

\bibitem[\protect\citeauthoryear{Riley et~al.}{Riley
  et~al.}{2021}]{Riley:2021pdl}
Riley T.~E.,  et~al., 2021, \mn@doi [Astrophys. J. Lett.]
  {10.3847/2041-8213/ac0a81}, 918, L27

\bibitem[\protect\citeauthoryear{Romani, Kandel, Filippenko, Brink  \&
  Zheng}{Romani et~al.}{2021}]{Romani:2021xmb}
Romani R.~W.,  Kandel D.,  Filippenko A.~V.,  Brink T.~G.,   Zheng W.,  2021,
  \mn@doi [Astrophys. J. Lett.] {10.3847/2041-8213/abe2b4}, 908, L46

\bibitem[\protect\citeauthoryear{Romani, Kandel, Filippenko, Brink  \&
  Zheng}{Romani et~al.}{2022}]{Romani:2022jhd}
Romani R.~W.,  Kandel D.,  Filippenko A.~V.,  Brink T.~G.,   Zheng W.,  2022,
  \mn@doi [Astrophys. J. Lett.] {10.3847/2041-8213/ac8007}, 934, L18

\bibitem[\protect\citeauthoryear{Roy et~al.}{Roy et~al.}{2015}]{Roy:2014cwa}
Roy J.,  et~al., 2015, \mn@doi [Astrophys. J. Lett.]
  {10.1088/2041-8205/800/1/L12}, 800, L12

\bibitem[\protect\citeauthoryear{Sandin \& Ciarcelluti}{Sandin \&
  Ciarcelluti}{2009}]{Sandin:2008db}
Sandin F.,  Ciarcelluti P.,  2009, \mn@doi [Astropart. Phys.]
  {10.1016/j.astropartphys.2009.09.005}, 32, 278

\bibitem[\protect\citeauthoryear{Schaffner-Bielich}{Schaffner-Bielich}{2005}]{Schaffner-Bielich2005}
Schaffner-Bielich J.,  2005, Journal of Physics G: Nuclear and Particle
  Physics, 31, S651

\bibitem[\protect\citeauthoryear{Schmid, Schwarz  \& Widerin}{Schmid
  et~al.}{1999}]{Schmid:1998mx}
Schmid C.,  Schwarz D.~J.,   Widerin P.,  1999, \mn@doi [Phys. Rev. D]
  {10.1103/PhysRevD.59.043517}, 59, 043517

\bibitem[\protect\citeauthoryear{Schnitzeler, Eatough, Ferri\`ere, Kramer, Lee,
  Noutsos  \& Shannon}{Schnitzeler et~al.}{2016}]{Schnitzeler:2016urr}
Schnitzeler D. H. F.~M.,  Eatough R.~P.,  Ferri\`ere K.,  Kramer M.,  Lee
  K.~J.,  Noutsos A.,   Shannon R.~M.,  2016, \mn@doi [Mon. Not. Roy. Astron.
  Soc.] {10.1093/mnras/stw841}, 459, 3005

\bibitem[\protect\citeauthoryear{{Scott} \& {Sivertsson}}{{Scott} \&
  {Sivertsson}}{2009}]{Scott2009}
{Scott} P.,  {Sivertsson} S.,  2009, \mn@doi [Phys. Rev. Lett.]
  {10.1103/PhysRevLett.103.211301}, \href
  {http://adsabs.harvard.edu/abs/2009PhRvL.103u1301S} {103, 211301}

\bibitem[\protect\citeauthoryear{Sedrakian}{Sedrakian}{2019}]{Sedrakian:2018kdm}
Sedrakian A.,  2019, \mn@doi [Phys. Rev. D] {10.1103/PhysRevD.99.043011}, 99,
  043011

\bibitem[\protect\citeauthoryear{Serylak et~al.}{Serylak
  et~al.}{2022}]{Serylak:2022kna}
Serylak M.,  et~al., 2022, \mn@doi [Astron. Astrophys.]
  {10.1051/0004-6361/202142670}, 665, A53

\bibitem[\protect\citeauthoryear{Shahbaz, Linares, Rodriguez-Gil  \&
  Casares}{Shahbaz et~al.}{2019}]{Shahbaz:2019lgw}
Shahbaz T.,  Linares M.,  Rodriguez-Gil P.,   Casares J.,  2019, \mn@doi [Mon.
  Not. Roy. Astron. Soc.] {10.1093/mnras/stz1652}, 488, 198

\bibitem[\protect\citeauthoryear{Shao, Tang, Jiang  \& Fan}{Shao
  et~al.}{2020}]{Shao:2020bzt}
Shao D.-S.,  Tang S.-P.,  Jiang J.-L.,   Fan Y.-Z.,  2020, \mn@doi [Phys. Rev.
  D] {10.1103/PhysRevD.102.063006}, 102, 063006

\bibitem[\protect\citeauthoryear{Shaw, Heinke, Steiner, Campana, Cohn, Ho,
  Lugger  \& Servillat}{Shaw et~al.}{2018}]{Shaw:2018wxh}
Shaw A.~W.,  Heinke C.~O.,  Steiner A.~W.,  Campana S.,  Cohn H.~N.,  Ho W.
  C.~G.,  Lugger P.~M.,   Servillat M.,  2018, \mn@doi [Mon. Not. Roy. Astron.
  Soc.] {10.1093/mnras/sty582}, 476, 4713

\bibitem[\protect\citeauthoryear{Silk \& Stebbins}{Silk \&
  Stebbins}{1993}]{Silk:1992bh}
Silk J.,  Stebbins A.,  1993, \mn@doi [Astrophys. J.] {10.1086/172846}, 411,
  439

\bibitem[\protect\citeauthoryear{Sirunyan et~al.}{Sirunyan
  et~al.}{2021}]{CMS:2020krr}
Sirunyan A.,  et~al., 2021, \mn@doi [JHEP] {10.1007/JHEP03(2021)011}, 03, 011

\bibitem[\protect\citeauthoryear{Smits, Tingay, Wex, Kramer  \& Stappers}{Smits
  et~al.}{2011}]{Smits:2011zh}
Smits R.,  Tingay S.~J.,  Wex N.,  Kramer M.,   Stappers B.,  2011, \mn@doi
  [Astron. Astrophys.] {10.1051/0004-6361/201016141}, 528, A108

\bibitem[\protect\citeauthoryear{Somiya}{Somiya}{2012}]{Somiya:2011np}
Somiya K.,  2012, \mn@doi [Class. Quant. Grav.]
  {10.1088/0264-9381/29/12/124007}, 29, 124007

\bibitem[\protect\citeauthoryear{Springel et~al.,}{Springel
  et~al.}{2008}]{Springel:2008cc}
Springel V.,  et~al., 2008, \mn@doi [Mon. Not. Roy. Astron. Soc.]
  {10.1111/j.1365-2966.2008.14066.x}, 391, 1685

\bibitem[\protect\citeauthoryear{Stairs, Thorsett, Taylor  \& Wolszczan}{Stairs
  et~al.}{2002}]{Stairs:2002cw}
Stairs I.~H.,  Thorsett S.~E.,  Taylor J.~H.,   Wolszczan A.,  2002, \mn@doi
  [Astrophys. J.] {10.1086/344157}, 581, 501

\bibitem[\protect\citeauthoryear{Stovall et~al.}{Stovall
  et~al.}{2019}]{Stovall:2018rvy}
Stovall K.,  et~al., 2019, \mn@doi [Astrophys. J.] {10.3847/1538-4357/aaf37d},
  870, 74

\bibitem[\protect\citeauthoryear{Strader, Li, Chomiuk, Heinke, Udalski,
  Peacock, Shishkovsky  \& Tremou}{Strader et~al.}{2016}]{Strader:2016qpu}
Strader J.,  Li K.-L.,  Chomiuk L.,  Heinke C.~O.,  Udalski A.,  Peacock M.,
  Shishkovsky L.,   Tremou E.,  2016, \mn@doi [Astrophys. J.]
  {10.3847/0004-637X/831/1/89}, 831, 89

\bibitem[\protect\citeauthoryear{Strader et~al.}{Strader
  et~al.}{2019}]{Strader:2018qbi}
Strader J.,  et~al., 2019, \mn@doi [Astrophys. J.] {10.3847/1538-4357/aafbaa},
  872, 42

\bibitem[\protect\citeauthoryear{Suleimanov, Poutanen  \& Werner}{Suleimanov
  et~al.}{2012}]{Suleimanov:2012sq}
Suleimanov V.,  Poutanen J.,   Werner K.,  2012, \mn@doi [Astron. Astrophys.]
  {10.1051/0004-6361/201219480}, 545, A120

\bibitem[\protect\citeauthoryear{Swiggum et~al.}{Swiggum
  et~al.}{2015}]{Swiggum:2015yra}
Swiggum J.~K.,  et~al., 2015, \mn@doi [Astrophys. J.]
  {10.1088/0004-637X/805/2/156}, 805, 156

\bibitem[\protect\citeauthoryear{Swihart et~al.}{Swihart
  et~al.}{2017}]{Swihart:2017yaz}
Swihart S.~J.,  et~al., 2017, \mn@doi [Astrophys. J.]
  {10.3847/1538-4357/aa9937}, 851, 31

\bibitem[\protect\citeauthoryear{Tan et~al.}{Tan
  et~al.}{2016}]{PandaX-II:2016vec}
Tan A.,  et~al., 2016, \mn@doi [Phys. Rev. Lett.]
  {10.1103/PhysRevLett.117.121303}, 117, 121303

\bibitem[\protect\citeauthoryear{Tauris \& Janka}{Tauris \&
  Janka}{2019}]{Tauris:2019sho}
Tauris T.~M.,  Janka H.-T.,  2019, \mn@doi [Astrophys. J. Lett.]
  {10.3847/2041-8213/ab5642}, 886, L20

\bibitem[\protect\citeauthoryear{Taylor \& Cordes}{Taylor \&
  Cordes}{1993}]{Taylor:1993my}
Taylor J.~H.,  Cordes J.~M.,  1993, \mn@doi [Astrophys. J.] {10.1086/172870},
  411, 674

\bibitem[\protect\citeauthoryear{Thorsett \& Chakrabarty}{Thorsett \&
  Chakrabarty}{1999}]{Thorsett:1998uc}
Thorsett S.~E.,  Chakrabarty D.,  1999, \mn@doi [Astrophys. J.]
  {10.1086/306742}, 512, 288

\bibitem[\protect\citeauthoryear{{Thorsett}, {Arzoumanian}  \&
  {Taylor}}{{Thorsett} et~al.}{1993}]{1993ApJ...412L..33T}
{Thorsett} S.~E.,  {Arzoumanian} Z.,   {Taylor} J.~H.,  1993, \mn@doi [\apjl]
  {10.1086/186933}, \href
  {https://ui.adsabs.harvard.edu/abs/1993ApJ...412L..33T} {412, L33}

\bibitem[\protect\citeauthoryear{Tolman}{Tolman}{1939}]{Tolman:1939jz}
Tolman R.~C.,  1939, \mn@doi [Phys. Rev.] {10.1103/PhysRev.55.364}, 55, 364

\bibitem[\protect\citeauthoryear{Tolos, Schaffner-Bielich  \& Dengler}{Tolos
  et~al.}{2015}]{Tolos:2015qra}
Tolos L.,  Schaffner-Bielich J.,   Dengler Y.,  2015, \mn@doi [Phys. Rev. D]
  {10.1103/PhysRevD.92.123002}, 92, 123002

\bibitem[\protect\citeauthoryear{Trigo, Boirin, Costantini, Mendez  \&
  Parmar}{Trigo et~al.}{2011}]{Trigo:2011mx}
Trigo M.~D.,  Boirin L.,  Costantini E.,  Mendez M.,   Parmar A.,  2011,
  \mn@doi [Astron. Astrophys.] {10.1051/0004-6361/201016200}, 528, A150

\bibitem[\protect\citeauthoryear{Trumper, Burwitz, Haberl  \& Zavlin}{Trumper
  et~al.}{2004}]{Trumper:2003we}
Trumper J.~E.,  Burwitz V.,  Haberl F.,   Zavlin V.~E.,  2004, \mn@doi [Nucl.
  Phys. B Proc. Suppl.] {10.1016/j.nuclphysbps.2004.04.094}, 132, 560

\bibitem[\protect\citeauthoryear{Tumasyan et~al.}{Tumasyan
  et~al.}{2022}]{CMS:2021dzg}
Tumasyan A.,  et~al., 2022, \mn@doi [JHEP] {10.1007/JHEP06(2022)156}, 06, 156

\bibitem[\protect\citeauthoryear{Ullio, Bergstrom, Edsjo  \& Lacey}{Ullio
  et~al.}{2002}]{Ullio:2002pj}
Ullio P.,  Bergstrom L.,  Edsjo J.,   Lacey C.~G.,  2002, \mn@doi [Phys. Rev.
  D] {10.1103/PhysRevD.66.123502}, 66, 123502

\bibitem[\protect\citeauthoryear{{Val Baker, A. K. F.}, {Norton, A. J.}  \&
  {Quaintrell, H.}}{{Val Baker, A. K. F.} et~al.}{2005}]{refId0}
{Val Baker, A. K. F.} {Norton, A. J.}  {Quaintrell, H.} 2005, \mn@doi [A\&A]
  {10.1051/0004-6361:20053074}, 441, 685

\bibitem[\protect\citeauthoryear{Venkatraman~Krishnan
  et~al.}{Venkatraman~Krishnan et~al.}{2020}]{VenkatramanKrishnan:2020pbi}
Venkatraman~Krishnan V.,  et~al., 2020, \mn@doi [Science]
  {10.1126/science.aax7007}, 367, 577

\bibitem[\protect\citeauthoryear{Verbiest et~al.,}{Verbiest
  et~al.}{2008}]{Verbiest:2008gy}
Verbiest J. P.~W.,  et~al., 2008, \mn@doi [Astrophys. J.] {10.1086/529576},
  679, 675

\bibitem[\protect\citeauthoryear{Verbiest, Weisberg, Chael, Lee  \&
  Lorimer}{Verbiest et~al.}{2012}]{Verbiest:2012kh}
Verbiest J. P.~W.,  Weisberg J.~M.,  Chael A.~A.,  Lee K.~J.,   Lorimer D.~R.,
  2012, \mn@doi [Astrophys. J.] {10.1088/0004-637X/755/1/39}, 755, 39

\bibitem[\protect\citeauthoryear{Walker, Combet, Hinton, Maurin  \&
  Wilkinson}{Walker et~al.}{2011}]{Walker:2011fs}
Walker M.~G.,  Combet C.,  Hinton J.~A.,  Maurin D.,   Wilkinson M.~I.,  2011,
  \mn@doi [Astrophys. J. Lett.] {10.1088/2041-8205/733/2/L46}, 733, L46

\bibitem[\protect\citeauthoryear{Walker et~al.}{Walker
  et~al.}{2016}]{Walker:2015twz}
Walker M.~G.,  et~al., 2016, \mn@doi [Astrophys. J.]
  {10.3847/0004-637X/819/1/53}, 819, 53

\bibitem[\protect\citeauthoryear{Wang, Bose, Frenk, Gao, Jenkins, Springel  \&
  White}{Wang et~al.}{2020a}]{Wang:2019ftp}
Wang J.,  Bose S.,  Frenk C.~S.,  Gao L.,  Jenkins A.,  Springel V.,   White S.
  D.~M.,  2020a, \mn@doi [Nature] {10.1038/s41586-020-2642-9}, 585, 39

\bibitem[\protect\citeauthoryear{Wang et~al.}{Wang
  et~al.}{2020b}]{Wang:2020uua}
Wang L.,  et~al., 2020b, \mn@doi [The Astrophysical Journal]
  {10.3847/1538-4357/ab76cc}, 892, 43

\bibitem[\protect\citeauthoryear{Wechsler, Bullock, Primack, Kravtsov  \&
  Dekel}{Wechsler et~al.}{2002}]{Wechsler:2001cs}
Wechsler R.~H.,  Bullock J.~S.,  Primack J.~R.,  Kravtsov A.~V.,   Dekel A.,
  2002, \mn@doi [Astrophys. J.] {10.1086/338765}, 568, 52

\bibitem[\protect\citeauthoryear{Weisberg \& Huang}{Weisberg \&
  Huang}{2016}]{Weisberg:2016jye}
Weisberg J.~M.,  Huang Y.,  2016, \mn@doi [Astrophys. J.]
  {10.3847/0004-637X/829/1/55}, 829, 55

\bibitem[\protect\citeauthoryear{Xiang, Jiang, Zhang  \& Yang}{Xiang
  et~al.}{2014}]{Xiang:2013xwa}
Xiang Q.-F.,  Jiang W.-Z.,  Zhang D.-R.,   Yang R.-Y.,  2014, \mn@doi [Phys.
  Rev.] {10.1103/PhysRevC.89.025803}, C89, 025803

\bibitem[\protect\citeauthoryear{{Yang} \& {Gao}}{{Yang} \&
  {Gao}}{2011}]{Yang2011}
{Yang} R.-J.,  {Gao} X.-T.,  2011, \mn@doi [Classical and Quantum Gravity]
  {10.1088/0264-9381/28/6/065012}, \href
  {http://adsabs.harvard.edu/abs/2011CQGra..28f5012Y} {28, 065012}

\bibitem[\protect\citeauthoryear{Yang, Zhang, Li, Wang, Pan, Lingfu  \&
  Zhou}{Yang et~al.}{2017}]{Yang_2017}
Yang Y.-Y.,  Zhang C.-M.,  Li D.,  Wang D.-H.,  Pan Y.-Y.,  Lingfu R.-F.,
  Zhou Z.-W.,  2017, \mn@doi [The Astrophysical Journal]
  {10.3847/1538-4357/835/2/185}, 835, 185

\bibitem[\protect\citeauthoryear{Yao, Manchester  \& Wang}{Yao
  et~al.}{2017}]{Yao_2017}
Yao J.~M.,  Manchester R.~N.,   Wang N.,  2017, \mn@doi [The Astrophysical
  Journal] {10.3847/1538-4357/835/1/29}, 835, 29

\bibitem[\protect\citeauthoryear{Zhang et~al.,}{Zhang
  et~al.}{2011}]{Zhang:2010qr}
Zhang C.~M.,  et~al., 2011, \mn@doi [Astron. Astrophys.]
  {10.1051/0004-6361/201015532}, 527, A83

\bibitem[\protect\citeauthoryear{Zhang et~al.}{Zhang
  et~al.}{2022}]{CDEX:2022kcd}
Zhang Z.~Y.,  et~al., 2022, \mn@doi [Phys. Rev. Lett.]
  {10.1103/PhysRevLett.129.221301}, 129, 221301

\bibitem[\protect\citeauthoryear{{Zheng} \& {Chen}}{{Zheng} \&
  {Chen}}{2016}]{Zheng2016}
{Zheng} H.,  {Chen} L.-W.,  2016, \mn@doi [\apj] {10.3847/0004-637X/831/2/127},
  \href {http://adsabs.harvard.edu/abs/2016ApJ...831..127Z} {831, 127}

\bibitem[\protect\citeauthoryear{{Zheng} \& {Huang}}{{Zheng} \&
  {Huang}}{2011}]{Zheng2011}
{Zheng} R.,  {Huang} Q.-G.,  2011, \mn@doi [\jcap]
  {10.1088/1475-7516/2011/03/002}, \href
  {http://adsabs.harvard.edu/abs/2011JCAP...03..002Z} {3, 2}

\bibitem[\protect\citeauthoryear{Zhu et~al.}{Zhu et~al.}{2019}]{Zhu:2019oax}
Zhu W.~W.,  et~al., 2019, \mn@doi [Astrophys. J.] {10.3847/1538-4357/ab2bef},
  881, 165

\bibitem[\protect\citeauthoryear{de Lavallaz \& Fairbairn}{de~Lavallaz \&
  Fairbairn}{2010}]{deLavallaz:2010wp}
de Lavallaz A.,  Fairbairn M.,  2010, \mn@doi [Phys. Rev.]
  {10.1103/PhysRevD.81.123521}, D81, 123521

\bibitem[\protect\citeauthoryear{van Leeuwen et~al.}{van Leeuwen
  et~al.}{2015}]{vanLeeuwen:2014sca}
van Leeuwen J.,  et~al., 2015, \mn@doi [Astrophys. J.]
  {10.1088/0004-637X/798/2/118}, 798, 118

\bibitem[\protect\citeauthoryear{Özel, Gould  \& Güver}{Özel
  et~al.}{2012}]{Ozel_2012}
Özel F.,  Gould A.,   Güver T.,  2012, \mn@doi [The Astrophysical Journal]
  {10.1088/0004-637X/748/1/5}, 748, 5

\makeatother
\end{thebibliography}




\appendix

\section{Neutron stars tables}

\begin{table*}
	\caption{
		List of the examined neutron stars, where 
		(a) NS-NS binary system,
		(b) NS in X-ray binaries,
		(c) radio MSP,
		(d1) WD-NS system,
		(d2) WD-NS GalCluster pulsar.
		The gray entities are shown for the cross-check, taken from the references used in \citep{Shao:2020bzt}. 	
		The calculation of the distance to the GC has been conducted under assumption that $Sgr A^{\star}$ is located in a nearly stationary position at the dynamical center of the Galaxy, with J2000 coordinates specified as $RA=17h45m40.0409s$ and $DEC=-29\degree 00'28.118''$ \citep{Reid:2004rd}. The Galactic longitude and latitude are defined as $l = 359.944\degree$ and $b = -0.046\degree$, respectively. Furthermore, we adopted the distance from the Sun to the Galactic center $8.122 \pm 0.031$ kpc \citep{GRAVITY:2018ofz}.				
	}
	\centering
	\scalebox{0.53}{		
\begin{tabular}{c|c|c|c|c|c|c|c|c}
			\hline
			\hline
			Name
			&Gal. latitude,
			&Gal. longitude,
			&distance to Sun,
			&distance to GC,
			&NS mass,
			&Comp. star 
			&NS radius,
			&Age $\tau_{c}$\\
			&$b$, [deg]
			&$l$, [deg]
			&[kpc]
			&[kpc]			
			&[$M_{\odot}$]
			&mass, [$M_{\odot}$]
			&[km]			
			&[Gyr]\\
			\hline
			$^{(b)}$ 4U 1820-30\citep{Guver:2010td}    &-7.9138  &2.7882  &8.4  
			&1.236216
			&1.58\textsupsub{\tiny{+0.06}}{\tiny{-0.06}} &-- &9.11\textsupsub{\tiny{+0.4}}{\tiny{-0.4}}  &--\\
			$^{(a)}$ B1534+12 \citep{Stairs:2002cw,Kiziltan:2013oja}    &48.34  &19.85  &1.051 
			&7.510862
			&1.3332\textsupsub{\tiny{+0.001}}{\tiny{-0.001}}    &1.3452\textsupsub{\tiny{+0.001}}{\tiny{-0.001}} &-- &0.248 \citep{Mignani:2012vg}\\	
			\rowcolor{Gray}
			$^{(a)}$ B1534+12 \citep{Fonseca:2014qla}    &48.34143147	&19.84752562  &1.051
			\textsupsub{\tiny{+0.005}}{\tiny{-0.005}}
			 &7.510771					 			
			 &1.333\textsupsub{\tiny{+0.0002}}{\tiny{-0.0002}}    &1.345\textsupsub{\tiny{+0.0002}}{\tiny{-0.0002}} &-- &--\\			
			$^{(a)}$ B1913+16 \citep{Weisberg:2016jye,Kiziltan:2013oja} &2.122  &49.968  &4.167 
			 &6.314654
			 &1.4398\textsupsub{\tiny{+0.0002}}{\tiny{-0.0002}}    &1.3886\textsupsub{\tiny{+0.0002}}{\tiny{-0.0002}} &-- &0.108\citep{Mignani:2012vg}\\
			$^{(a)}$ B2127+11C \citep{Jacoby:2006dy,Kiziltan:2013oja}   &-27.32  &65.03  &10.4 
			 &10.527893
			 &1.358\textsupsub{\tiny{+0.01}}{\tiny{-0.01}}    &1.354\textsupsub{\tiny{+0.01}}{\tiny{-0.01}} &-- &0.097\citep{Mignani:2012vg}\\
			$^{(a)}$ B2303+46 \citep{Zhang:2010qr}    &-12.021  &104.931  &3.159  
			&9.429559
			&1.38\textsupsub{\tiny{+0.06}}{\tiny{-0.1}}    &1.4\textsupsub{\tiny{+0.1}}{\tiny{-0.1}} &-- &0.0297\citep{Mignani:2012vg}\\
\hline
\hline
\end{tabular}
\label{tab:NSlist_part1}
}	
\end{table*}

\begin{table*}
	\caption{
		Continuation of Table~\ref{tab:NSlist_part1}.
	}
	\centering
	\scalebox{0.53}{		
	\begin{tabular}{c|c|c|c|c|c|c|c|c}
		\hline
		\hline
		Name
		&Gal. latitude,
		&Gal. longitude,
		&distance to Sun,
		&distance to GC,
		&NS mass,
		&Comp. star 
		&NS radius,
		&Age $\tau_{c}$\\
		&$b$, [deg]
		&$l$, [deg]
		&[kpc]
		&[kpc]			
		&[$M_{\odot}$]
		&mass, [$M_{\odot}$]
		&[km]			
		&[Gyr]\\
		\hline
			$^{(d1)}$ J1125+7819 \citep{Lynch:2018zxo}  &37.895  &128.289  &0.903 
			&8.600628
			&1.35\textsupsub{\tiny{+0.1}}{\tiny{-0.1}}  &0.29\textsupsub{\tiny{+0.1}}{\tiny{-0.1}} &-- &--\\
			$^{(d1)}$ J1641+8049 \citep{Lynch:2018zxo}  &31.763  &113.840  &3.035 
			&9.600967
			&1.35\textsupsub{\tiny{+0.1}}{\tiny{-0.1}}  &0.04\textsupsub{\tiny{+0.1}}{\tiny{-0.1}} &-- &--\\
			$^{(d1)}$ J1938+6604 \citep{Lynch:2018zxo}  &20.030  &97.946  &3.380 
			&9.197132
			&1.35\textsupsub{\tiny{+0.1}}{\tiny{-0.1}}  &0.87\textsupsub{\tiny{+0.1}}{\tiny{-0.1}} &-- &--\\
			$^{(d1)}$ J0751+1807 \citep{Fortin:2014ufa,Freire:2007xg}  &21.09  &202.73  
			&1.07\textsupsub{\tiny{+0.24}}{\tiny{-0.1}} \citep{Guillot:2019vqp}
			&9.059141			
			&1.26\textsupsub{\tiny{+0.14}}{\tiny{-0.12}}  
			&0.12\textsupsub{\tiny{+0.02}}{\tiny{-0.02}} 
			&13.6\textsupsub{\tiny{+1.0}}{\tiny{-1.0}}   
			&8.0\\
			\rowcolor{Gray}
			$^{(d1)}$ J0751+1807 \citep{EPTA:2016ndq}  &21.09  &202.73  
			&1.07\textsupsub{\tiny{+0.24}}{\tiny{-0.1}} \citep{Guillot:2019vqp}     
			&9.059141						
			&1.64\textsupsub{\tiny{+0.15}}{\tiny{-0.15}}  
			&--
			&--
			&--\\
			$^{(d1)}$ J0955-6150 \citep{Serylak:2022kna} &-5.737093  &283.684986  &4.04  
			&8.172637
			&1.71\textsupsub{\tiny{+0.02}}{\tiny{-0.02}}    &0.254\textsupsub{\tiny{+0.002}}{\tiny{-0.002}} &-- &2.1\\
\hline
\hline
\end{tabular}
\label{tab:NSlist_part2}
}	
\end{table*}

\newpage
\section{Neutron Stars mass report race.}

We have revised our plots to address the missing NSs and incorporate updated mass reports for previously examined NSs. Additional entities have been taken from Ref. \cite{Shao:2020bzt}, see gray lines in Tables \ref{tab:NSlist_part1}-\ref{tab:NSlist_part2}. 
Importantly, it is noteworthy that the inclusion of additional NSs has not made a significant influence on the overall findings and outcomes presented in our paper.
\begin{figure}
	\centering
	\includegraphics[scale=0.42]{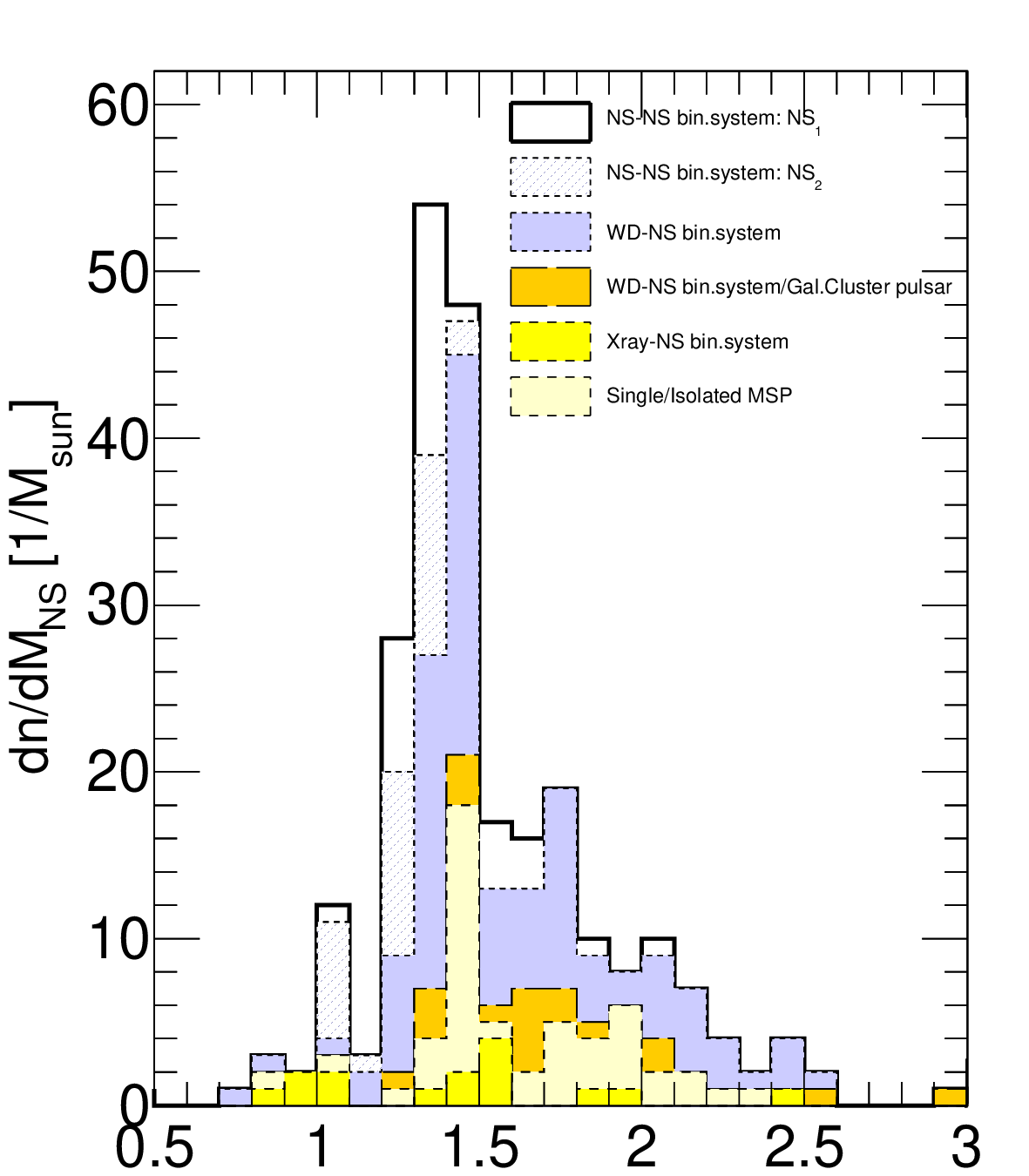}	
	\includegraphics[scale=0.42]{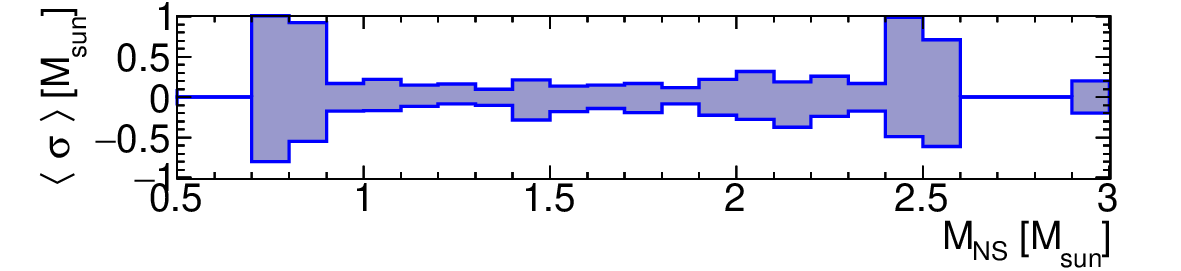} 
	\caption{
		Mass distribution of the examined NSs. Full set includes binary systems such as NS-NS, WD-NS, Xray-NS, and single Milisecond Pulsars (MSPs). Bottom pad shows the average uncertainty in the NS mass measurements.
	}
	\label{fig:YiZhong_NSs_massDistr}		
\end{figure}

\begin{figure*}
	\centering
	\includegraphics[scale=0.29]{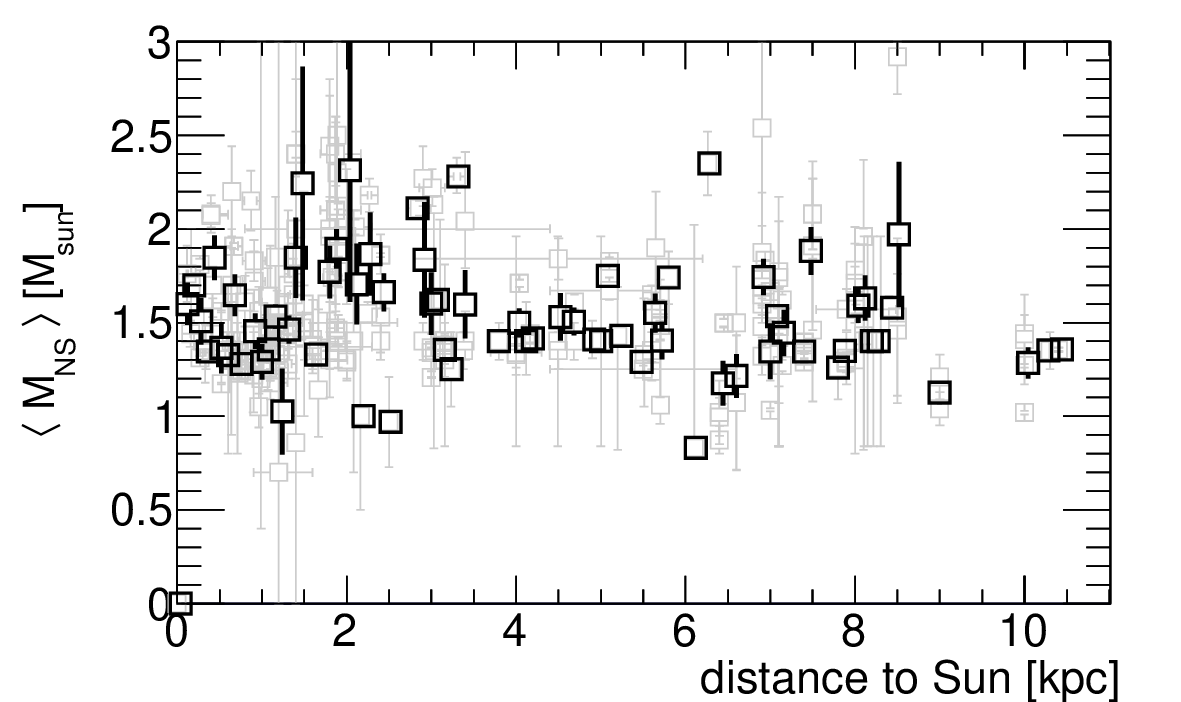}
    \includegraphics[scale=0.29]{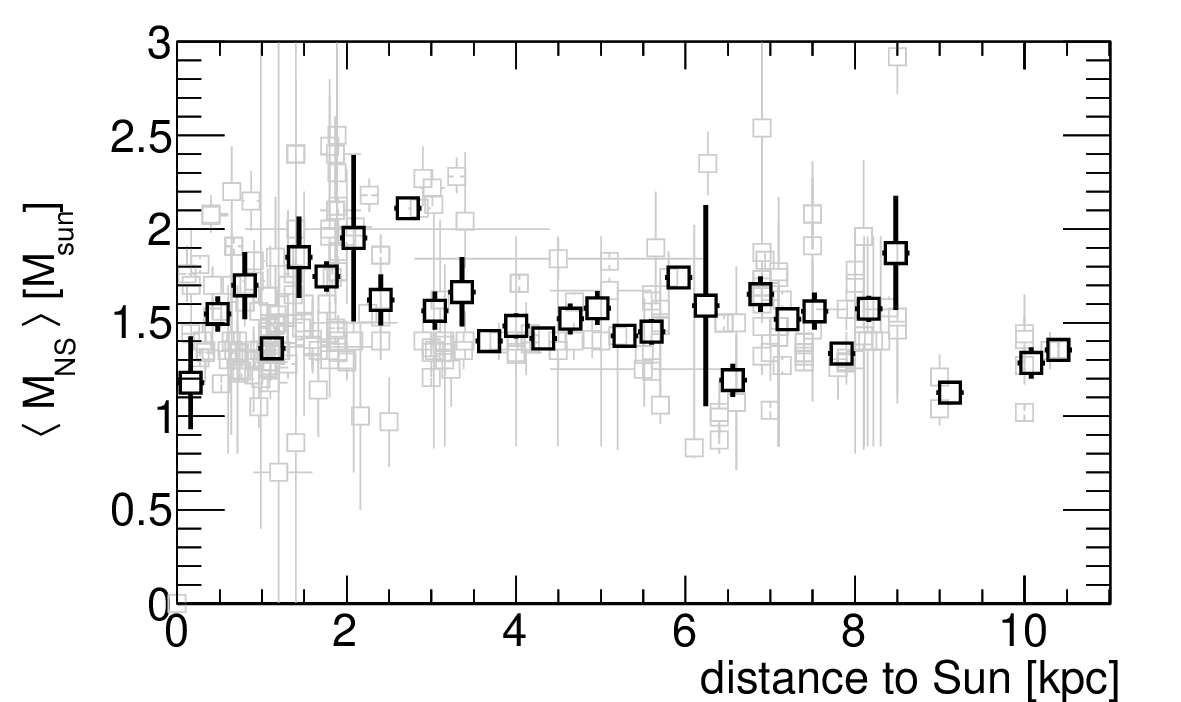}
	\includegraphics[scale=0.29]{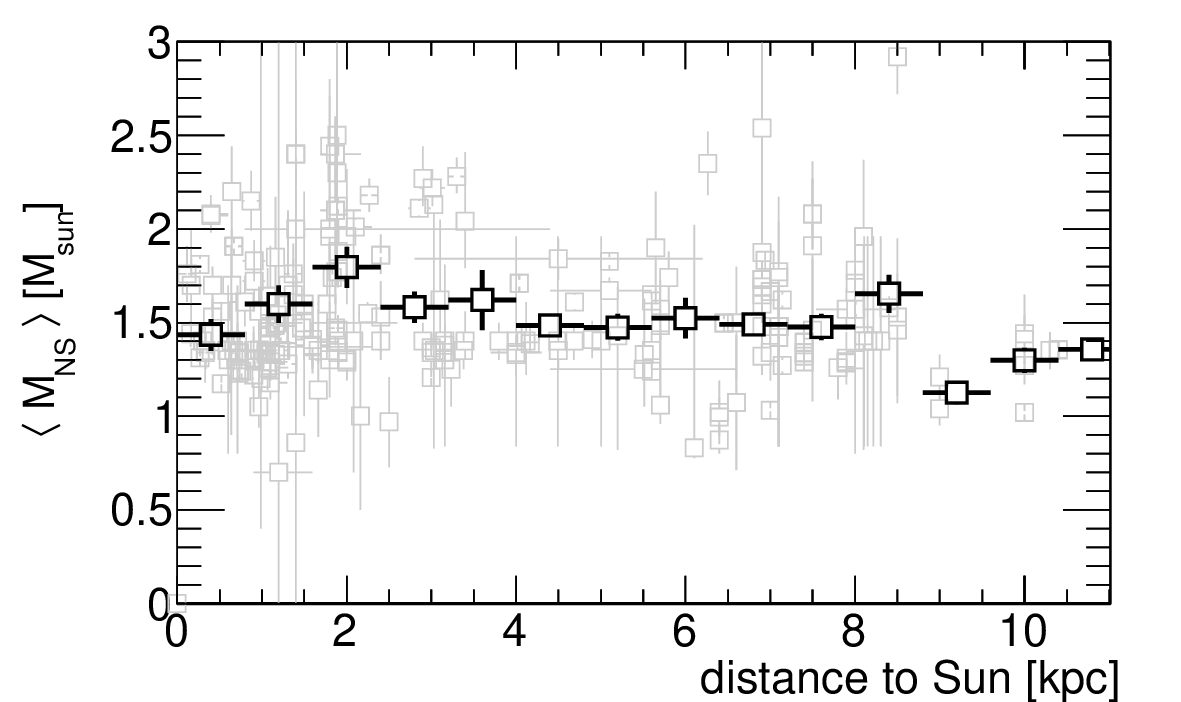}

	\includegraphics[scale=0.29]{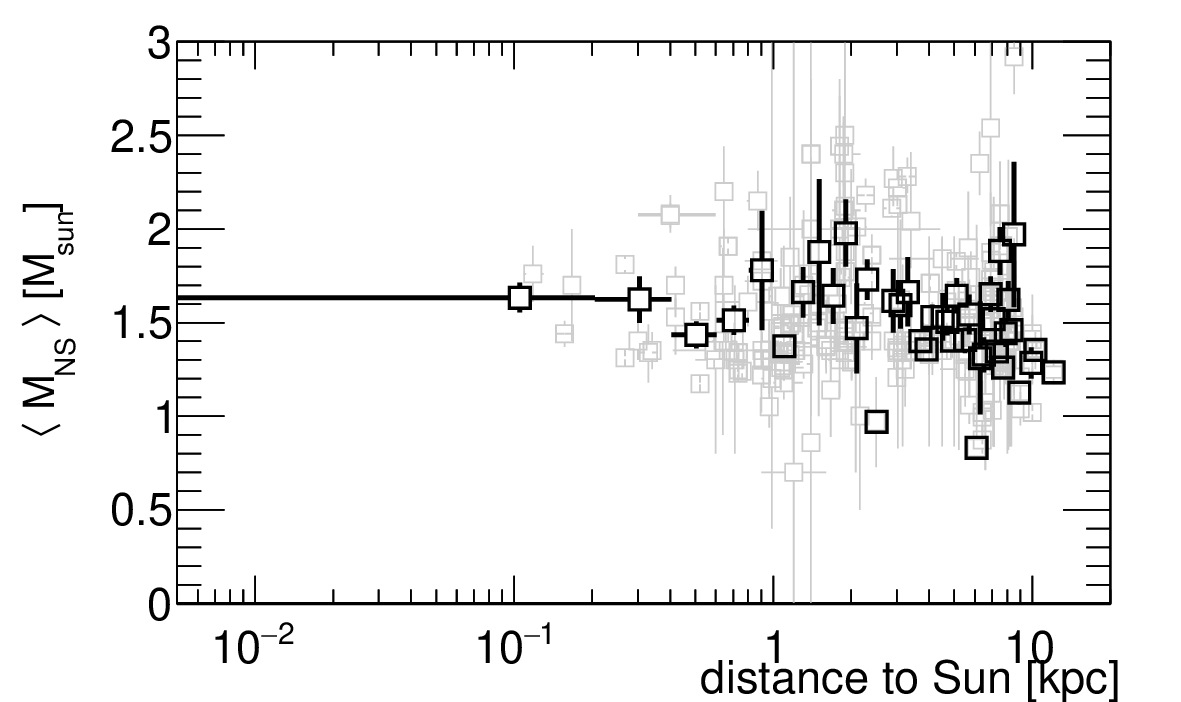}
	\includegraphics[scale=0.29]{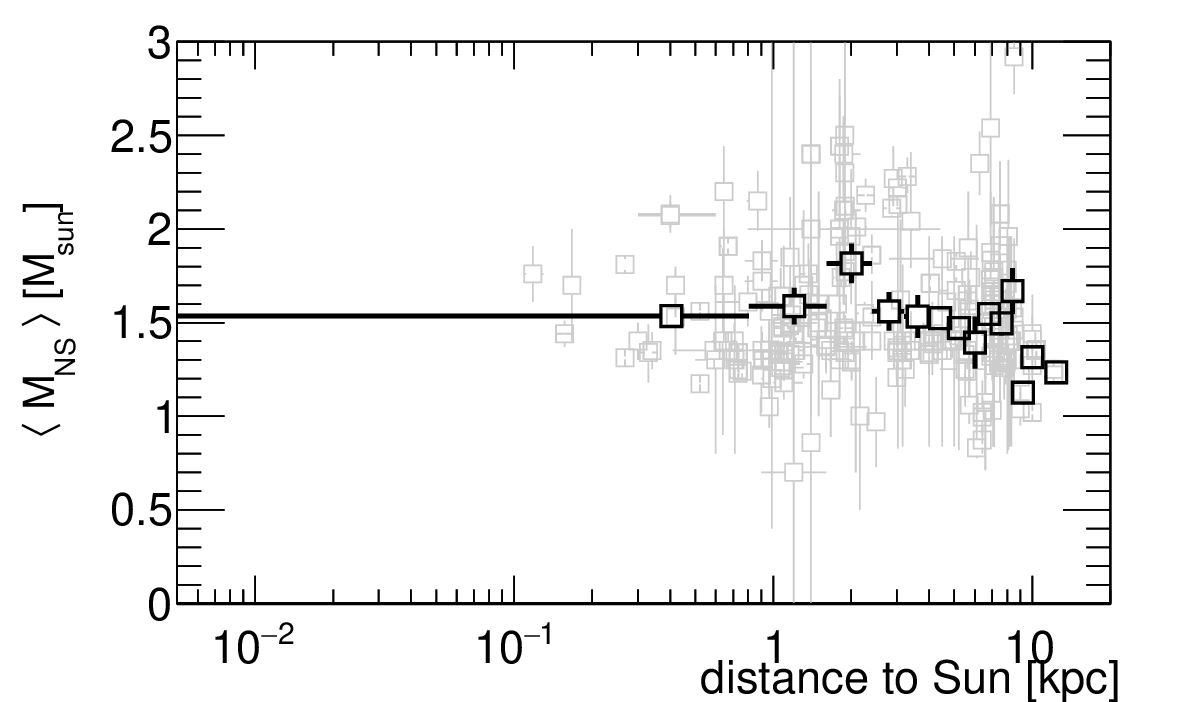}
	\includegraphics[scale=0.29]{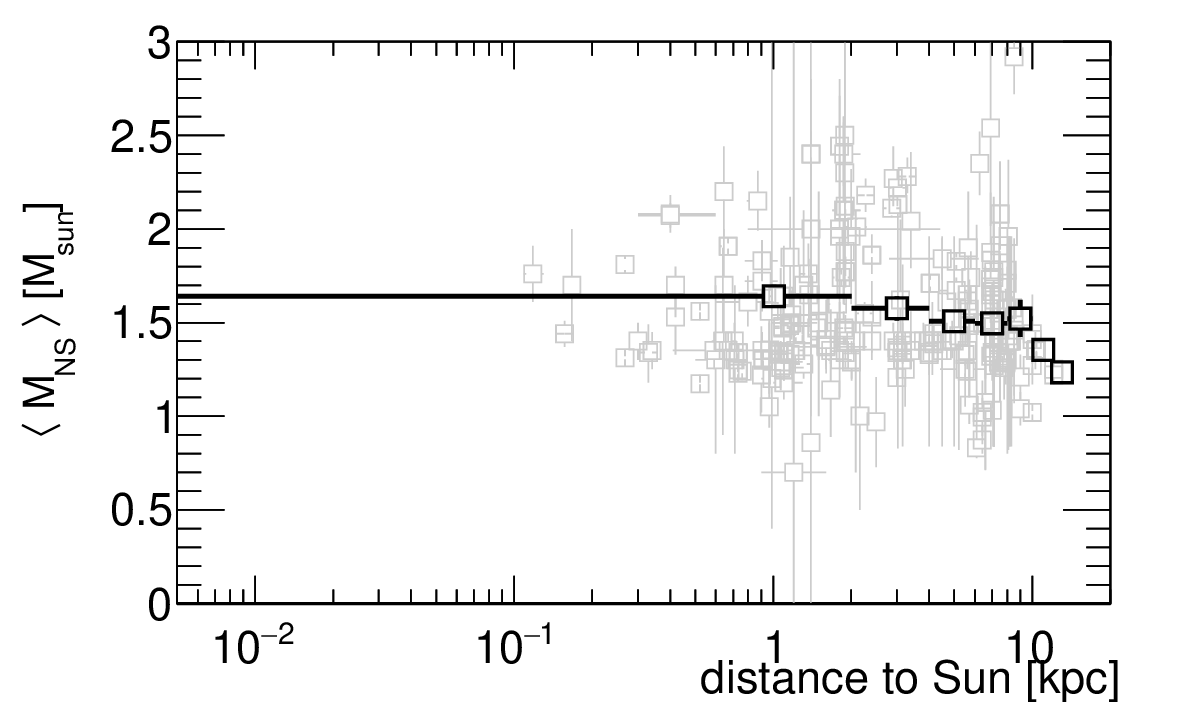}
	\caption{
		\textbf{Mass distribution of the examined NSs as a function of the distance to Sun.} 
		Black boxes - the average mass and radius, the vertical error bars simply reflect the statistical uncertainty at the given bin; light grey boxes - the measured masses and radii of the NSs with the measured uncertainties, see Tables \ref{tab:NSlist_part1}-\ref{tab:NSlist_part2}.	
		We show three different profiles using 2000, 500, and 200 bins from 0 to 20 kpc (from top to bottom) to plot the average mass distribution. Top: distance showen in linear scale. Bottom: in log scale.
	}
	\label{fig:YiZhong_GalaxyMap_NS}		
\end{figure*}

\begin{figure*}
	\centering	
	\includegraphics[scale=0.29]{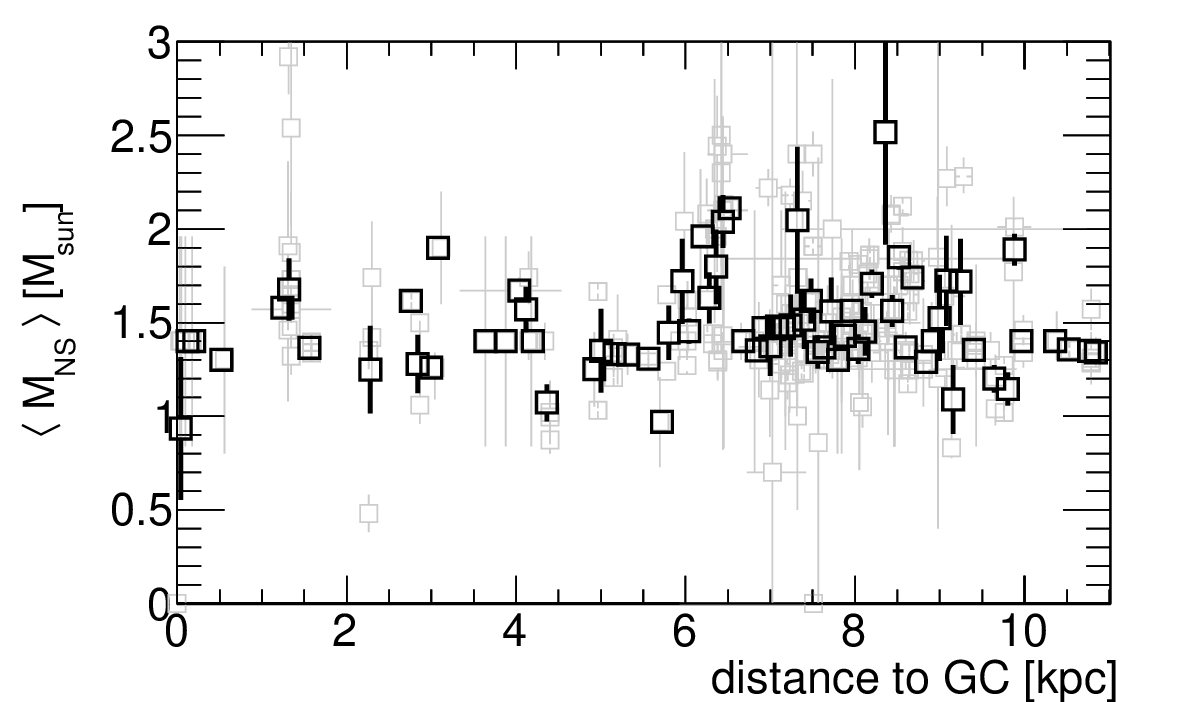}
	\includegraphics[scale=0.29]{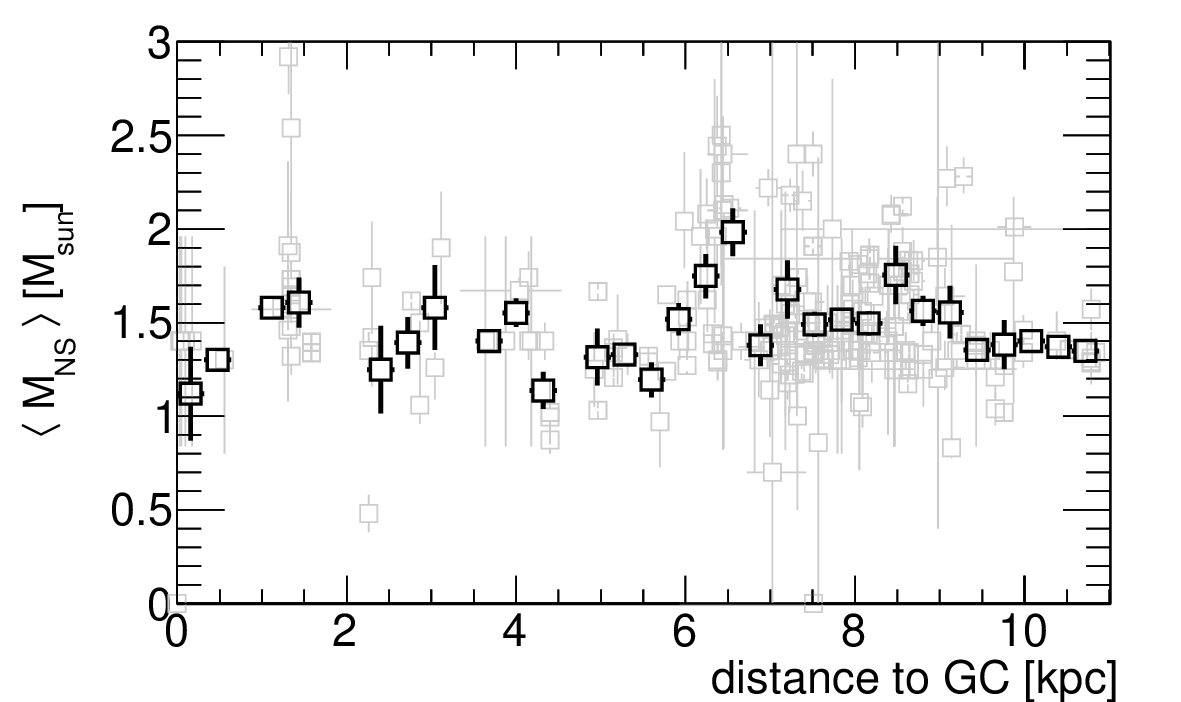}
	\includegraphics[scale=0.29]{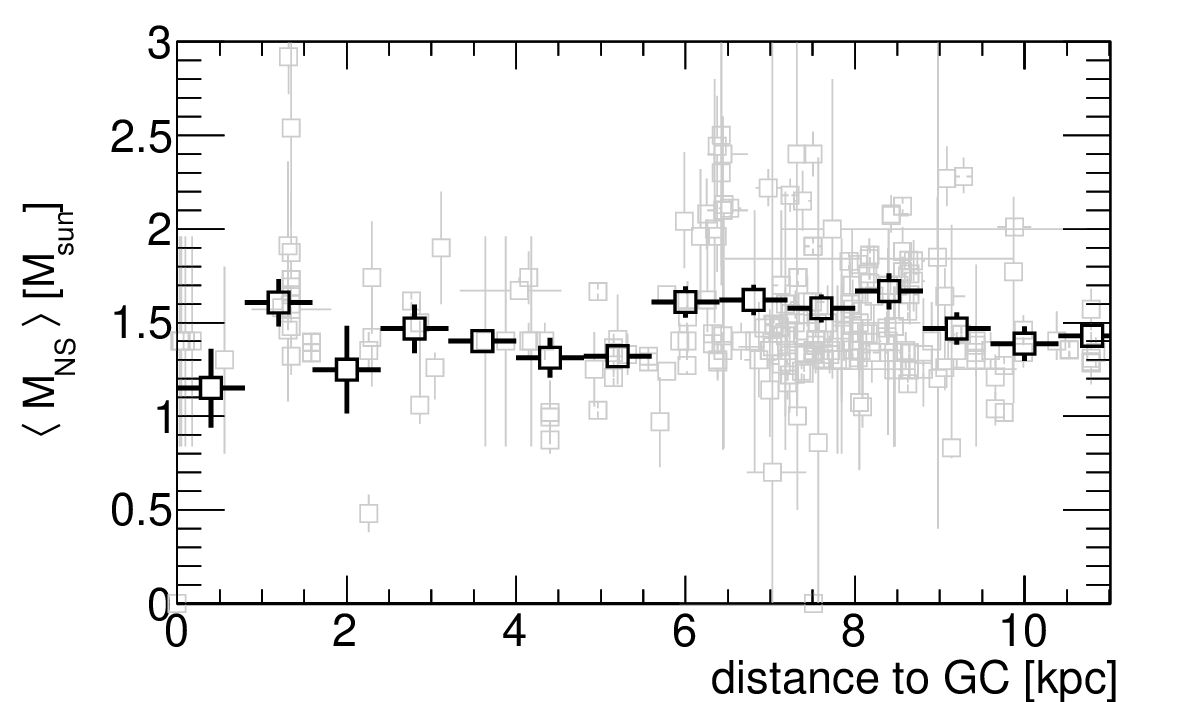}
	
	\includegraphics[scale=0.29]{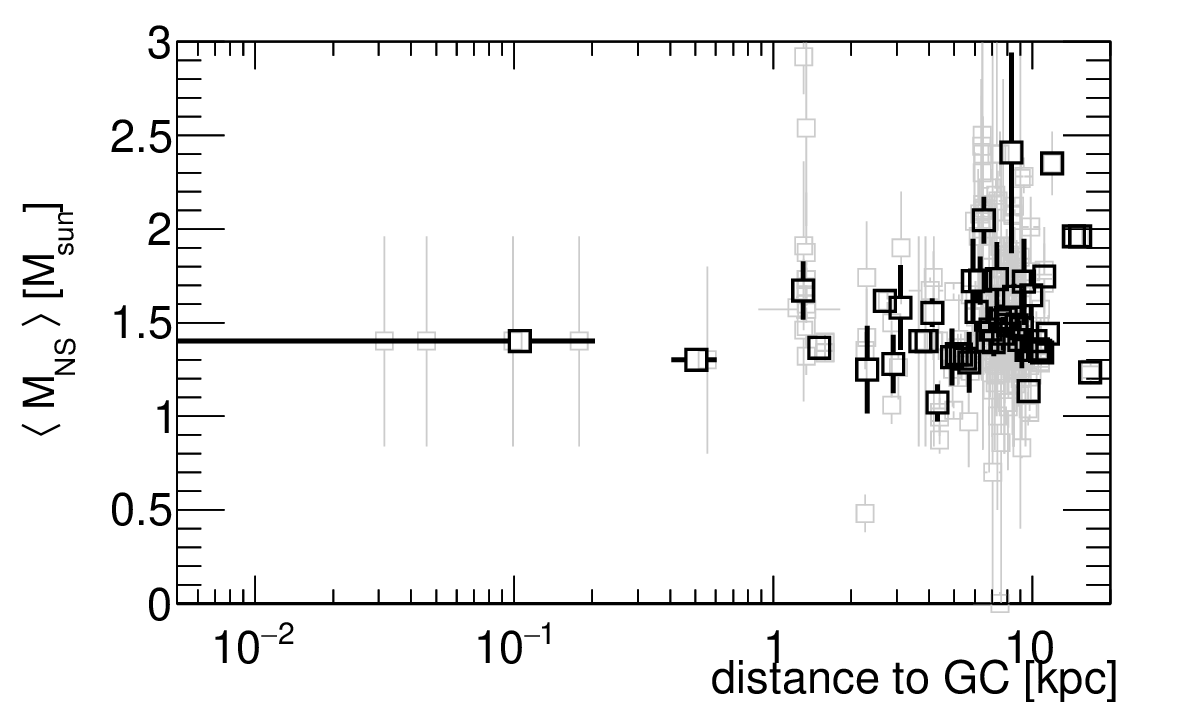}
	\includegraphics[scale=0.29]{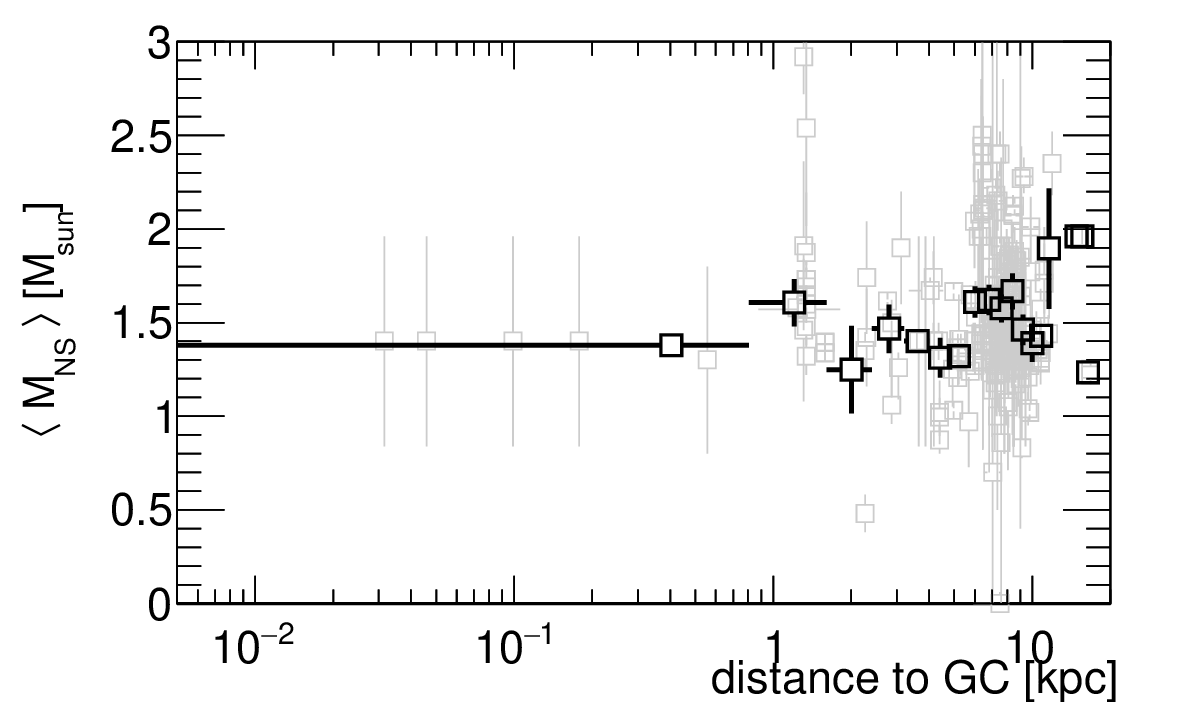}
	\includegraphics[scale=0.29]{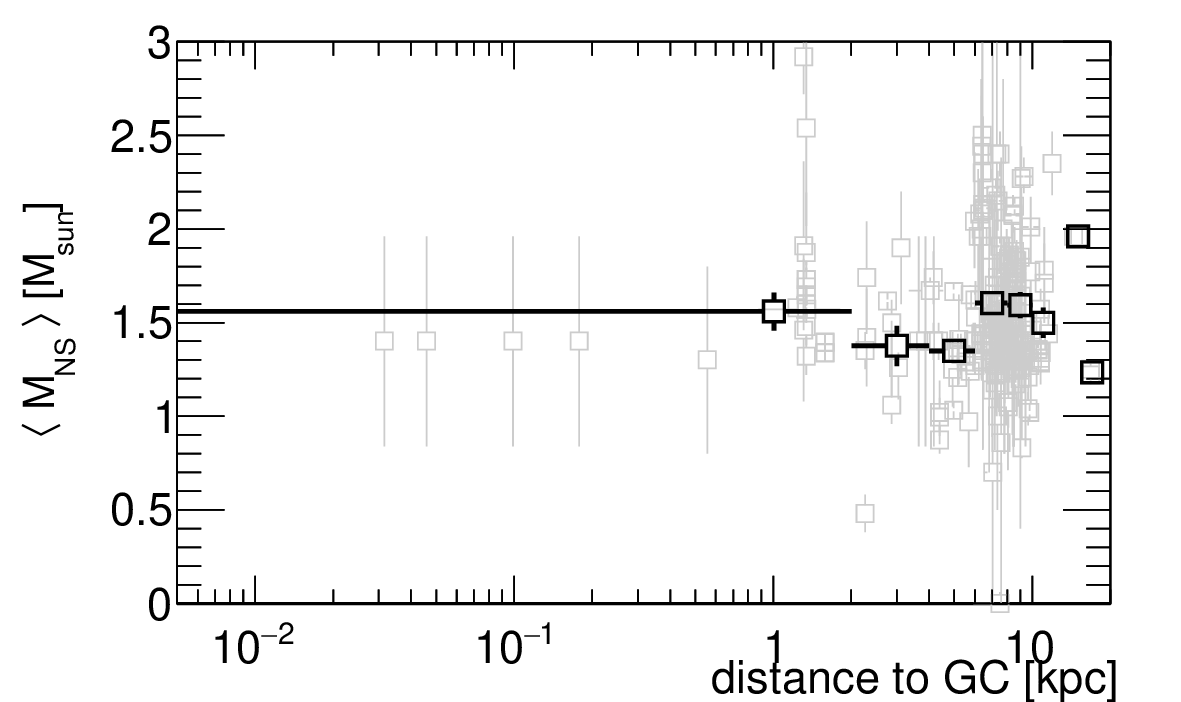}
	\caption{
		\textbf{Mass distribution of the examined NSs as a function of the distance to Galactic centre.}
		Black boxes - the average mass and radius, the vertical error bars simply reflect the statistical uncertainty at the given bin; light grey boxes - the measured masses and radii of the NSs with the measured uncertainties, see Tables \ref{tab:NSlist_part1}-\ref{tab:NSlist_part2}.	
		We show three different profiles using 2000, 500, and 200 bins from 0 to 20 kpc (from top to bottom) to plot the average mass distribution. Top: distance showen in linear scale. Bottom: in log scale.
	}
	\label{fig:YiZhong_dist2GC_corr_GalaxyMap_NS}		
\end{figure*}

\begin{figure}
	\centering
	\includegraphics[scale=0.45]{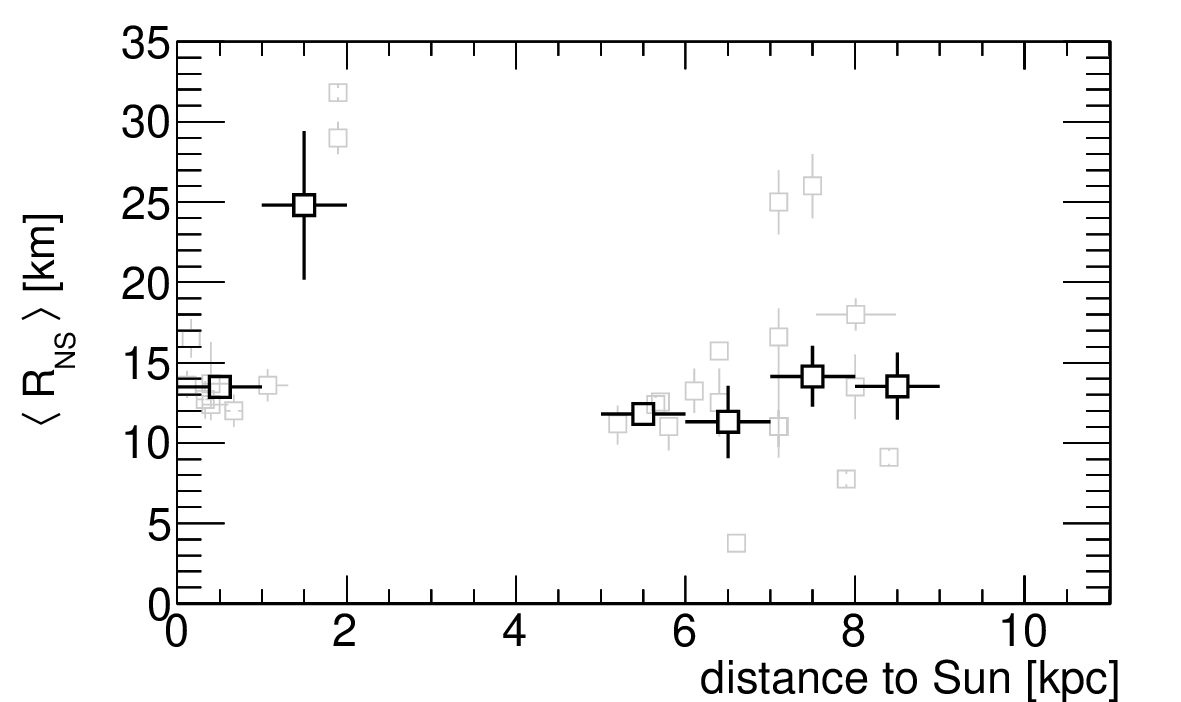}
	\includegraphics[scale=0.45]{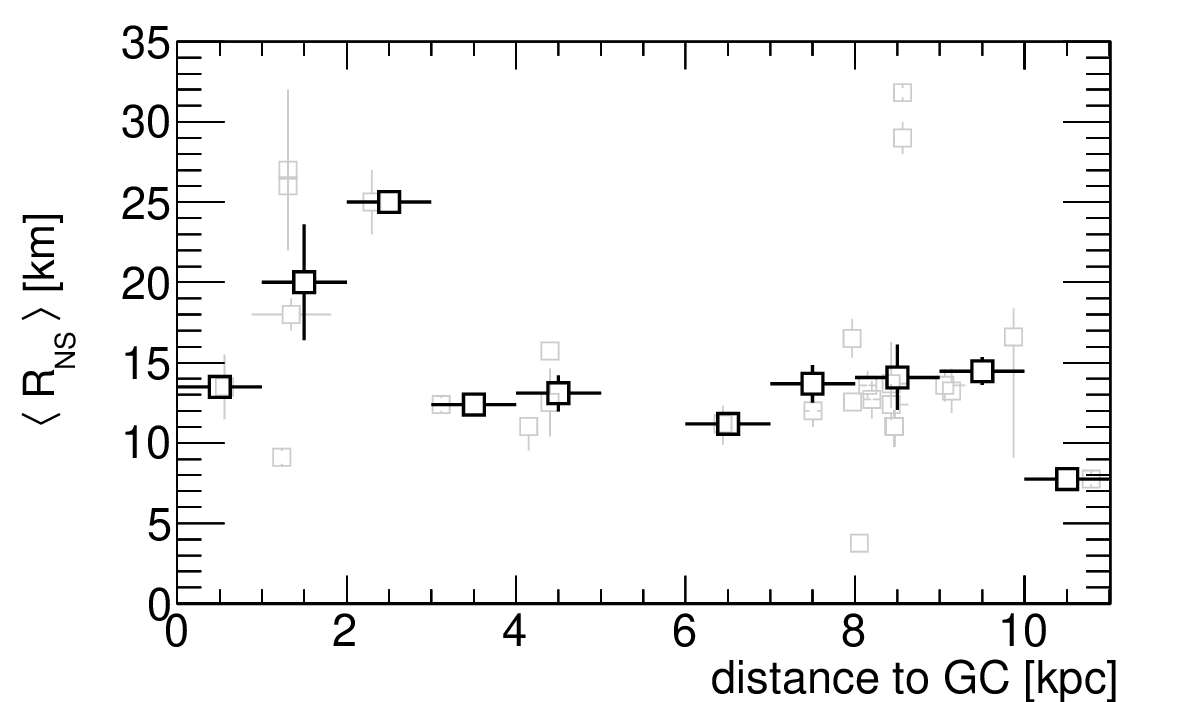}
	\caption{
		Radius distribution of the examined NSs as a function of the distance to Galactic centre. 
		Black boxes - the average mass and radius, the vertical error bars simply reflect the statistical uncertainty at the given bin; light grey boxes - the measured masses and radii of the NSs with the measured uncertainties, see Tables \ref{tab:NSlist_part1}-\ref{tab:NSlist_part2}. 
		Top: The distribution plotted with respect to the distance to Sun; 
		Bottom: same but plotted with respect to the distance to GC.
 }
\label{fig:YiZhong_NS_radius}		
\end{figure}

\begin{figure}
	\centering
	\includegraphics[scale=0.45]{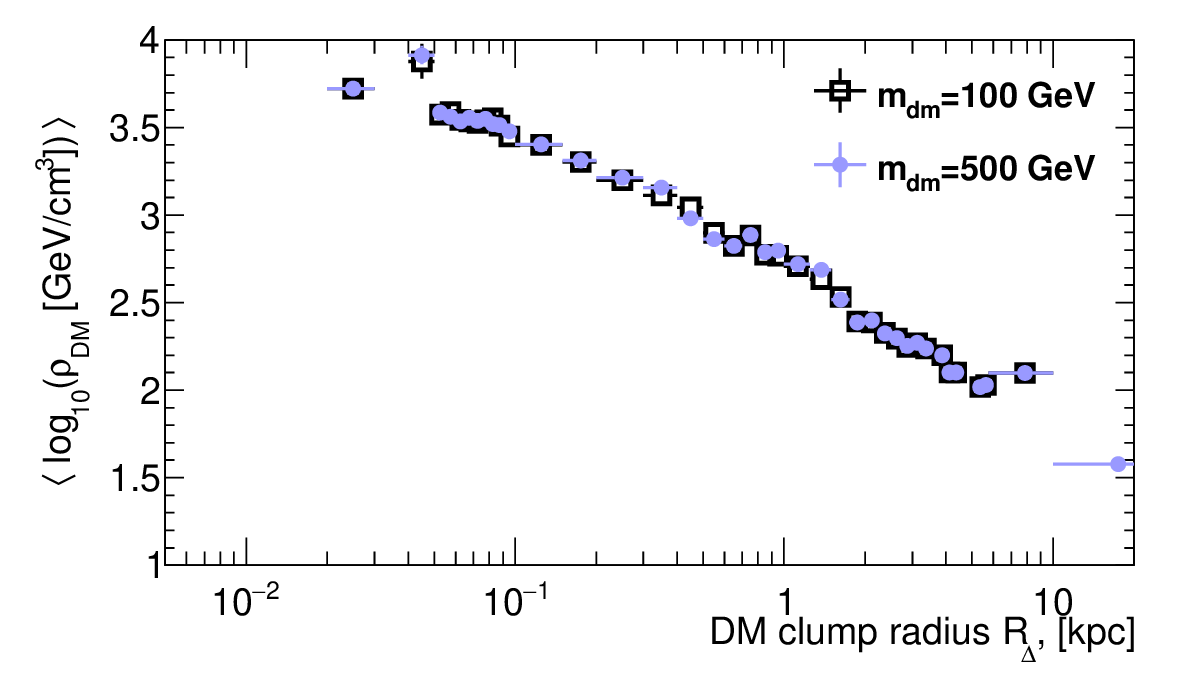}
	\includegraphics[scale=0.45]{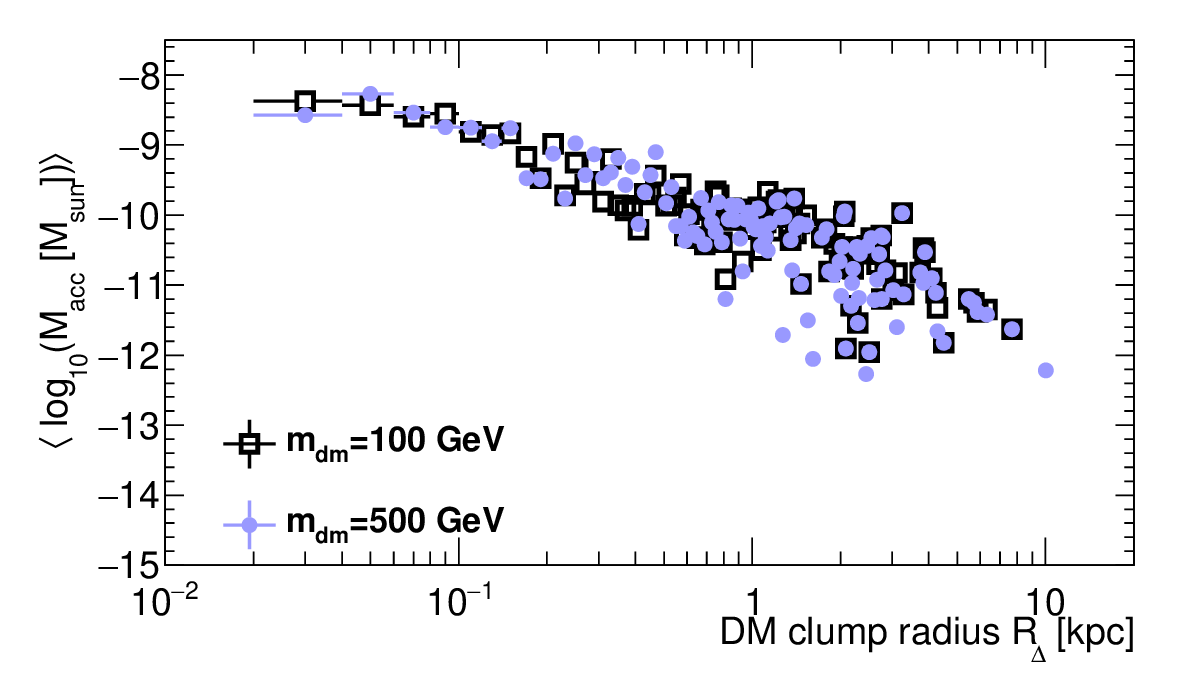}	
	\caption{
		Top: DM density distribution in a minihalo/clump. 
		Bottom: Accreted DM mass in a minihalo/clump, computed with the help of Eq.\eqref{eq:Kouv2013}, where DM density distribution in a minihalo/clump, was computed from the Einasto profile using the $\rho_{s}$ and $r_{s}$ values for the selected clump associated to the NS taken from \texttt{CLUMPY} simulation. 
		The results are shown for galaxy models assuming $m_{\rm{dm}}=100$ and 500 GeV. We used only those clumps that are associated with the NS.		
	}
	\label{fig:YiZhong_minihalo_dens}
\end{figure}

\begin{figure}
	\centering
	\includegraphics[scale=0.45]{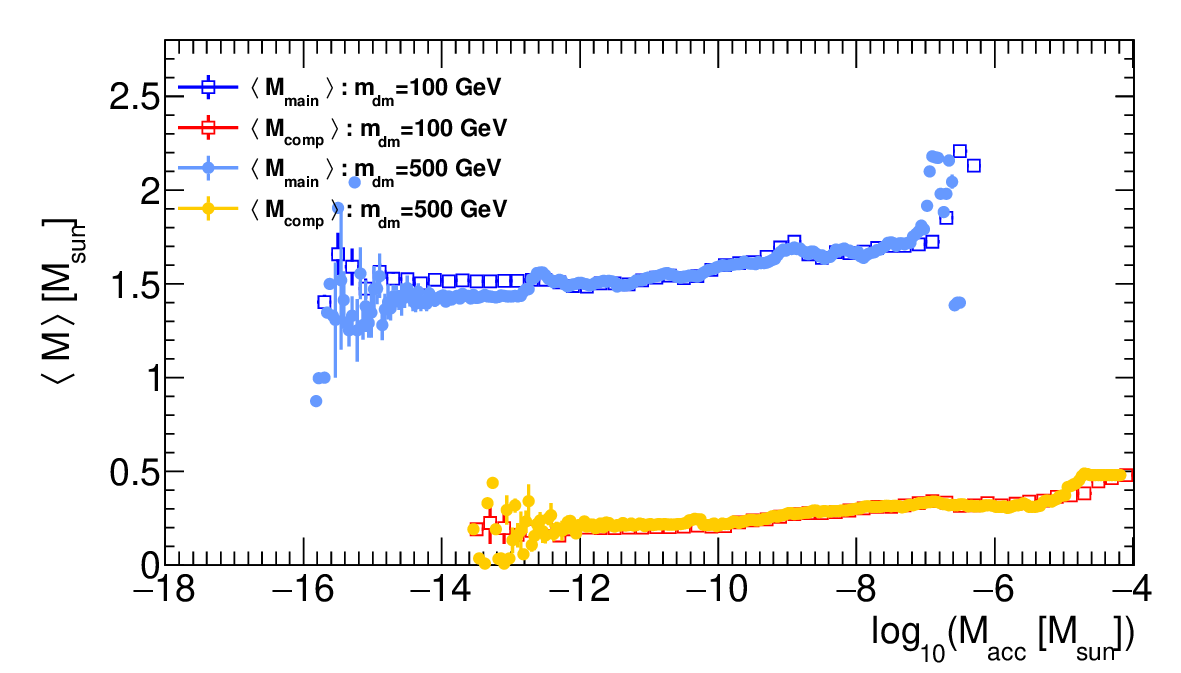}	
	\caption{	
		Change of the average star (NS/WD) mass with respect to the DM mass accreted inside the selected clumps, where the accreted mass was computed with the help of Eq.\eqref{eq:Kouv2013}. Results for the clumps obtained assuming particle mass $m_{\rm{md}}=100$ GeV (box) and 500 GeV (circles). 
		The profiles are shown for the main star (azure,blue) and companion star separately (red, orange).
	}
	\label{fig:YiZhong_DM_mass_accretion_and_NS}		
\end{figure}

\begin{figure*}
	\centering
	\includegraphics[scale=0.42]{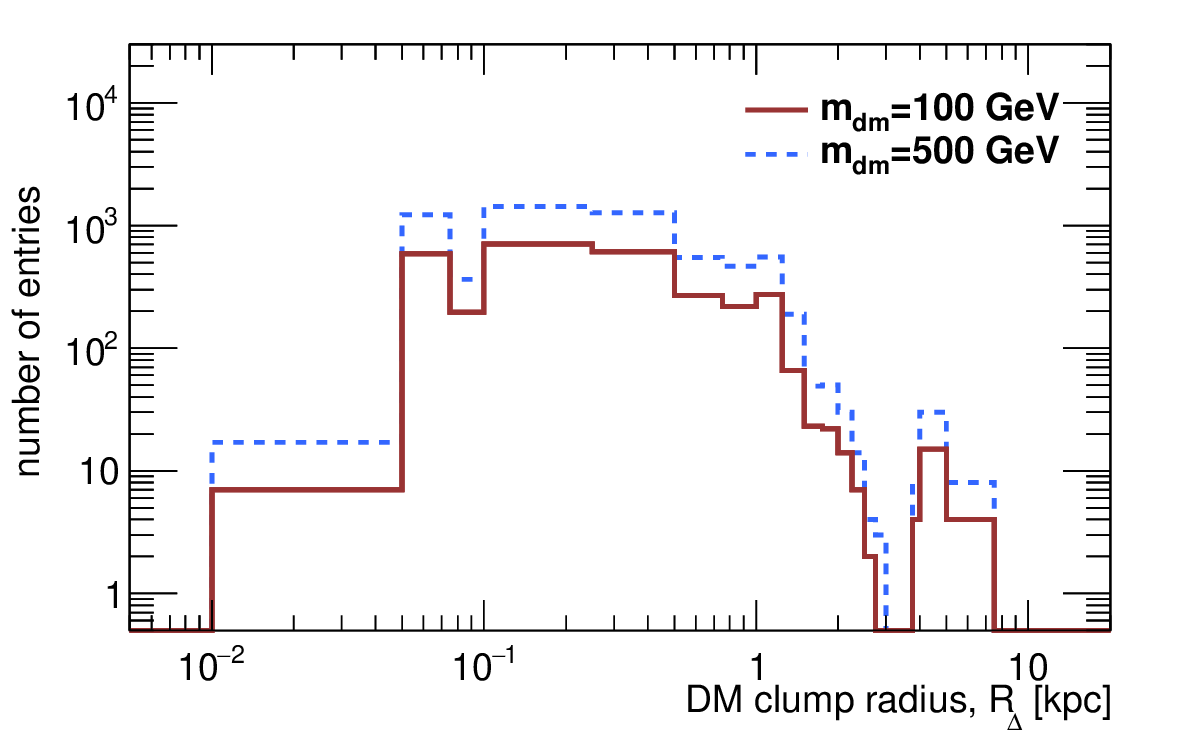}
	\includegraphics[scale=0.42]{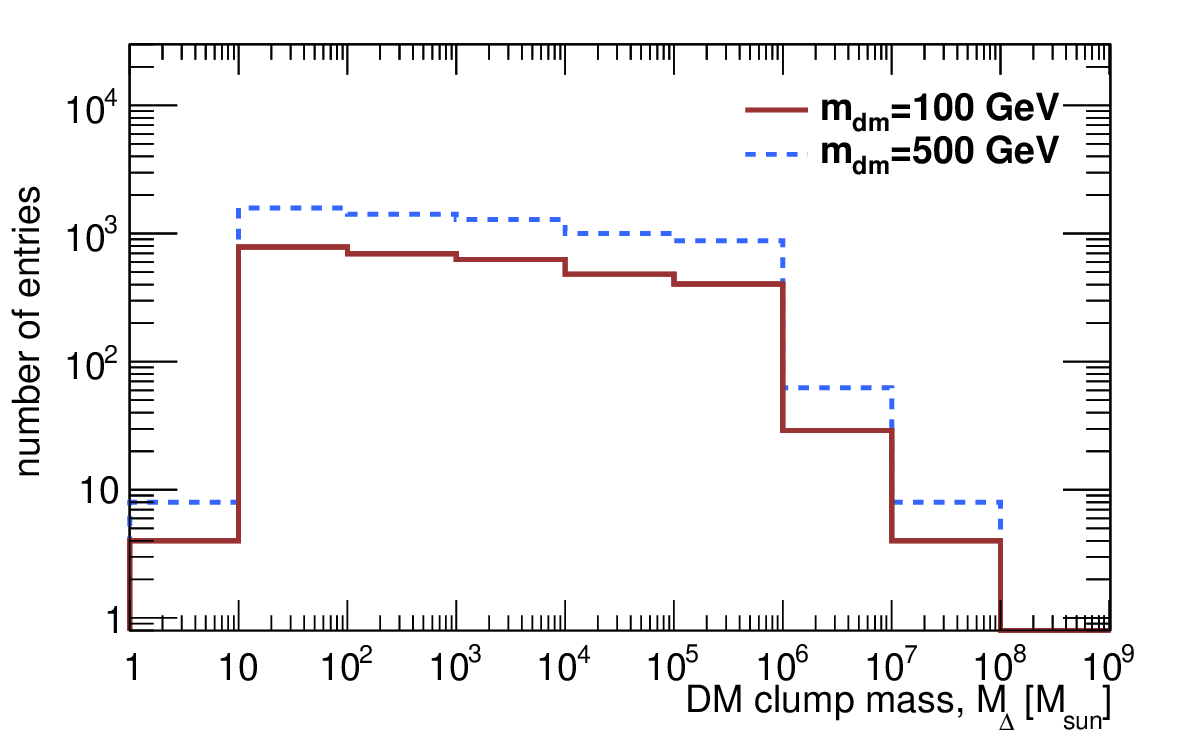}
	\caption{
		(left): The outer bound of the DM-clumps in the vicinity of the examined. 
		(right): The mass of the DM clumps in the vicinity of the examined NSs in absolute units. 
		The real-space distance between NSs and DM clumps coordinates should be less than 0.01, 0.05, 0.1, 0.25, 0.5, 0.75, 1, 2, and 5 kpc, while the scale of the selected clumps should be greater than the distance.		
	}
	\label{fig:YiZhong_DMclump_Rtidal}		
\end{figure*}

\begin{figure*}
	\centering
	\includegraphics[scale=0.42]{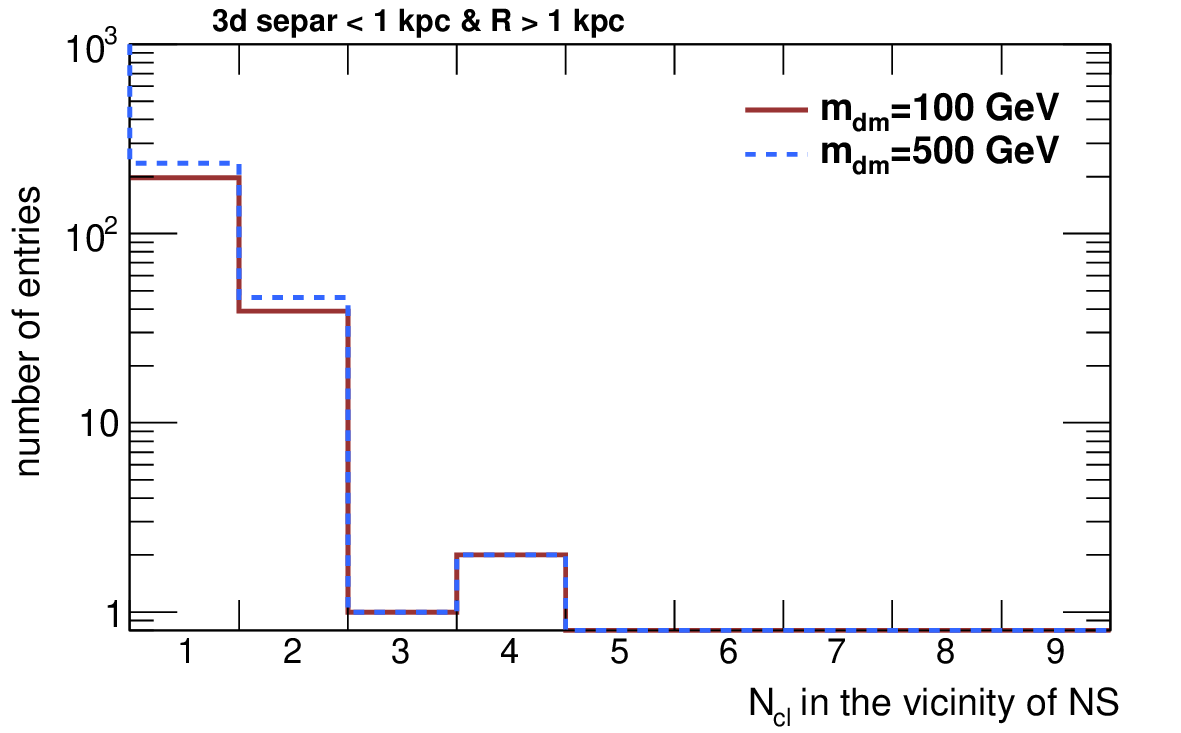}
	\includegraphics[scale=0.42]{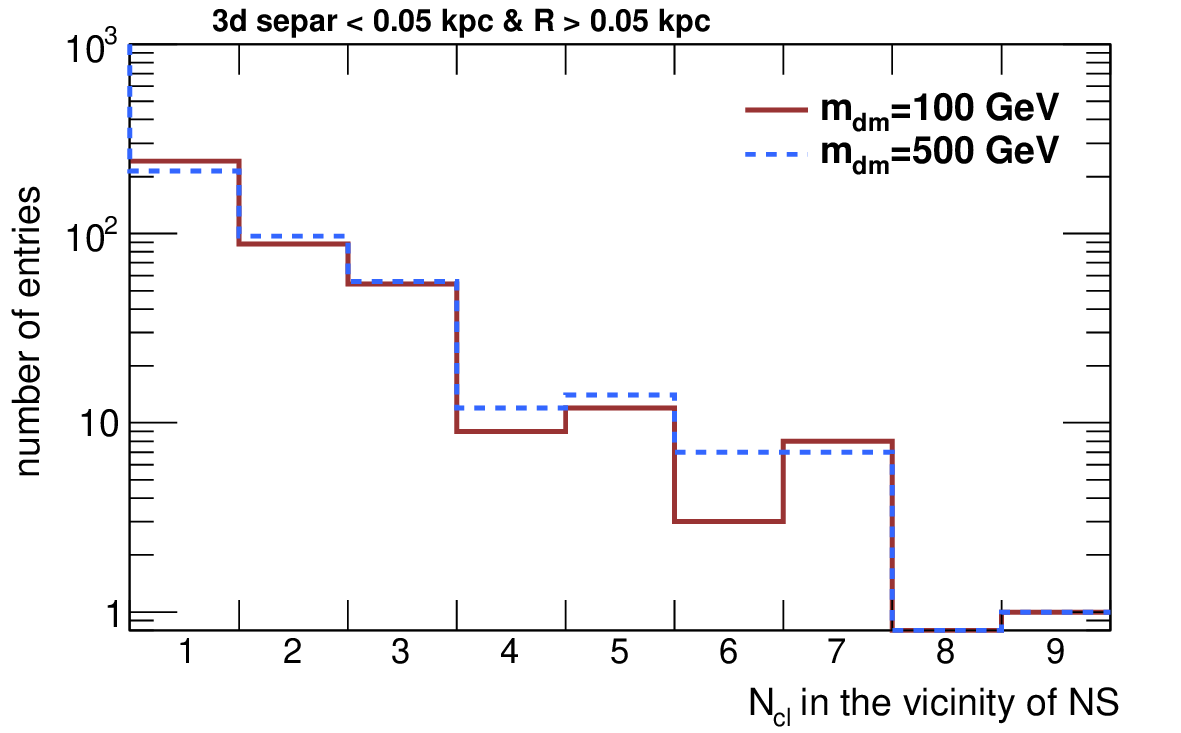}
	\caption{
		The number of DM clumps in the vicinity of the examined NSs in absolute units. 
		(left): The multiplicity of DM clumps in the vicinity of the NSs when the real space separation distance between DM clumps and NSs coordinates should be less than 1 kpc, while the scale of the selected clumps should be greater than 1 kpc. (right): The distance between NSs and DM clumps coordinates should be less than 0.05 kpc, while the scale should be greater than 0.05 kpc.		
	}
	\label{fig:YiZhong_DMclump_multiplicity}		
\end{figure*}

\begin{figure*}
	\centering		
	\includegraphics[scale=0.44]{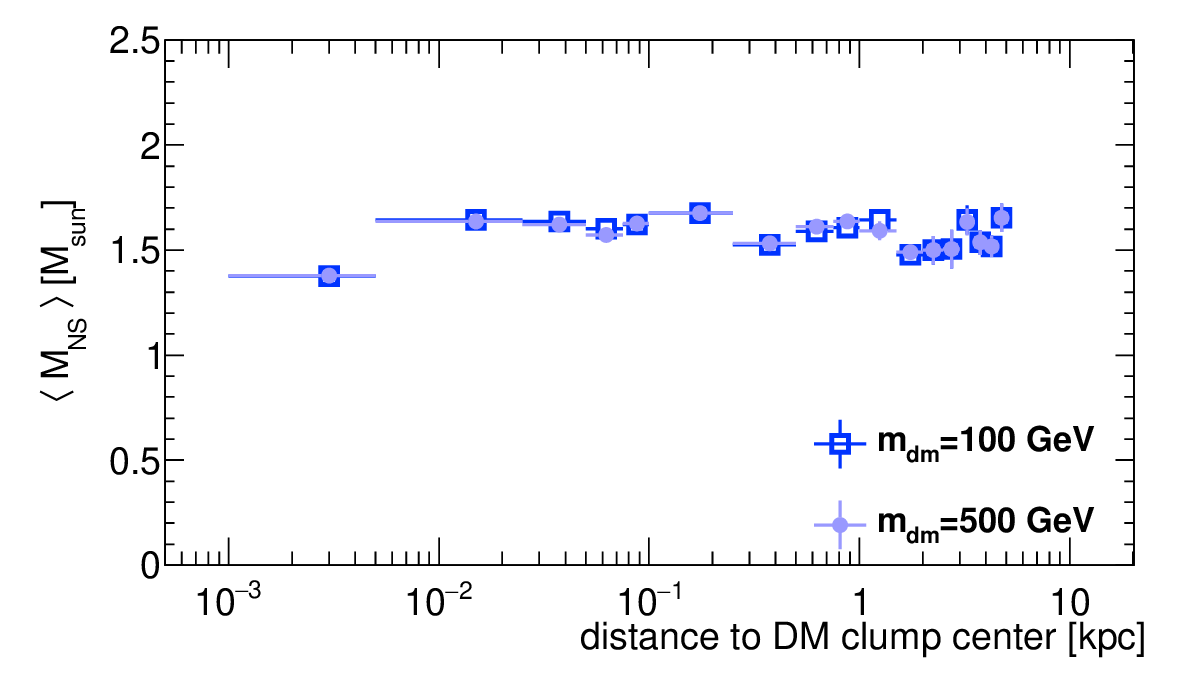}
	\includegraphics[scale=0.44]{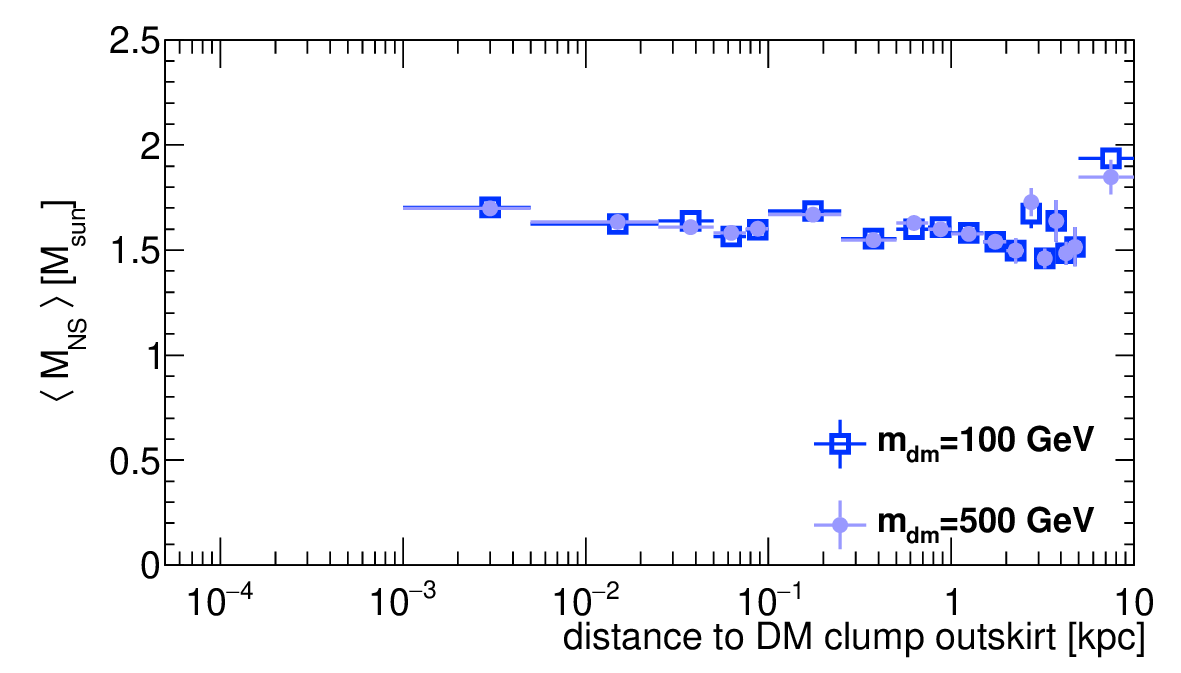}	
	\caption{
		Left: The distribution of NS mass inside the DM clumps as function of distance to the clump center.
		Right: The distribution of NS mass inside the DM clumps as function of distance to the clump outskirt.
		Results are shown for the DM clumps produced with the help of \texttt{CLUMPY} assuming $m_{\rm{md}}=100$ GeV (boxes); $m_{\rm{md}}=500$ GeV (circles).		
	}
	\label{fig:YiZhong_NSmass_vs_dist2ClumpCenter}
\end{figure*}

\begin{figure*}
	\centering		
	\includegraphics[scale=0.54]{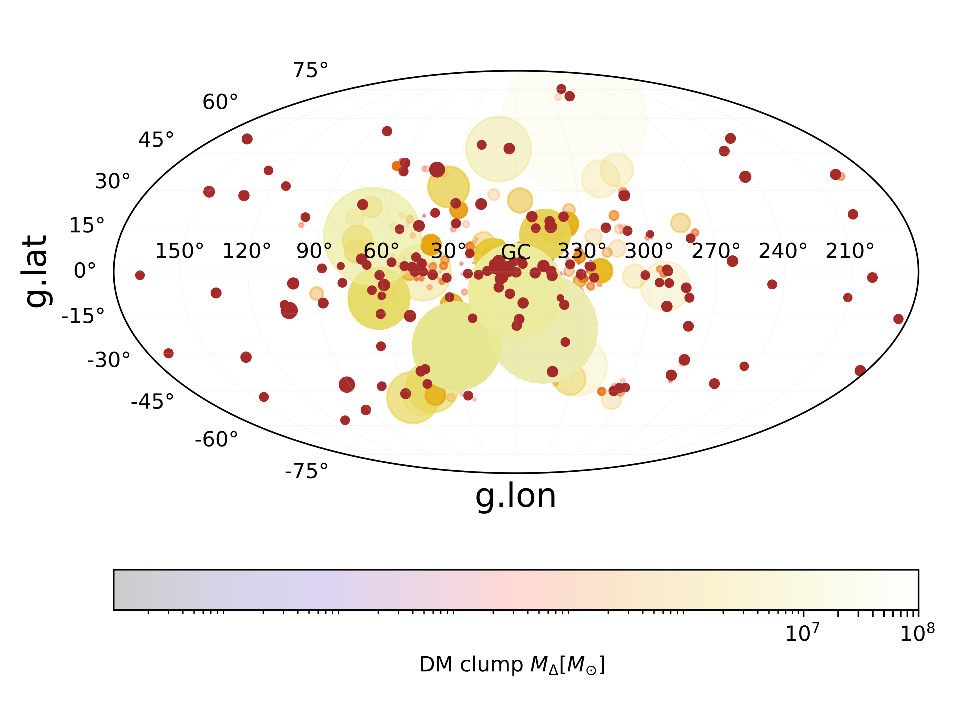}	
	\includegraphics[scale=0.54]{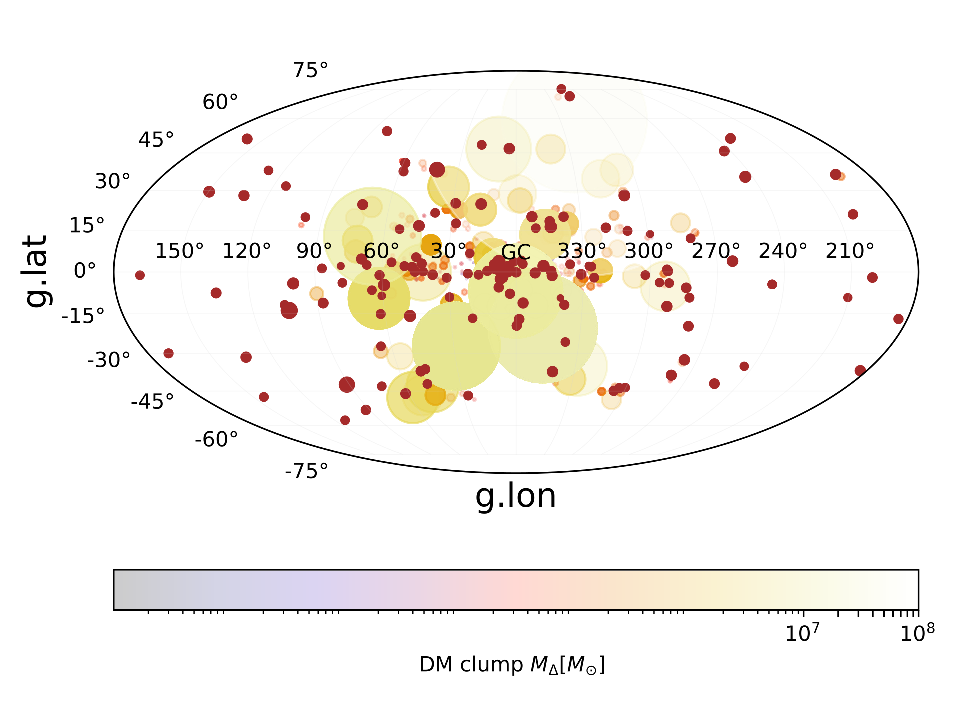}

	\includegraphics[scale=0.54]{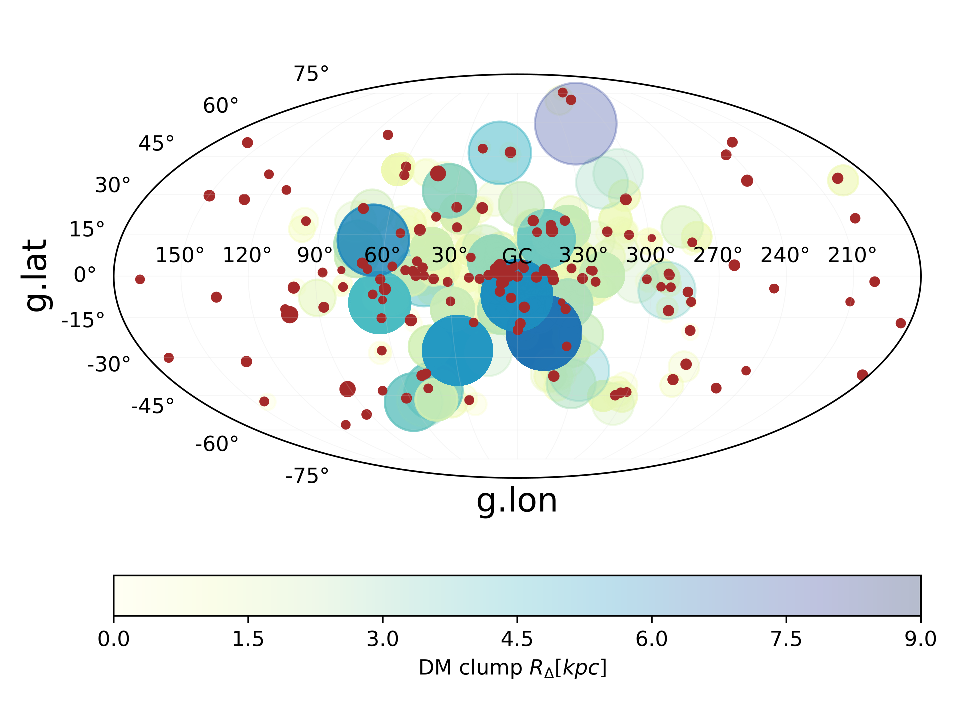}
	\includegraphics[scale=0.54]{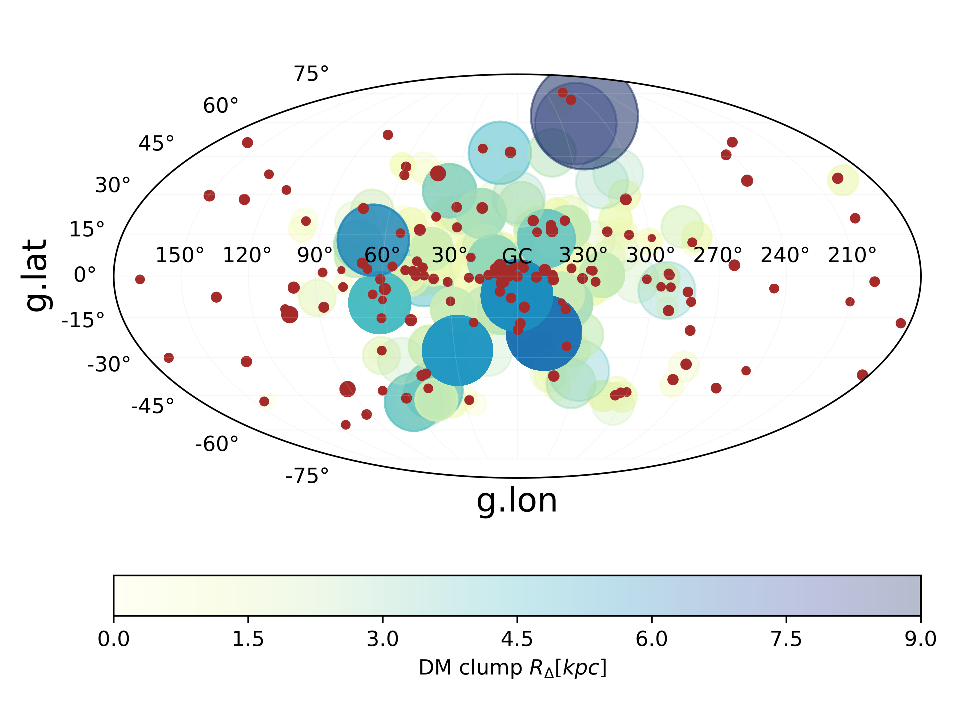}		 
	\caption{
		The distribution of DM clumps mass (top) and outer bound radius (bottom), as seen projected on the sky in Galactic coordinates. 
		DM clumps were selected with respect to the 3d-distance and scale. Namely, the 3d-separation distance should be less than 0.05 kpc, and the radius of the selected clump should be greater than 0.05 kpc.  
		The circle size indicates the mass of the DM clump (top), and the scale of the DM clumps (bottom). Red data points indicate known NSs. 
		(left) \texttt{CLUMPY} simulation using $m_{\rm{md}}=100$ GeV; (right) $m_{\rm{md}}=500$ GeV.		
}
\label{fig:YiZhong_GalaxyMap_DMclumps}		
\end{figure*}

\begin{figure*}
	\centering		
	\includegraphics[scale=0.44]{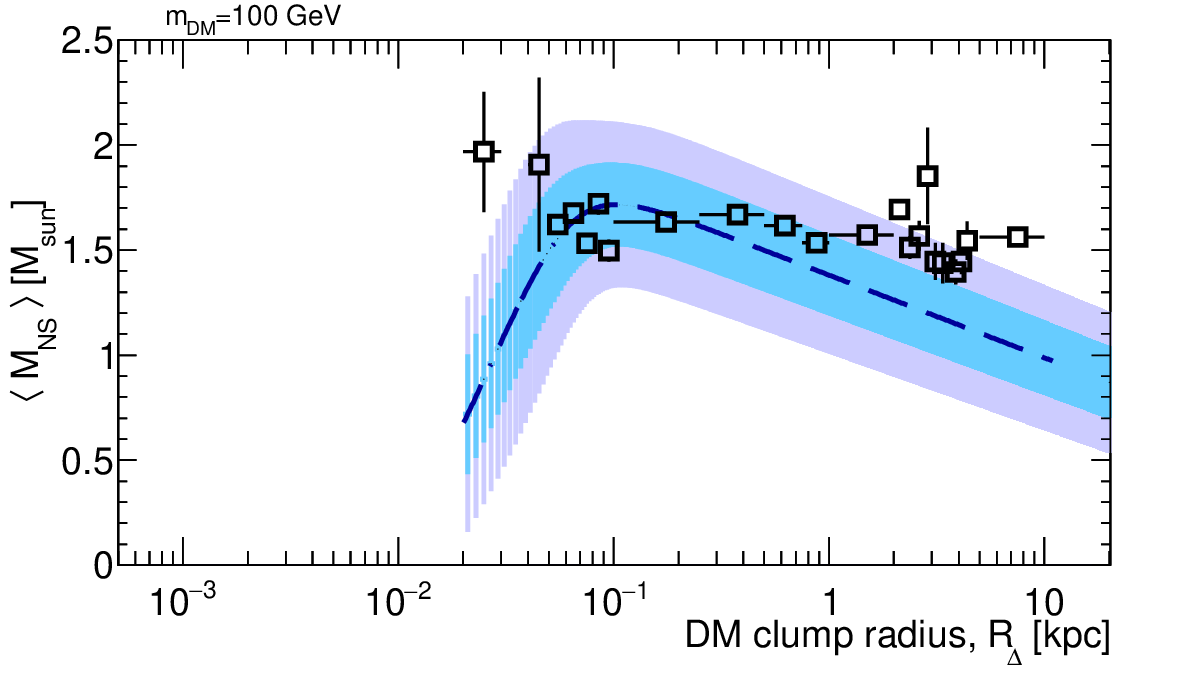}	
	\includegraphics[scale=0.44]{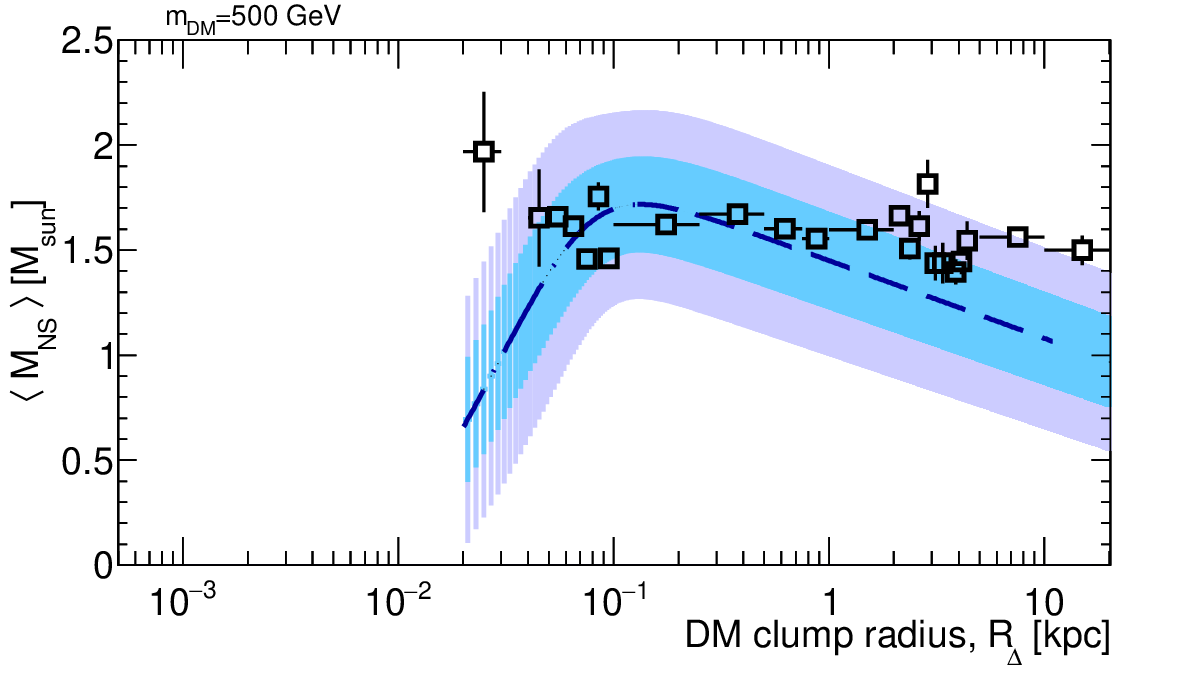}	

	\includegraphics[scale=0.44]{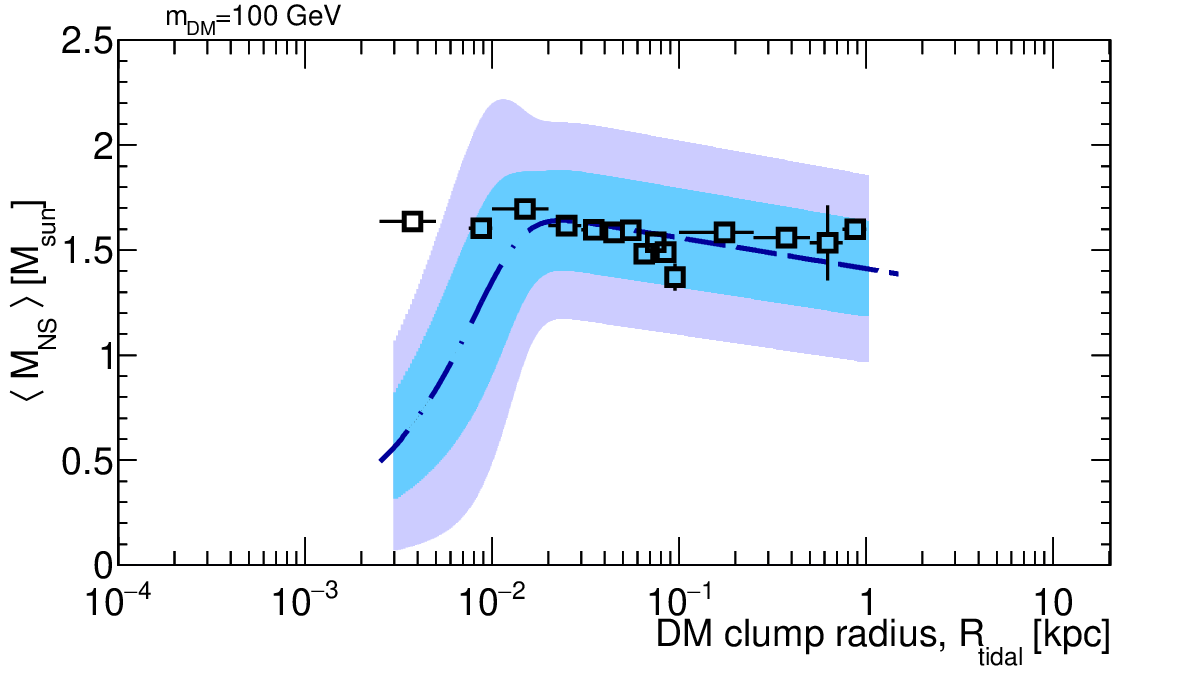}
	\includegraphics[scale=0.44]{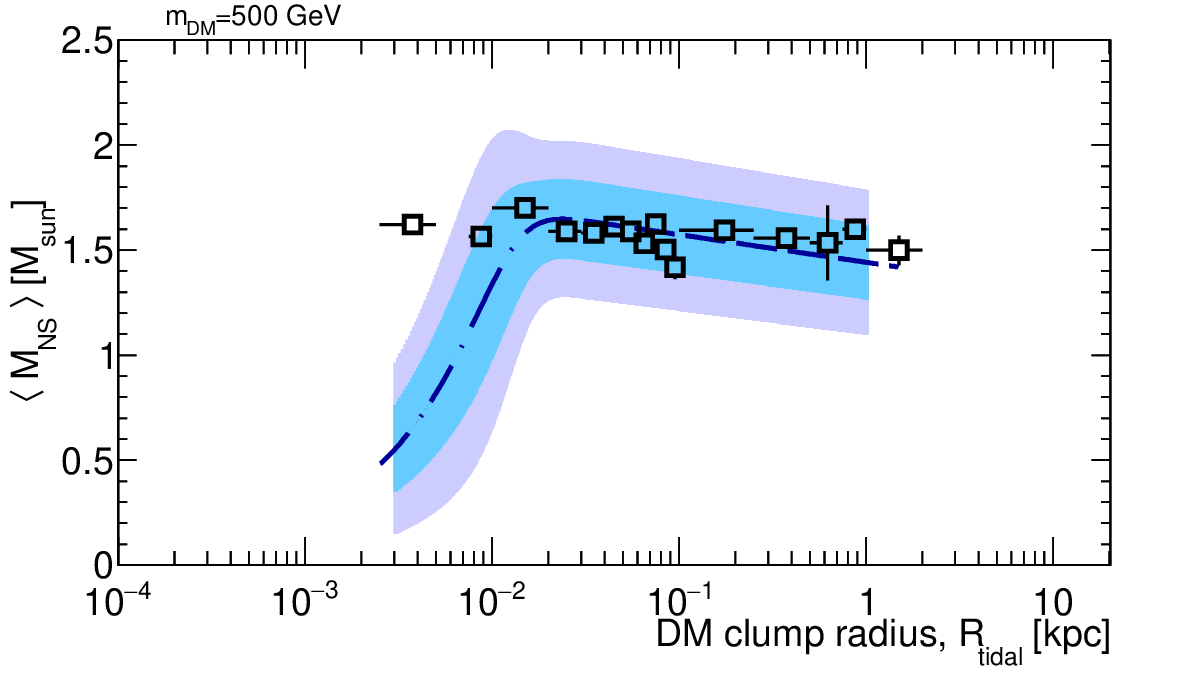}	
	\caption{
		Fits to the NSs mass as a function of the clump radius (top) and tidal clump radius (bottom). Clumps were selected in the vicinity of 0.01, 0.05, 0.1, 0.25, 0.5, 0.75, 1, 2, and 5 kpc to the NSs, clumps scale was requested to be greater than this separation distance. Left: using the DM mass $m_{\rm{dm}}=100$ GeV. 
		Right: using the DM mass $m_{\rm{dm}}=500$ GeV.		
	}
\label{fig:YiZhong_fits_NSmass_evolution}		
\end{figure*}

\begin{figure*}
	\centering	
	\includegraphics[scale=0.44]{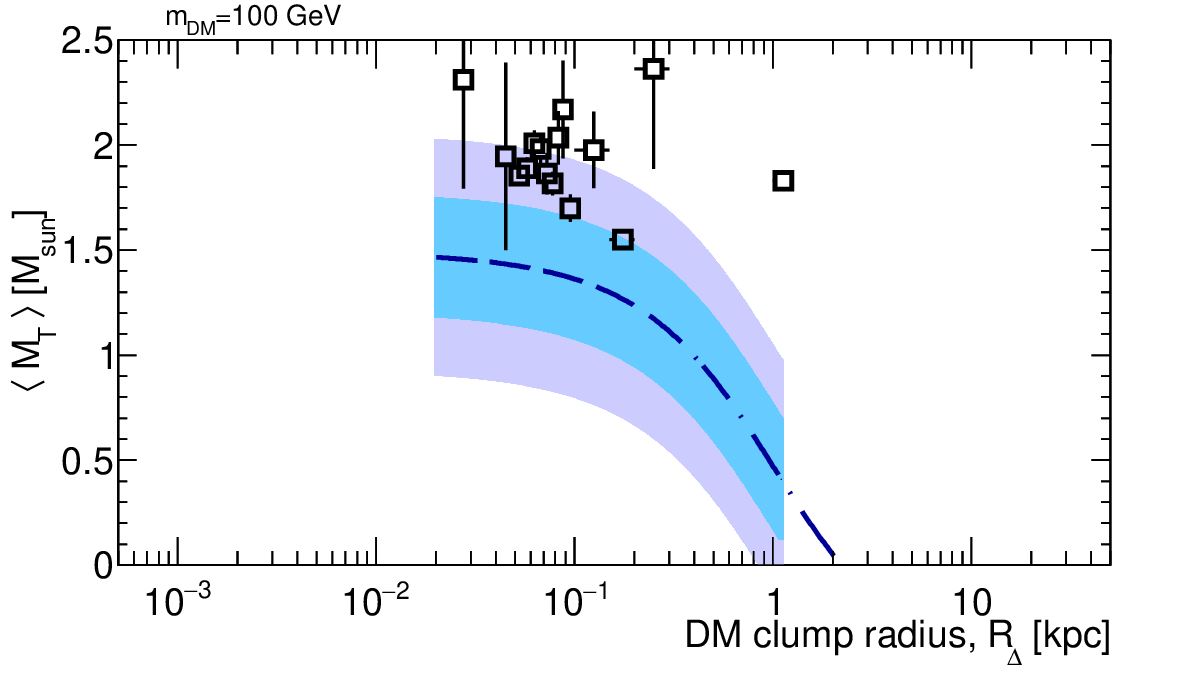}
	\includegraphics[scale=0.44]{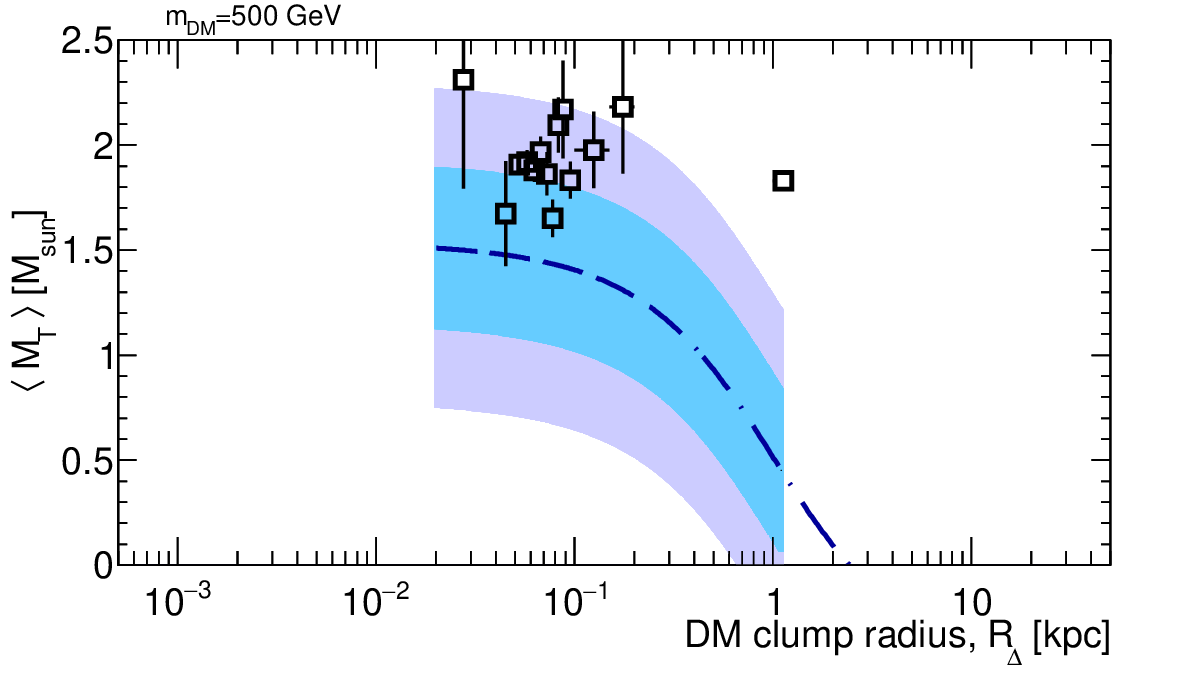}

	\includegraphics[scale=0.44]{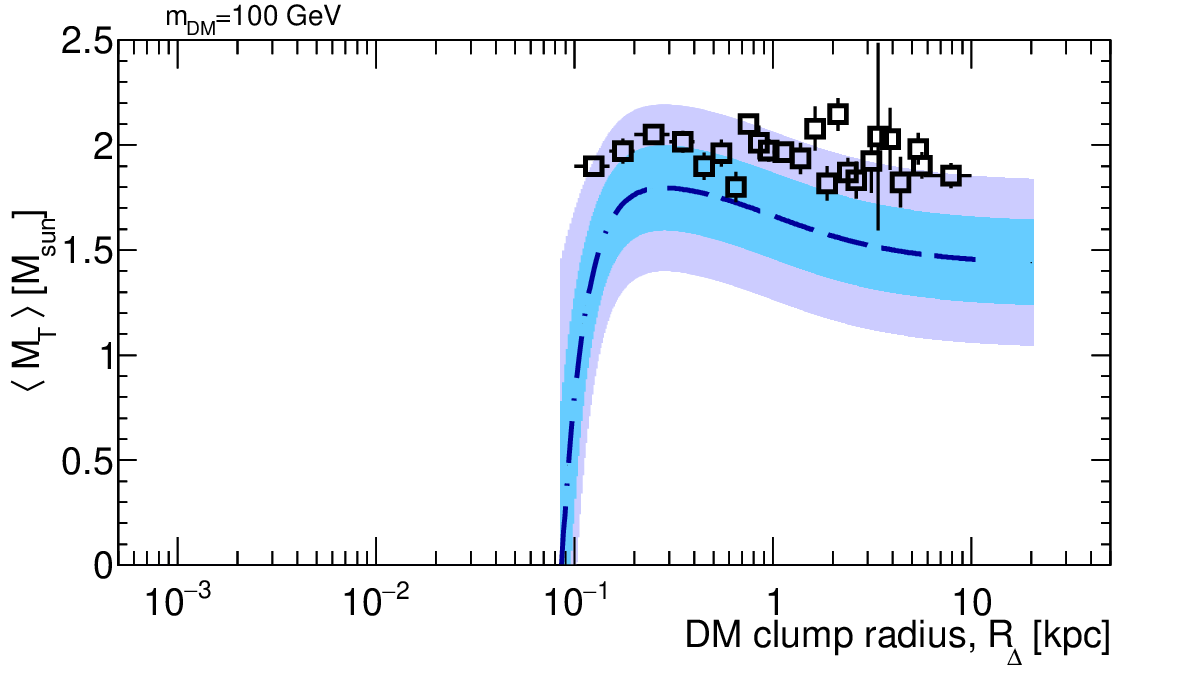}
	\includegraphics[scale=0.44]{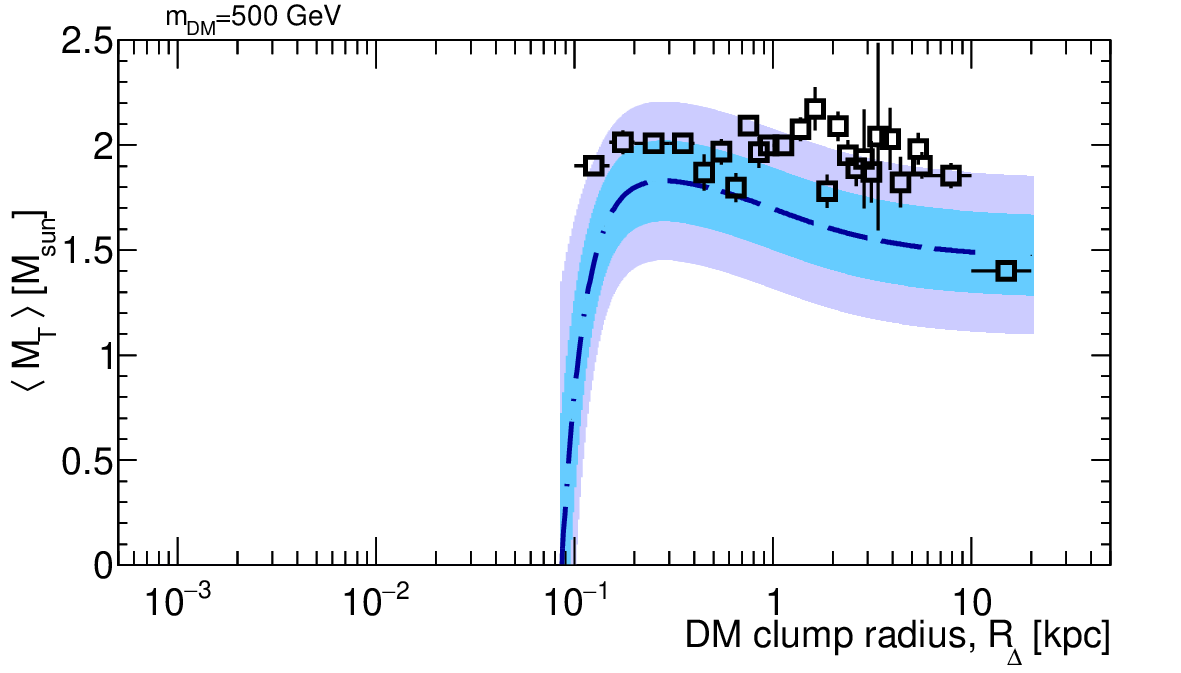}	
	\caption{
		Fits to the total mass of the binary systems as a function of the clump radius. Top: Clumps were selected in the vicinity of 0.05 kpc to the NSs, clumps scale was requested to be greater than this separation distance ($> 0.05$ kpc). Bottom: Clumps were selected in the vicinity of 0.1 kpc to the NSs, clumps scale was requested to be greater than this separation distance ($> 0.1$ kpc). 
		Left: using the DM mass $m_{\rm{dm}}=100$ GeV. 
		Right: using the DM mass $m_{\rm{dm}}=500$ GeV. Fits performed with the help of the power law function.		
}
\label{fig:YiZhong_fits_Totalmass_evolution}		
\end{figure*}

\bsp	
\label{lastpage}
\end{document}